\documentclass[aip,graphicx]{revtex4-1}
\usepackage{amsthm}
\usepackage{amsmath}
\usepackage{graphicx}
\usepackage{dcolumn}
\usepackage{bm}

\usepackage[utf8]{inputenc}
\usepackage[T1]{fontenc}
\usepackage{mathptmx}
\usepackage{multirow}
\usepackage{color}
\usepackage{float}
\usepackage[caption=false]{subfig}
\usepackage{caption}
\draft 

\begin{document}
	
	\title[Particle-laden turbulent Couette flow]{Dynamics  of particle-laden turbulent Couette flow. Part1: Turbulence modulation by inertial particles}
	
	\author{S.Ghosh}
	\altaffiliation[Also at ]{Department of Chemical Engineering,
		\\Indian Institute of Technology, Bombay}
	\email{swagnikg90@gmail.com}
	\author{P.S.Goswami}%
	\email{(for correspondence) psg@iitb.ac.in}
	\affiliation{Department of Chemical Engineering,
		\\Indian Institute of Technology, Bombay
		\\Powai, Mumbai-400-076, India
	}%
	
	\date{\today}
	
\begin{abstract}
In particle-laden turbulent flows it is established that the turbulence in carrier fluid phase gets affected by the dispersed particle phase for volume fraction above $10^{-4}$ and hence reverse coupling or two-way coupling becomes relevant in that volume fraction regime. 
Owing to their greater inertia, larger particles tend to change either the mean flow or the intensity of fluid phase fluctuations. In a very recent study \citep{muramulla2020disruption} a discontinuous decrease of turbulence intensity is observed in a vertical particle-laden turbulent channel-flow for a critical volume fraction O($10^{-3}$) for particles with varying Stokes Number in the range of 1-420 based on the fluid-integral time-scales. The collapse of turbulent intensity is found out to be a result of ’catastrophic reduction of turbulent energy production rate’. Mechanistically, particle-fluid coupling in particle-laden turbulent Couette-flow differs from that in a closed channel flow due to different fluid coherent structures and different particle clustering behaviours and these act as the motivation in investigating the existence of continuous or discontinuous transition in turbulence modulation by inertial particles in particle-laden turbulent Couette flow. In this article, the turbulence modulation in the fluid phase by inertial particles is explored through two-way coupled DNS of particle-laden sheared turbulent suspension where particle volume fraction ($\phi$) is varied from $1.75\times10^{-4}$ – $1.05\times10^{-3}$ and Reynolds Number based on half-channel width ($\delta$) and wall velocity ($U$) ($Re_{\delta}$) is kept at 750. The particles are heavy point particles with $St\sim367$ based on fluid integral time-scale represented by $\delta/U$. A discontinuous decrease of fluid turbulence intensity, mean square velocity and Reynolds stress is observed beyond a critical volume fraction $\phi_{cr}\sim7.875\times10^{-4}$. The drastic reduction of shear production of turbulence and in turn the reduction of viscous dissipation of turbulent kinetic energy are two important phenomena for the occurrence of discontinuous transition similar to channelflow. The step-wise particle injection and step-wise removal study confirms that it is the presence of particles which is majorly behind this discontinuous transition. Additionally, the effect of inelastic collisions are explored and found to increase the $\phi_{cr}$ slightly, although the nature of turbulence modulation remained similar to when the collisions are elastic. The explicit role of the inter-particle collisions is studied by simulating a hypothetical case where only inter-particle collisions are kept switched off. In this case, $\phi_{cr}$ increases more than the inelastic case. The turbulence modulation carries the signatures of transition from sheared turbulence to particle-driven fluid fluctuation. The increase in $\phi_{cr}$ for different collisional cases are found out to be the result of decrease in reverse force acting on the fluid at a fixed volume fraction less than $\phi_{cr}$.
	\end{abstract}
	
	
	\maketitle 
	
\section {Introduction}
   In particle-laden turbulent flow, the strong  coupling that exists between particles
   and the fluid phase, \textcolor{black}{has enormous impact on controlling the dynamics of both the phases. To understand the transport problems like} 
   stresses, heat-transfer and mass-transfer, it is essential to investigate the dynamics of both the phases. Fluctuations along with the mean flow play an important role in \textcolor{black}{determining the transport coefficients} . In the backdrop of wall-bounded turbulent flows, fluctuating kinetic energy fluxes and hence the energy transfer \textcolor{black}{has been investigated} at different wall-normal locations of the geometry in physical space \citep*{marati2004energy} and in spectral domain \citep*{andrade2018analyzing}. 
   In the two-phase interaction, \textcolor{black}{besides the interaction between mean flows, interaction between fluctuating velocities of both the phases include additional complexities.}
   \\ Turbulent fluid velocity and vorticity fluctuations and the inter-particle collisions determine the translational and rotational velocity fluctuations in the particle phase. 
\textcolor{black}{When the particle inertia is very low, they follow the fluid streamlines and behave like passive tracer. At an intermediate inertia, preferential concentration of the particles happens at low vorticity and high strain zones \citep{squires1991preferential,fessler1994preferential}}  Particles with high Stokes number do not tend to follow local streamlines rather they move across streamlines because of their inertia being greater than that of fluid. Due to this, larger particles tend to change either the mean flow or the intensity of fluid phase fluctuations.
   \\\textcolor{black}{Turbulent Couette flow is a protoype of canonical wall-bounded flow.} 
   Mechanistically, particle-fluid coupling in particle-laden turbulent Couette-flow differs from that in a closed-channel flow due to different fluid coherent structures and different particle clustering behaviours. \textcolor{black}{\citet*{pirozzoli2011large} performed DNS simulations of unladen turbulent Couette flow in high Reynolds number regime ($Re_c$
   based on wall-velocity and half-channel-width: 3000-21333). Their work showed that for Couette flow, at extreme Reynolds number, a log-law layer could not be found unlike channel flow where log-law layer exists at infinite Reynolds number. Also a unique feature observed in Couette flow was the existence of a secondary peak of streamwise velocity fluctuation profile in the outer-layer which corresponds to the elongated streak and roller structure with dimensions that of the channel-geometry}. Using two-way coupled DNS of turbulent Couette flow, \citet*{wang2019modulation} investigated the effect of particle inertia on turbulence modulation and on the turbulence regeneration cycle \textcolor{black}{for $Re_b=500$ (Reynolds number based on half the relative velocity of the walls and half channel-width), volume fraction ($\phi_v$) less than $10^{-3}$ and Stokes number $St_{turb}$ (Stokes number based on the turnover time of the large scale vortex ) in the range 0.056-5.56}. Streamwise coupling of the phases along with spatial distribution of high-inertial particle played a key-role in stabilizing the turbulence. The turbulence attenuation was mainly a result of increased momentum dissipation due to particle-phase drag disrupting the large vortical structures. Low-inertial particles were observed to trigger laminar-turbulence transition through strengthening of the large-scale vortical structure. In a turbulent Couette-Poiseuille flow configuration, with zero skin friction, in absence of inter-particle collisions and two-way coupling, inertial particles were found to cluster up differently near the two walls as a result of vorticity-strain rate selection mechanism  \citep*{yang2017preferential}. 
   Randomly oriented particle clusters are observed near the moving wall similar to Isotropic turbulence . On the other-hand streaky particle structures, signature of wall-trubulence, were observed near the stationary wall. Anisotropy in fluctuations were due to different coherent flow structures observed in different walls.
   In a very recent study of particle laden turbulent channel-flow, a discontinuous decrease of turbulence intensity was observed for a critical volume fraction O($10^3$) \citep{muramulla2020disruption}. The phenomenon was observed for particles with varying Stokes Number in the range of 1-420 based on the fluid-integral time-scales. The collapse of turbulent intensity was found out to be a result of 'catastrophic reduction of turbulent energy production rate'.
   \\In vertical channel-flow, at higher volume loading, the drag force exerted by the particles to the fluid alters the fluid pressure field and hence the net driving force. The horizontal turbulent shear flow (present case), unlike vertical channel-flow, pressure drop does not act as a driving force. Along with these differences, the very mechanistic difference of vortical coherent structures coupled with clustering behaviours of inertial particles act as the motivation in investigating the existence of continuous or discontinuous modulation in turbulence modulation by inertial particles in particle-laden turbulent Couette flow.
  \\ In this article we \textcolor{black}{investigate} the effect of particle feedback force on the fluid phase turbulence through two-way coupled DNS of particle-laden turbulent shear flow. The nature of the modulation is explored through analyzing fluid phase velocity statistics. An attempt \textcolor{black}{to understand the underlying mechanism} of the modulation, is made by detailed analysis of the various terms of mean momentum, mean kinetic energy and turbulent kinetic energy budget equations along with the analysis of streamwise velocity and vorticity fluctuation field. Additionally, the effect of inter-particle collision on the turbulence modulation is studied. In this regard, the ideal elastically colliding particle-laden turbulent suspension is compared with the suspension having inter-particle and wall-particle collisions \textcolor{black}{which are} slightly inelastic. The explicit role of the inter-particle collisions is studied by simulating a hypothetical case where only inter-particle collisions are switched off.
  
  
  \section{Simulation Methodology}
 \label{sec:sim_meth}
\begin{figure}[h]
		\centering
\includegraphics[scale=0.8]{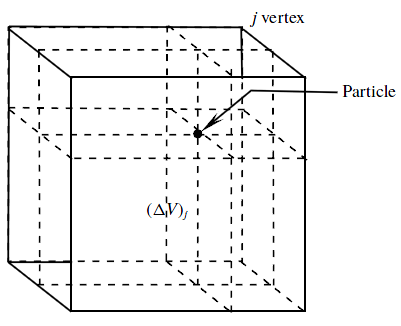} 
 \caption{The Projection of Nearest Neigbour (PNN) method for projecting the feed-back force of a particle located within the volume $\Delta V$ onto the vertices of the volume of a grid $^4$}
 \label{PNN}
 \end{figure}

   The flow system is simulated in Euler-Lagrange framework by two-way coupled Direct Numerical Simulation.  
  \begin{figure}[h]
  	\centering
  	\includegraphics[scale=0.8]{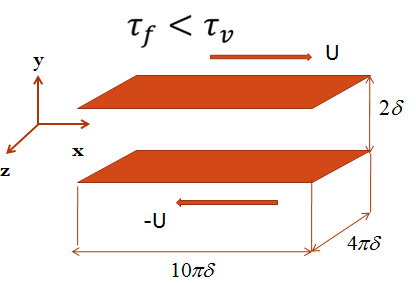}
  	\caption{Schematic of the system}
  	\label{fig:schematic1}
  \end{figure}
   The particles are considered to be spherical, sub-Kolomogorov sized heavy inertial with diameter
  of $39\mu m$ having a density of 4000 kg/$m^3$ with a Stokes Number St about $367$ based on the fluid integral time scale $\delta/U$. The dimension of the simulation box is $10\pi\delta\times2\delta\times4\pi\delta$ where $\delta$ is the half-channel-width. Upper and the lower walls of the Couette-flow move with positive and negative velocities in the x-directions respectively. The  y-axis is along the cross-stream (
  \textcolor{black}{wall-normal}) direction, and z-axis indicates the spanwise or vorticity direction (Fig. \ref{fig:schematic1}). The origin is placed at the bottom-wall such that the upper wall remains at $y=2\delta$. The number of grids used in x, y, and z directions are 120, 55, and 90  respectively with the grid resolution of  $13.8\times1.92\times7.36$ in viscous \textcolor{black}{unit}. Here, the fluid phase is considered to be air at ambient condition. The Reynolds Number based on half-channel width and wall-velocity is kept at $750$. The simulations are performed in moderately dense limit by varying the volume fractions from $1.75\times10^{-4}$ to $1.05\times10^{-3}$ by changing the number of particles from $2000$ and $12000$ in the system. The assumption of binary inter-particle collision remains valid in this volume fraction regime.
  \\ Fluid phase is described by continuity  and the Navier Stokes equations (equations \ref{eq_cont} \& \ref{eq_NS}). Both the equations are solved in Eulerian grids.	
   \begin{equation} 
	\nabla\cdot\mathbf{u}=0
	\label{eq_cont}
	\end{equation}
	\begin{equation}
	\frac{\partial\mathbf{u}}{\partial t}+\mathbf{u}\cdot\nabla \mathbf{u}=-\frac{1}{\rho_{f}}\nabla p+ \nu \nabla^2 \mathbf{u} + \frac{1}{\rho_f}\mathbf{f}(\mathbf{x},t)
	\label{eq_NS}
	\end{equation}
	Here, $\mathbf{u}(\mathbf{x},t)$ represents three dimensional instantaneous fluid velocity field as function of position $\mathbf{x}$. $p(\mathbf{x},t)$ denotes the  instantaneous pressure field, $\nu$ and $\rho_f$ are the kinematic viscosity and the density of the fluid respectively. In our isothermal simulations fluid is considered to be incompressible. $\mathbf{f(\mathbf{x},t)}$ represents the reverse feed-back force on the fluid due to \textcolor{black}{the presence of} particles.  
	The flow-field is considered to be periodic along x and z direction, whereas along y-direction is bounded by the walls where no-slip and no-penetration boundary condition is applied (Fig. \ref{fig:schematic1}). The fluid flow field is  solved using Direct Numerical Simulation (DNS) in a  pseudo-spectral framework. Detailed numerical scheme, interpolation method for fluid velocity at the particle location, and correction of the velocity field for the calculation of the drag on the particle has been described in \textcolor{black}{\citet{goswami2010particle1}, \citet{muramulla2020disruption} and \citet{Pradeep_2022}}  .
  \\The reverse feed-back force by the particle on the fluid is computed using Projection of Nearest Neighbour method (PNN) as represented in figure \ref{PNN}.
  Point-particle forcing is used to represent the feed-back force. The feed-back due to single particle is distributed only to the eight nearest neighbour points of the fluid grids occupying the vertices of a parallelepiped in a volume weighted method discussed in \citet{muramulla2020disruption}. For a steady flow the reverse force can be mathematically represented as:
  \begin{equation}
    \mathbf{f(\mathbf{x})}=-\frac{1}{\Delta V}\sum_{j=1}^{8}(\mathbf{F_i^D}+\mathbf{F_i^L})(\Delta V)^{\prime}_j\delta(\mathbf{x}-\mathbf{x_j})  
  \end{equation}
  $F_i^D$ and $F_i^L$ represents the drag and the lift forces acting on the 'i'th particle respectively. In this work, the effect of lift force is neglected since it has a very little contribution in changing the second moments of the velocity fluctuations (\citet{muramulla2020disruption}). The drag force acting on the 'i' th particle is represented by Stokes Drag Law as follows:
  \begin{equation}
      \mathbf{F_i^D}=3\eta_f d_p(\mathbf{u}_f(\mathbf{x}_i)-\mathbf{v}_i)
      \label{eq:stokes_drag}
  \end{equation}
  \textcolor{black}{where, $\eta_f$ is the viscosity of air at ambient conditions.} 
  The detail account of the collision rule used here can be found in \citet{ghosh2022statistical}.
  In the two-way coupled DNS, the particle loading brings about change in the fluid flow characteristics; alters the friction velocity as well. 
\section{Fluid Phase Statistics and Turbulence Modulation}
	\label{sec:Fluid_phase_stats}
\begin{figure*}[h!]		
		\includegraphics[width=1.0\textwidth]{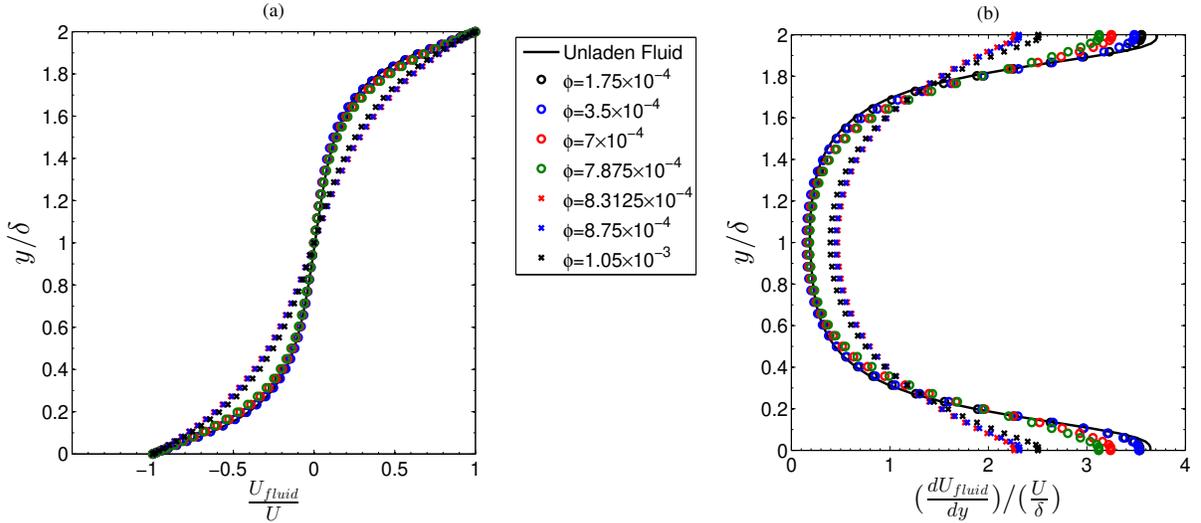}
	    \caption{Effect of particle volume fraction on (a) mean fluid velocity $U_{fluid}$ (b) mean fluid velocity gradient $\frac{dU_{fluid}}{dy}$}  
	    \label{fig:mean_fluid}     
    \end{figure*}

\begin{figure*}[h!]
  	\includegraphics[width=1.0\textwidth]{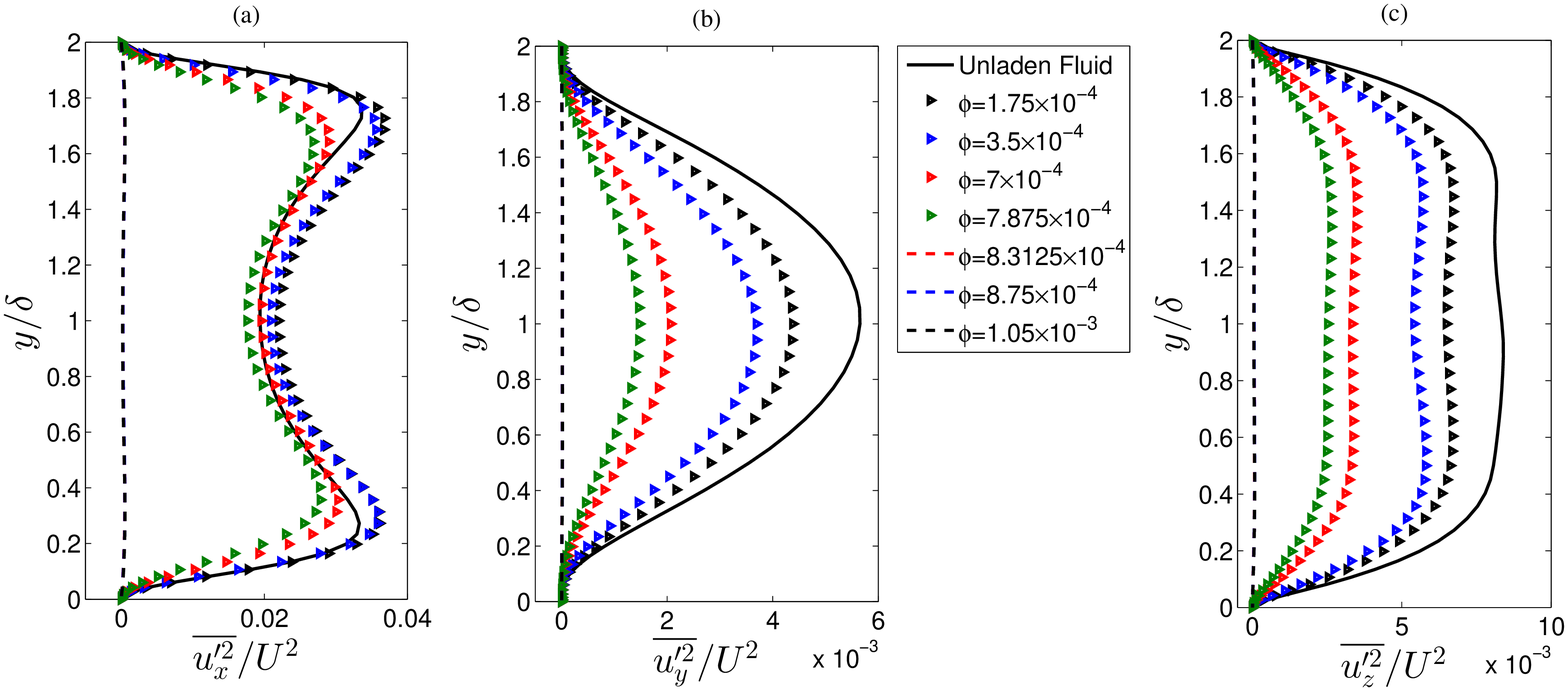}
  	\caption*{(i)}
  	\includegraphics[width=1.0\textwidth]{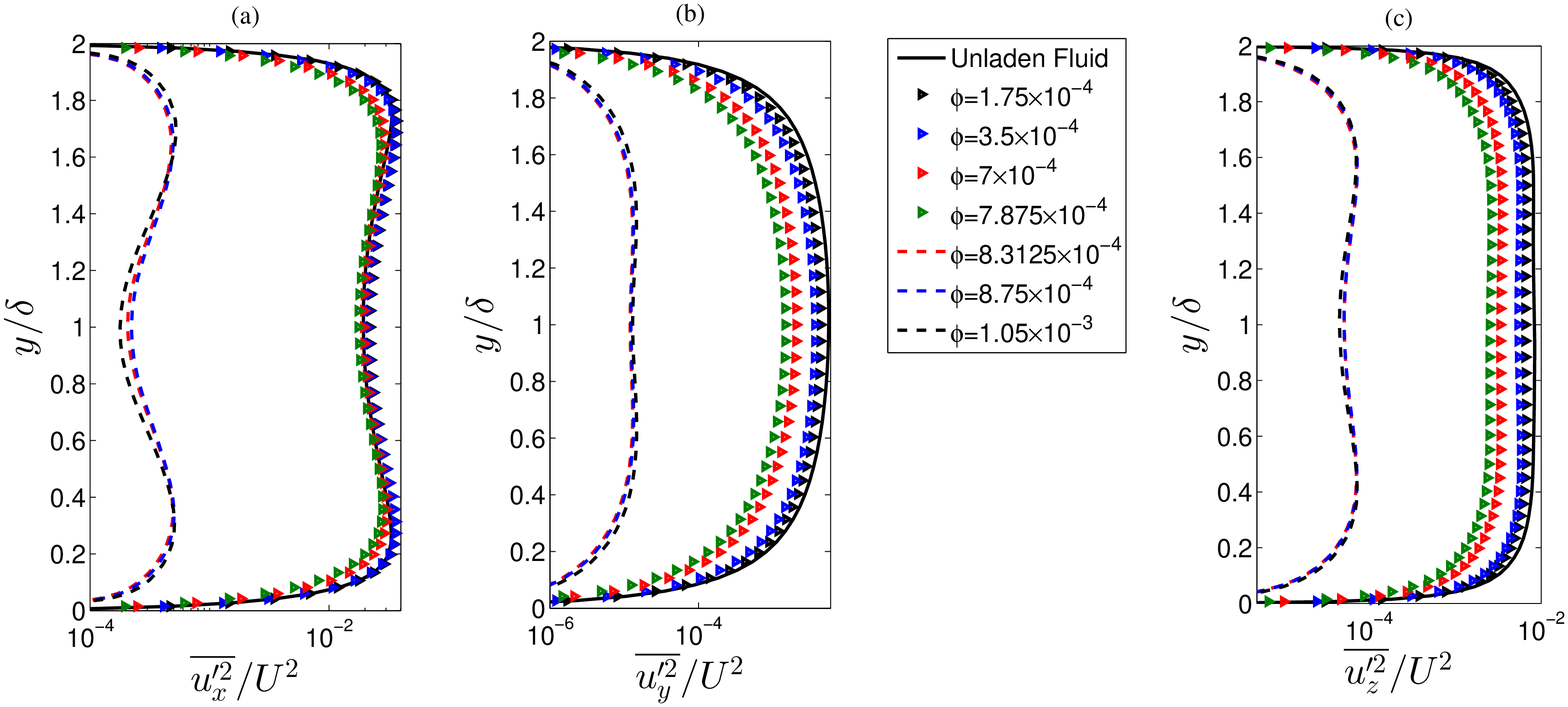}
  	\caption*{(ii)}
  	\caption{Effect of particle volume fraction on mean square fluid velocity (a) $\overline{u_x'^2}$, (b) $\overline{u_y'^2}$ and (c) $\overline{u_z'^2}$ plotted in (i) in linear scale and (ii) log-linear scale)}
  	\label{fig:ms_fluid}
    \end{figure*}

\begin{figure*}[h!]
    	{\includegraphics[width=1.0\textwidth]{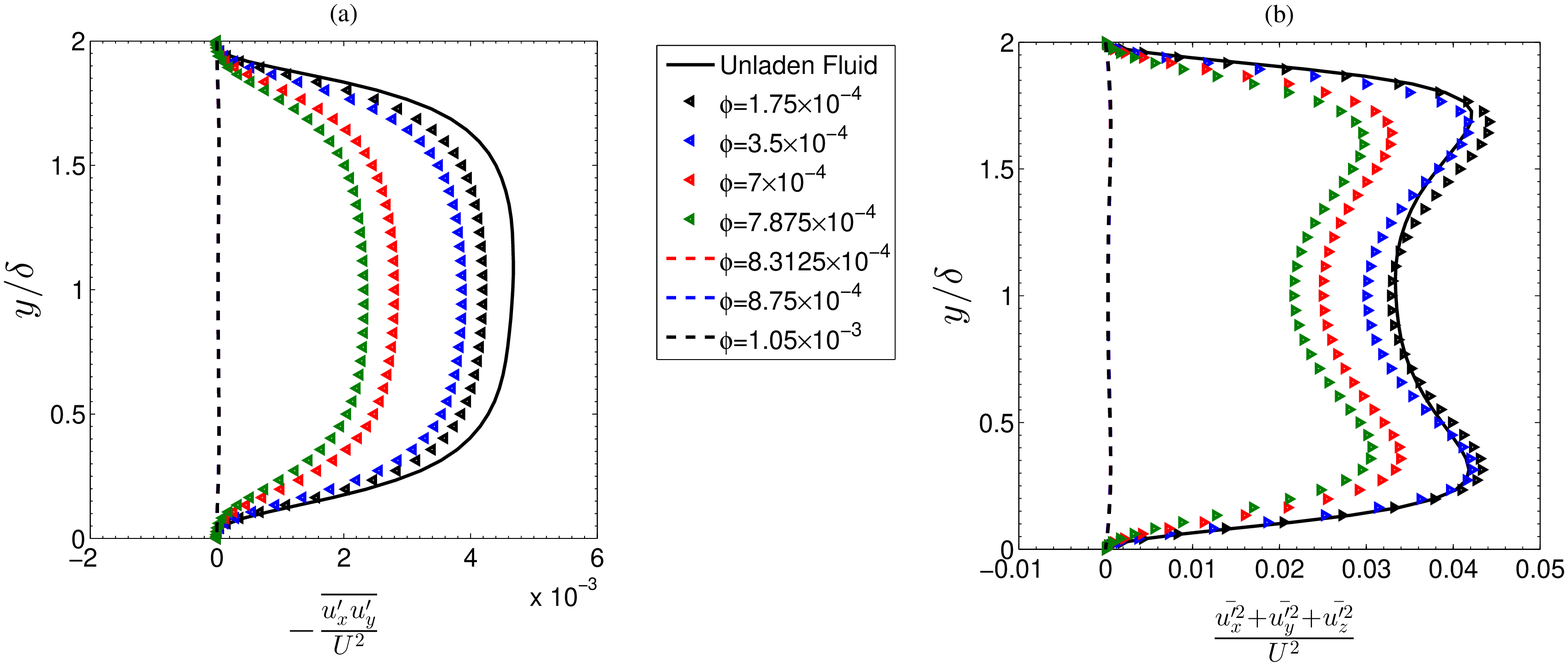}}
    	\caption*{(i)}
  	{\includegraphics[width=1.0\textwidth]{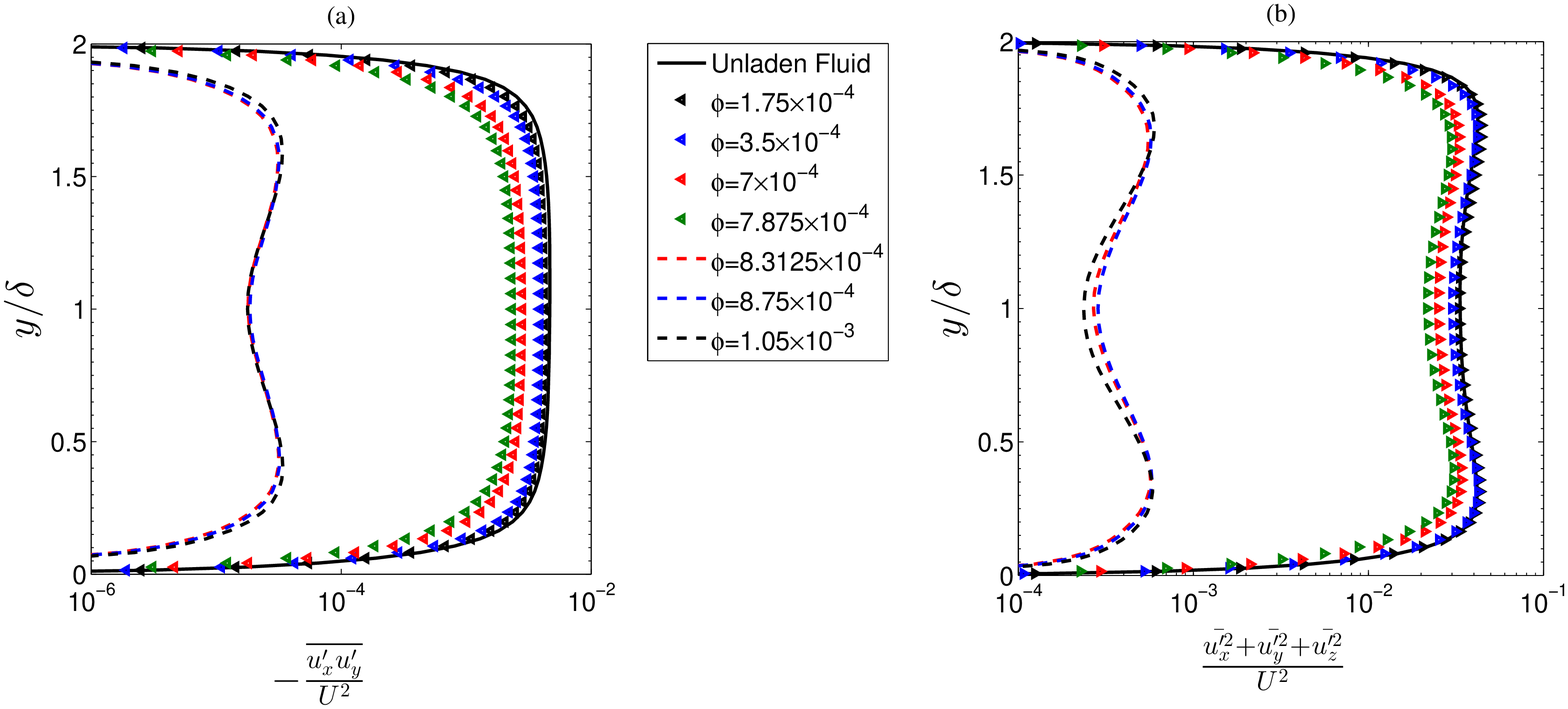}}
  	\caption*{(ii)}
  	\caption{Effect of particle volume fraction on (a) fluid turbulent kinetic energy and (b) Reynolds stress plotted in (i) in linear scale and (ii) in log-linear scale)}
  	\label{fig:tke_rs_fluid}
    \end{figure*}

	The fluid phase mean velocity and mean-square velocity statistics are \textcolor{black}{presented here in} figures \ref{fig:mean_fluid} to \ref{fig:tke_rs_fluid}.
    Mean velocity profiles $U_{fluid}$ in figure \ref{fig:mean_fluid} (a) show two distinct \textcolor{black}{regimes}. For unladen fluid and fluid with particle volume fraction ($\phi$) upto $\phi=7.875\times10^{-4}$, the mean velocity profile remains unchanged. For $\phi=8.3125\times10^{-4}$ and higher, the mean velocity follows a different profile which remains almost unchanged till $\phi=1.05\times10^{-3}$, \textcolor{black}{the maximum volume fraction reported here}. The similar trend is observed for mean fluid velocity gradient also. In this case, beyond particle volume fraction $\phi=7.875\times10^{-4}$ the mean fluid velocity gradient becomes slightly flatter with a \textcolor{black}{distinct} decrease in the velocity gradient at the wall, shown in figure \ref{fig:mean_fluid} (b). 
    The figure \ref{fig:ms_fluid} shows the effect of particle volume fraction on fluid mean square velocity. It is observed that with increase in volume loading the mean-square velocities in all the directions decrease. This decrease is continuous from $\phi=1.75\times10^{-4}$ to $\phi=7.875\times10^{-4}$ across all the directions. However, beyond $\phi=7.875\times10^{-4}$ the fluctuations decrease drastically for all the three directions. Figure \ref{fig:ms_fluid} (ii) shows that this sudden decrease in mean-square velocity of the fluid is about two-orders of magnitude. It is to be mentioned that a slight increase in $\overline {u_x'^2}$ is observed from unladen fluid to fluid with $\phi=3.5\times10^{-4}$;  this is in contrary to what is observed in other two directions (Fig. \ref{fig:ms_fluid} (i)).    
 Fluid Reynolds stress $\overline{u_x'u_y'}$ is found to decrease continuously with increase in volume loading from $\phi=0$ to $\phi=7.875\times10^{-4}$, beyond which drastic decrease of about two orders of magnitude (Fig. \ref{fig:tke_rs_fluid} (ii)) is observed. Similar trend is observed for fluid turbulent kinetic energy with one exception that from $\phi=0$ to $\phi=1.75\times10^{-4}$ the turbulent kinetic energy marginally increases, especially away from walls similar to the trend \textcolor{black}{in} $\overline{u_x'^2}$ profile. \textcolor{black}{In summary,} the results in this section show an abrupt change in the profile of mean fluid velocity and mean fluid velocity gradient accompanied with drastic reduction of second moments of fluid velocity fluctuations and fluid turbulent kinetic energy beyond a volume fraction of $\phi=7.875\times10^{-4}$. \textcolor{black}{A similar trend has also been reported in case of turbulent channel flow by \citet{muramulla2020disruption} with increase in particle loading at a constant volumtric flow-rate for the gas phase. It is worth noting that the present study differs from the previous one in the following aspects. Wall shear is the main driving force instead of the pressure gradient (for channel flow), gravity does not play any role on particle dynamics, and the modification of mean flow profile is associated with the change in volumetric flow-rate of fluid (in the half of the Couette domain) instead of the constant volumetric flow rate in the earlier work of \citet{muramulla2020disruption}}.
 \\ \textcolor{black}{To understand the abrupt change in the fluid phase dynamics, we perform a detailed analysis of momentum and kinetic energy budget, which are reported in the following sections.} 
\section{Momentum and Kinetic Energy Budget of the Fluid Phase}
The budget equations of momentum and kinetic energy of the fluid phase is quite important to understand the turbulence modulation.
   \subsection{Fluid phase mean momentum budget equation}
   \label{sec:mean_momentum_budget}
\begin{figure*}[htbp]
   \begin{minipage}{0.45\textwidth}
   \includegraphics[width=1.0\linewidth]{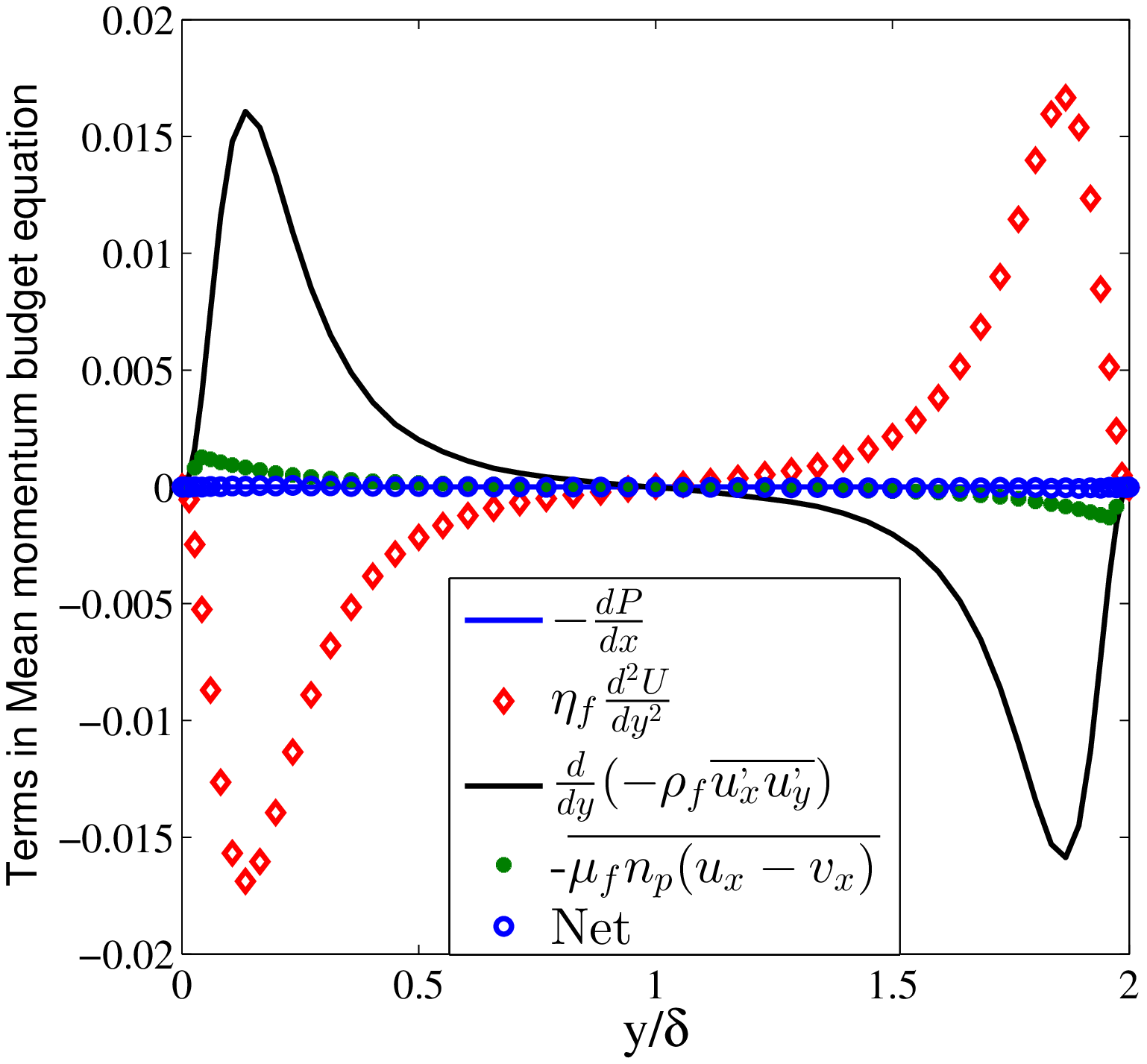}
   \caption*{(a)}
   \end{minipage}\hfill
   	\begin{minipage}{0.45\textwidth}
   	{\includegraphics[width=1.0\linewidth]{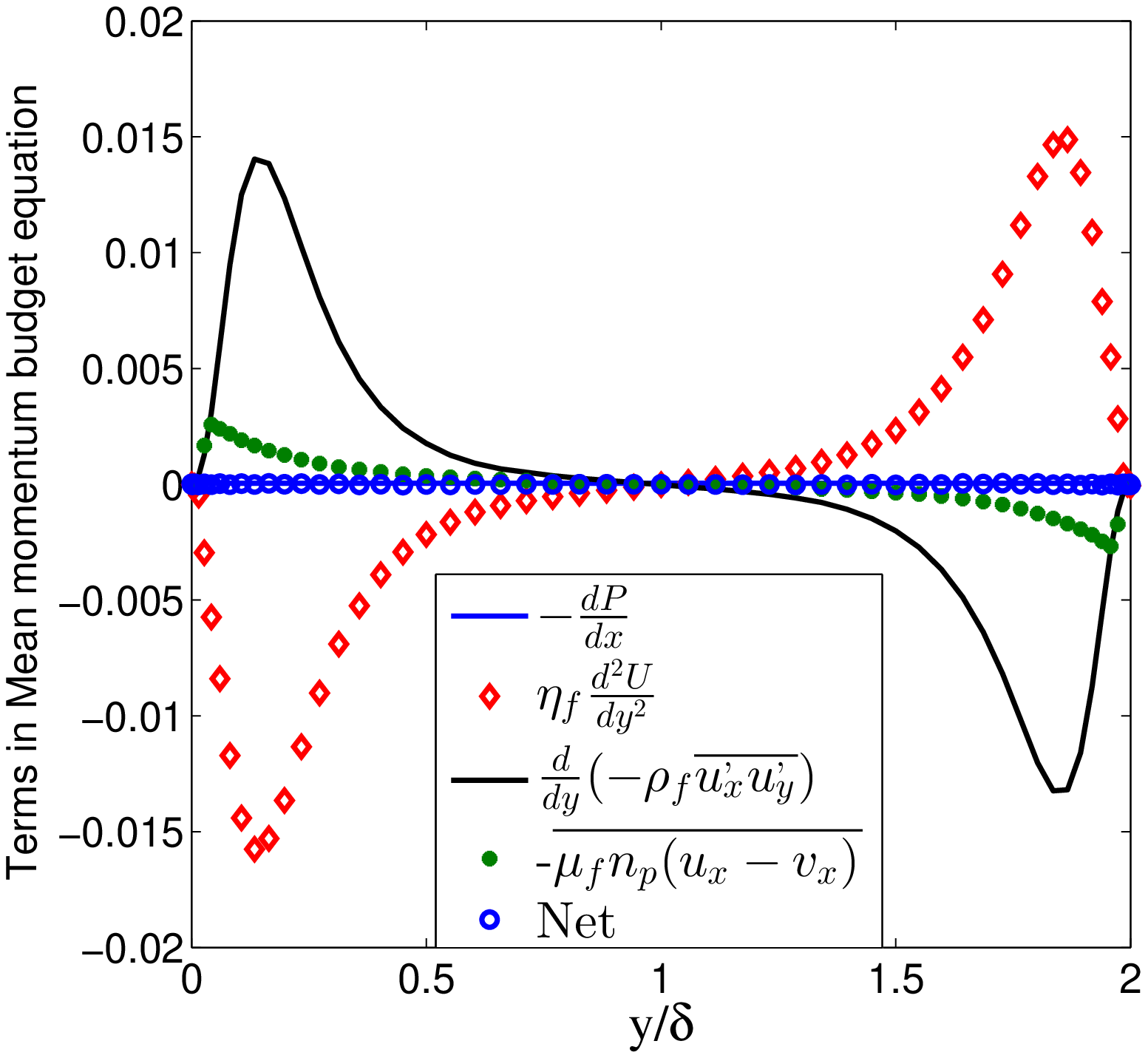}}
   	\caption*{(b)}
   	\end{minipage}\hfill
   	\begin{minipage}{0.45\textwidth}
   	{\includegraphics[width=1.0\linewidth]{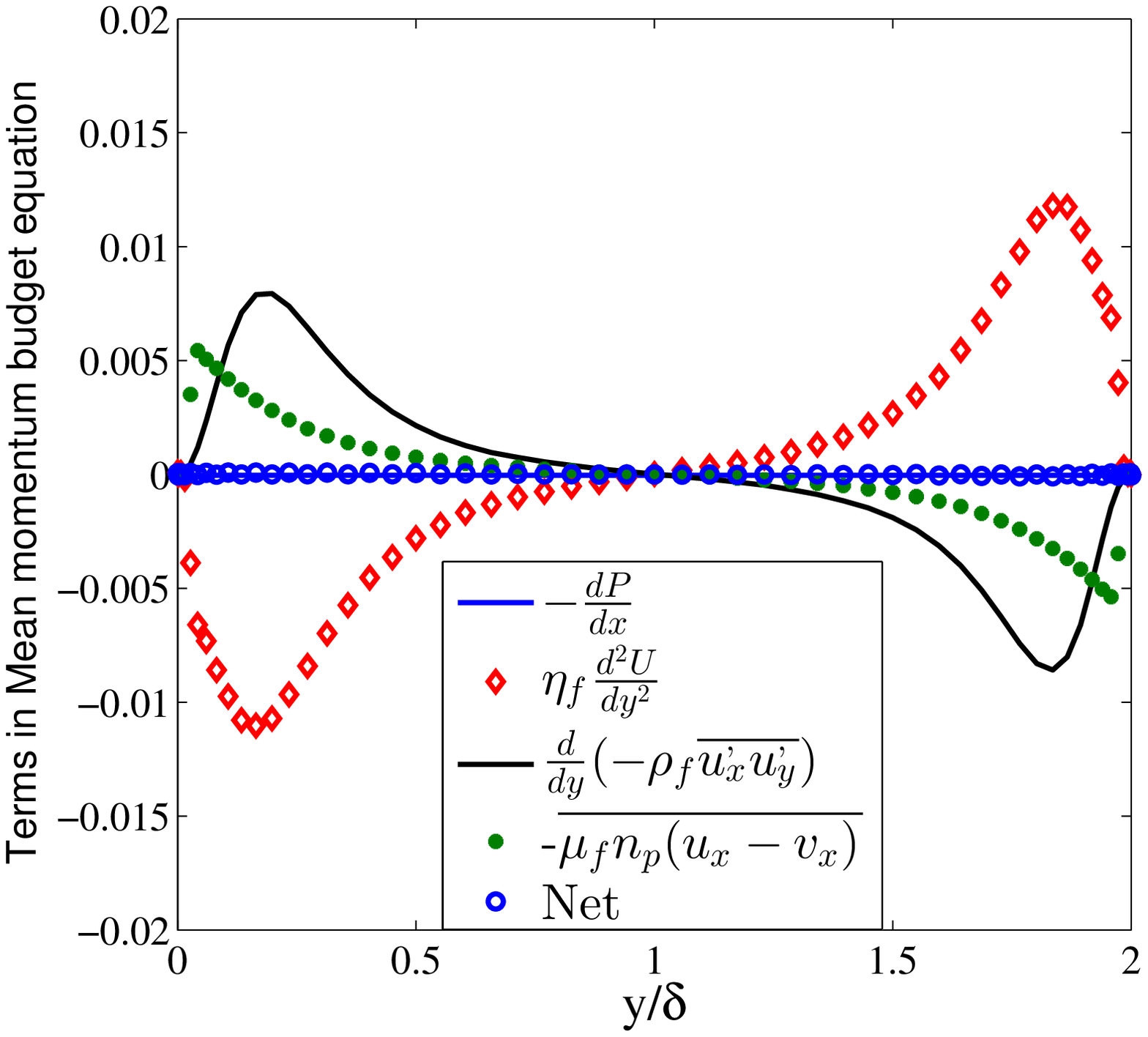}}
   	\caption*{(c)}
   	\end{minipage}\hfill
   	\begin{minipage}{0.45\textwidth}
   	{\includegraphics[width=1.0\linewidth]{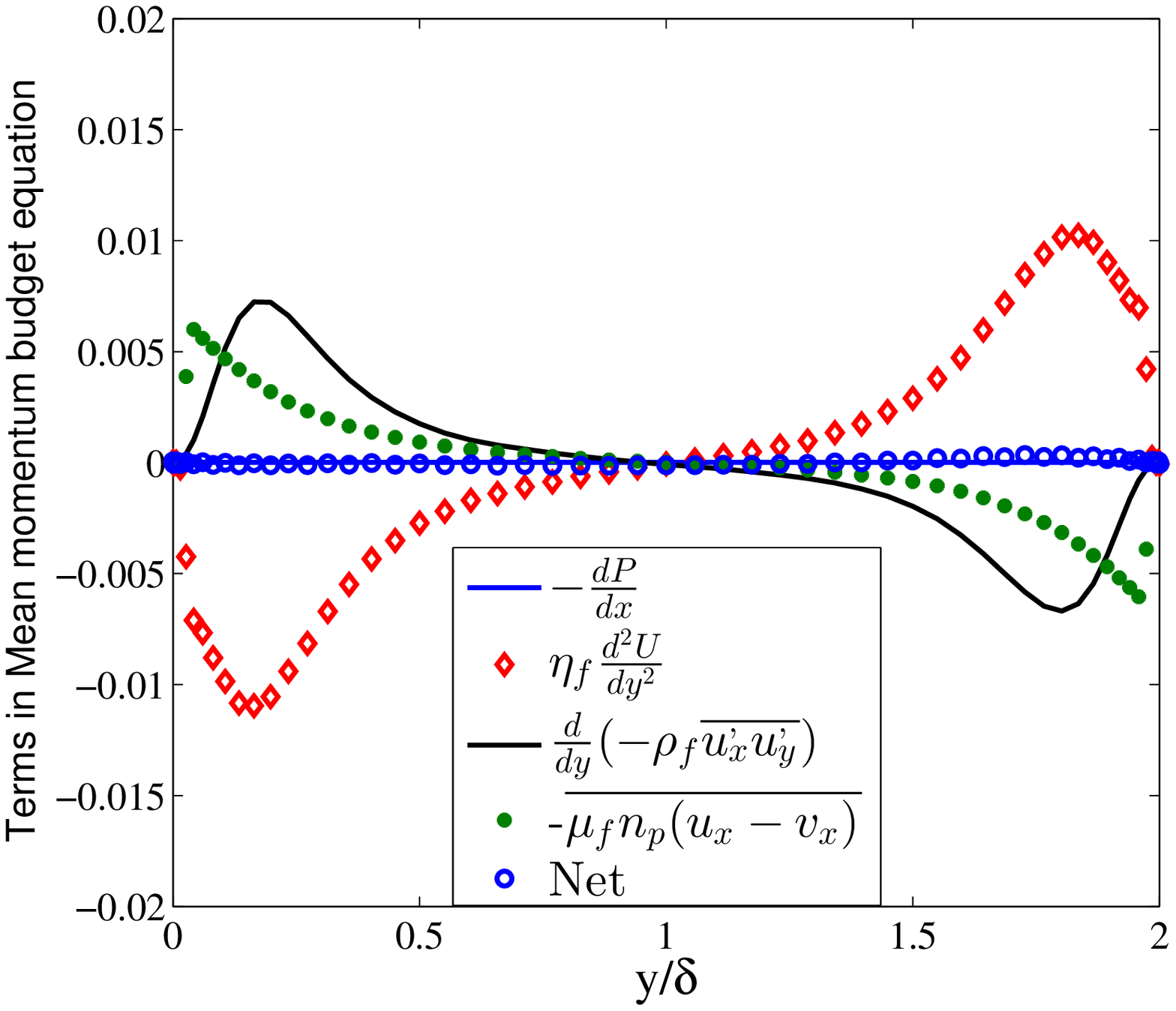}}
   	\caption*{(d)}
   	\end{minipage}\hfill
   	\begin{minipage}{0.45\textwidth}
   	\includegraphics[width=\linewidth]{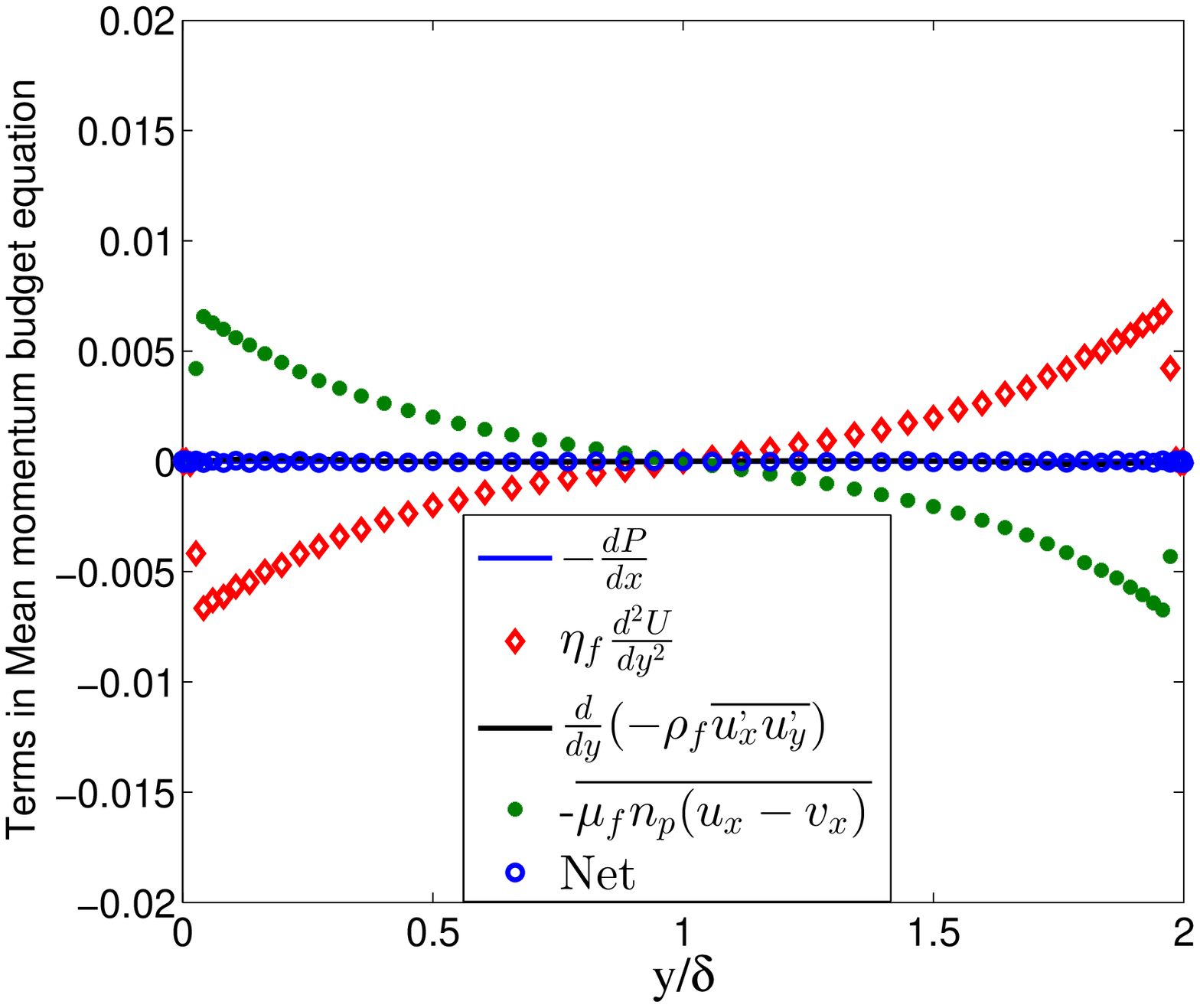}
 	\caption*{(e)}
 	\end{minipage}\hfill
 	\begin{minipage}{0.45\textwidth}
    \includegraphics[width=1.0\linewidth]{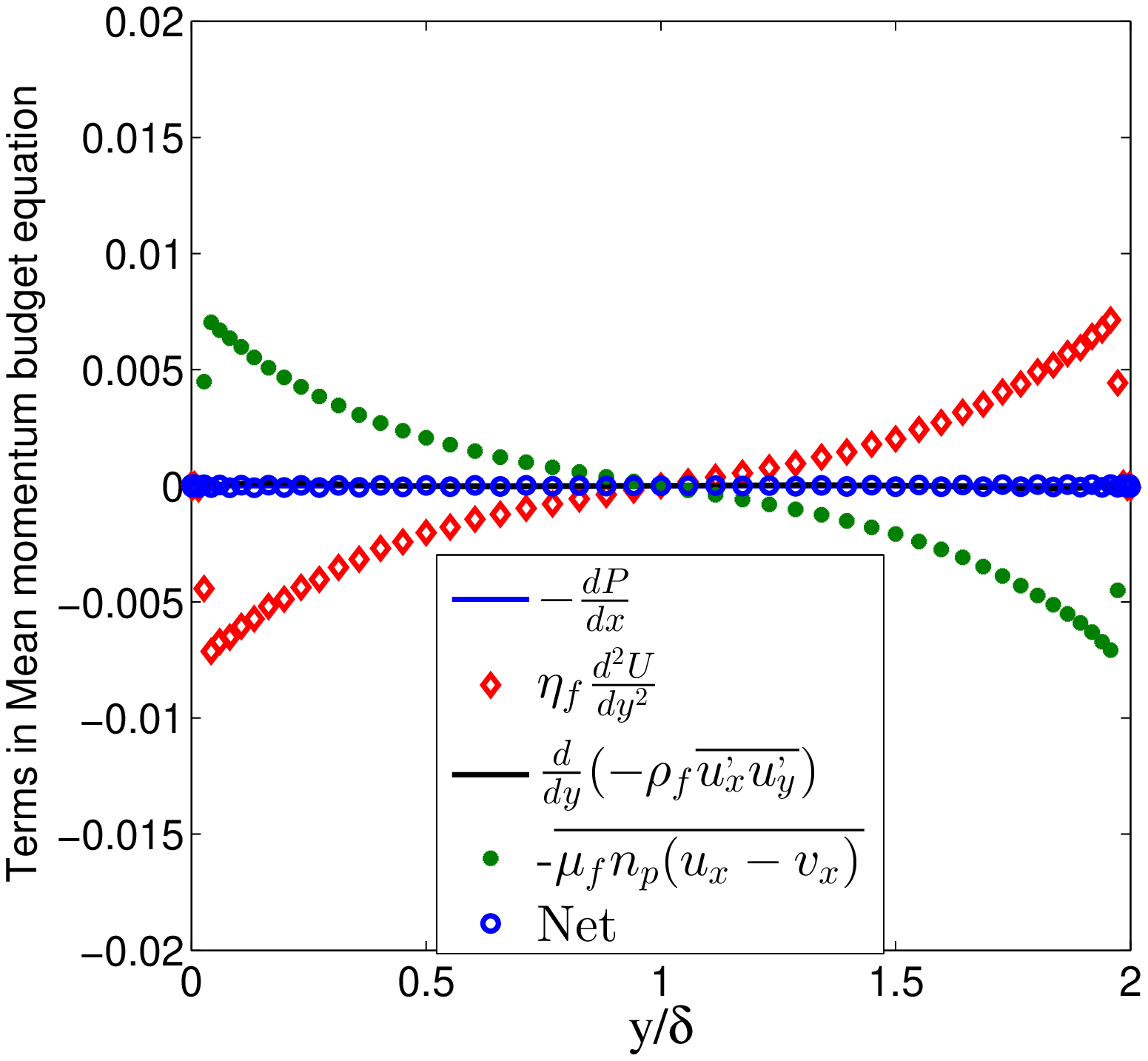}
 	\caption*{(f)}
 	\end{minipage}\hfill
  \caption{Fluid phase mean momentum budget at various volume fractions (a) $\phi=1.75\times10^{-4}$, (b) $\phi=3.5\times10^{-4}$, (c) $\phi=7\times10^{-4}$, (d) $\phi=7.875\times10^{-4}$, (e) $\phi=8.3125\times10^{-4}$ and (f) $\phi=8.75\times10^{-4}$}	
  \label{fig:mom_balance}
\end{figure*}

   The Mean Momentum Budget of the fluid phase along x direction can be written as: 
      \begin{equation}
   \underbrace{\eta_f\frac{d^2\bar{u_x}}{dy^2}}_{1}+\underbrace{\frac{d\tau_{xy}^{Rf}}{dy}}_{2}-\underbrace{\overline{\mu_fn_p(u_x-v_x)}}_{3}=0
   	\label{eq:momentum balance}
   \end{equation}
The terms in the mean momentum budget equation are    
   \begin{itemize}
   	\item 1: Momentum transfer due to viscous stress
   	\item 2: Momentum transfer by fluid Reynolds stress $\tau_{xy}^{Rf}=-\rho_f\overline{u_x'u_y'}$
   	\item 3: Momentum transfer due to particle feedback force
   \end{itemize}
   The effect of particle volume fraction on fluid phase mean momentum budget is shown in figure \ref{fig:mom_balance}.
In figure \ref{fig:mom_balance}, it is observed that the momentum transfer term due to Reynolds Stress and the transfer of momentum due to viscous forces mainly dominate at lower volume fractions and act in the opposite direction. It is also observed from figures \ref{fig:mom_balance}(a) to (d) that the momentum transfer due to Reynolds Stress term gradually decreases with the increase in $\phi$, whereas the momentum transfer due to particle reverse drag increases. At higher volume fraction $\phi=7\times10^{-4}$ and $7.875\times10^{-4}$, term (2) and term (3) of equation \ref{eq:momentum balance} together balance out term (1). It is observed that the peak values of term (3) occur nearer to the wall due to the very high slip velocity \textcolor{black}{between particle and fluid phase}. \\ 
A different phenomenon is observed for volume fractions greater than $\phi=7.875\times10^{-4}$. In this regime (Fig. \ref{fig:mom_balance} (e) and (f)) the momentum transfer term due to Reynolds Stress term  decreases drastically and the momentum transfer due to particle reverse drag balances out the momentum transfer term due to viscous stresses.  Above the volume fraction of $\phi=7.875\times10^{-4}$, the drastic decrease of \textcolor{black}{divergence of Reynolds stress may lead to a decrease in turbulence production, which can be expressed through energy budget}. The budget study of mean fluid kinetic energy and fluid turbulent kinetic energy is discussed below.  

\subsection{Fluid phase mean kinetic energy budget equation}
\label{sec:mean_ke_budget}
\begin{figure*}[!h]
    \begin{minipage}{0.5\textwidth}
 	{\includegraphics[width=1.0\linewidth]{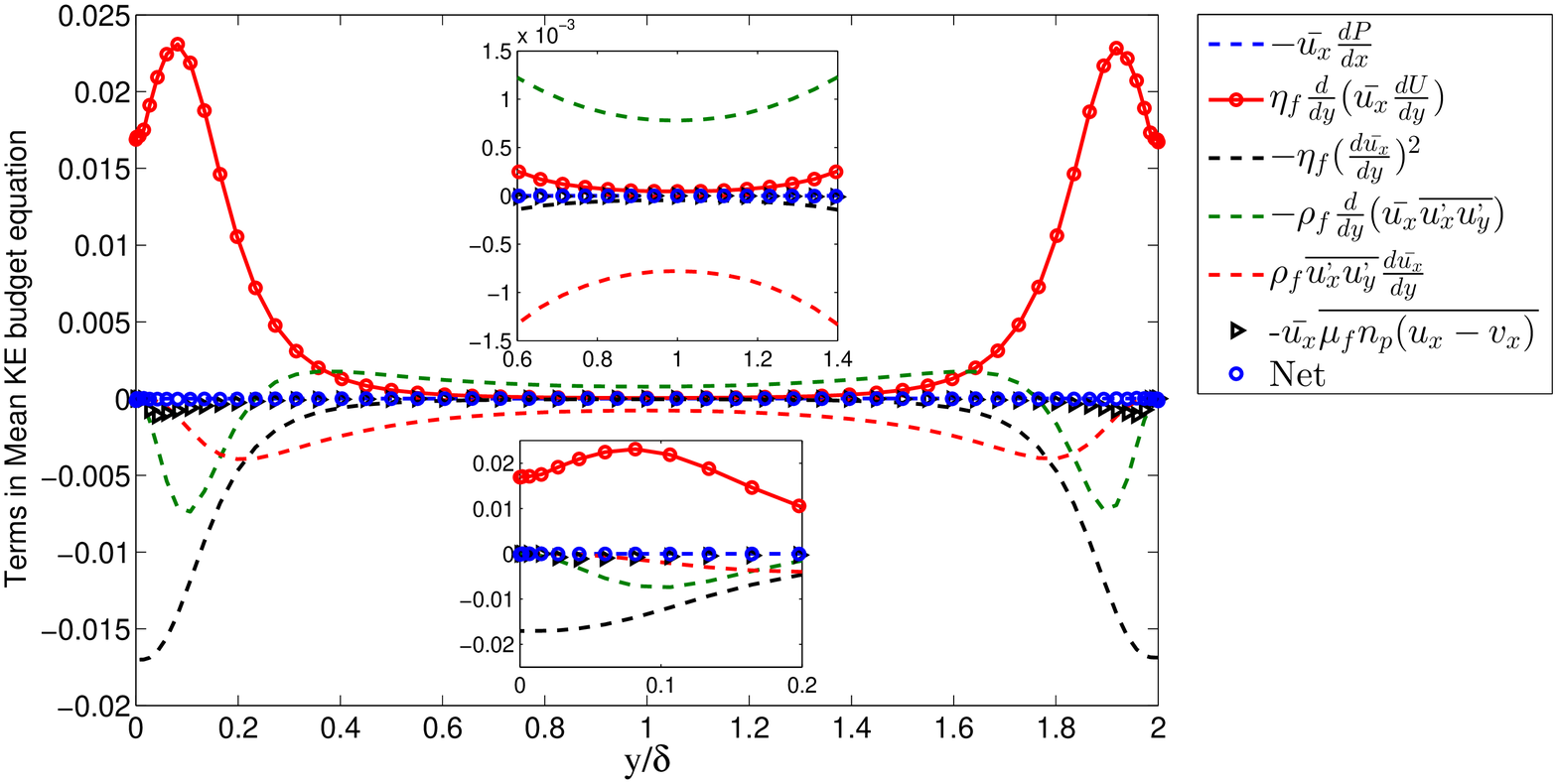}}
 	\caption*{(a)}
 	\end{minipage}\hfill
    \begin{minipage}{0.5\textwidth}
 	{\includegraphics[width=1.0\linewidth]{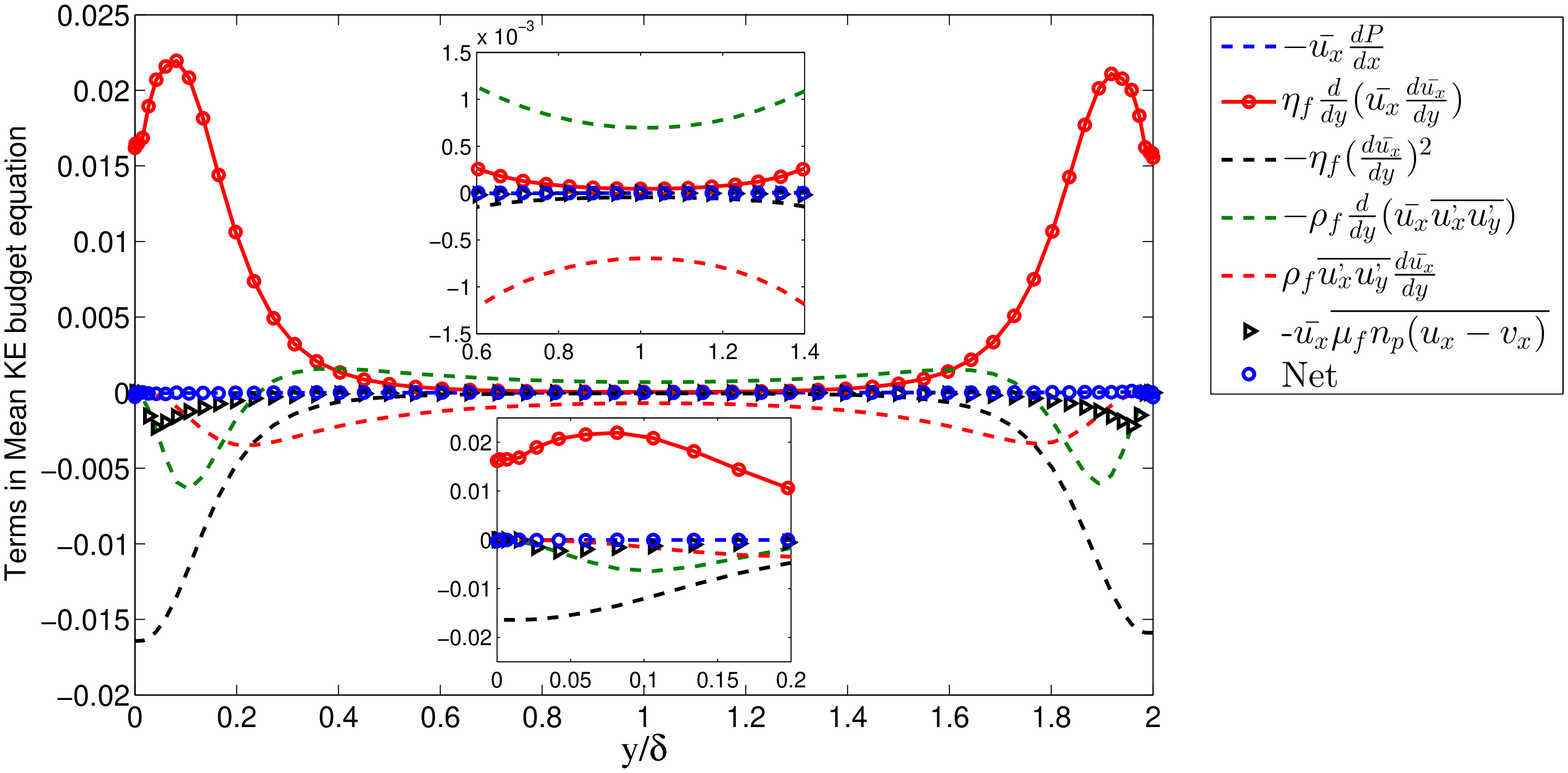}}
 	\caption*{(b)}
 	\end{minipage}\hfill
    \begin{minipage}{0.5\textwidth}
 	{\includegraphics[width=1.0\linewidth]{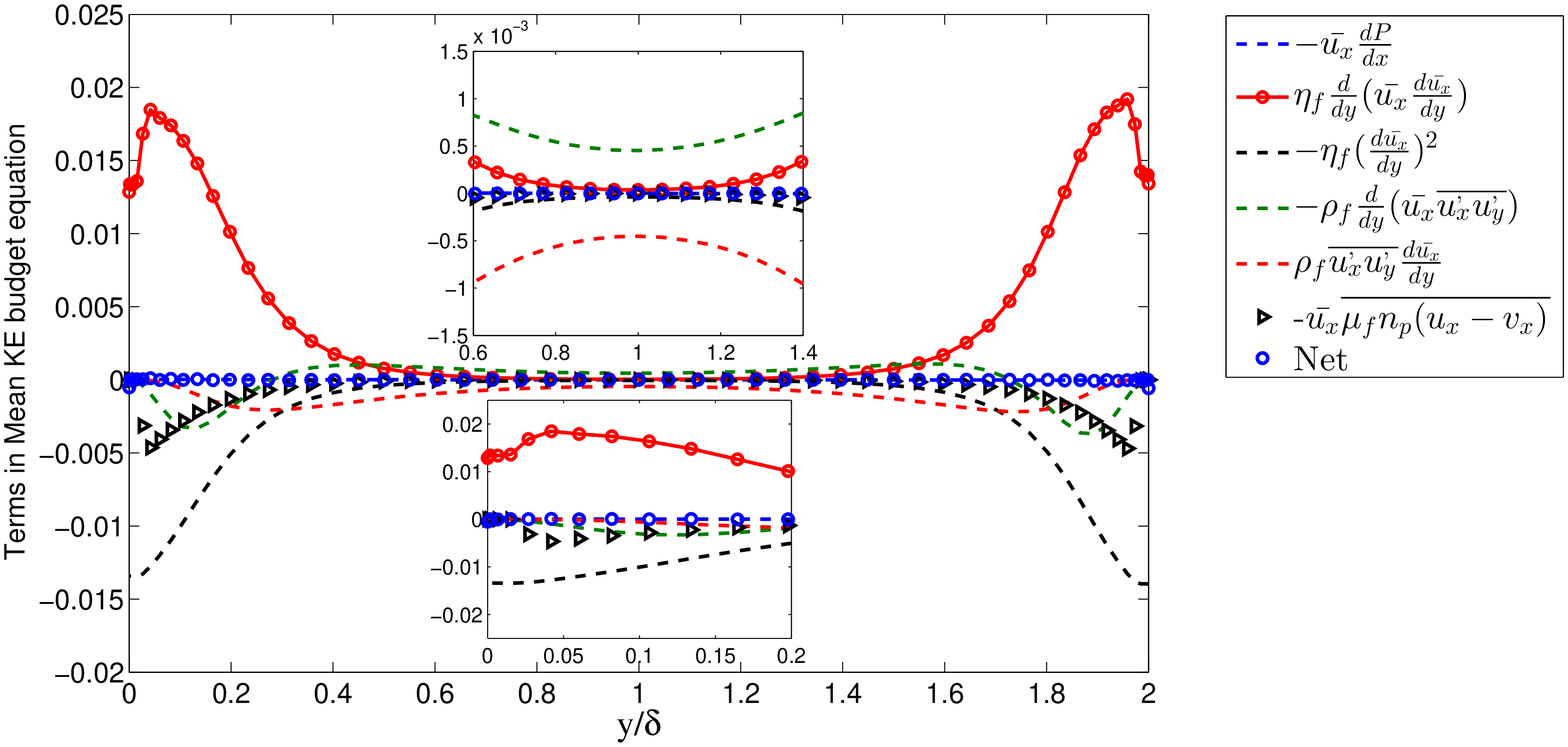}}
 	\caption*{(c)}
 	\end{minipage}\hfill
 	\begin{minipage}{0.5\textwidth}
 	{\includegraphics[width=1.0\linewidth]{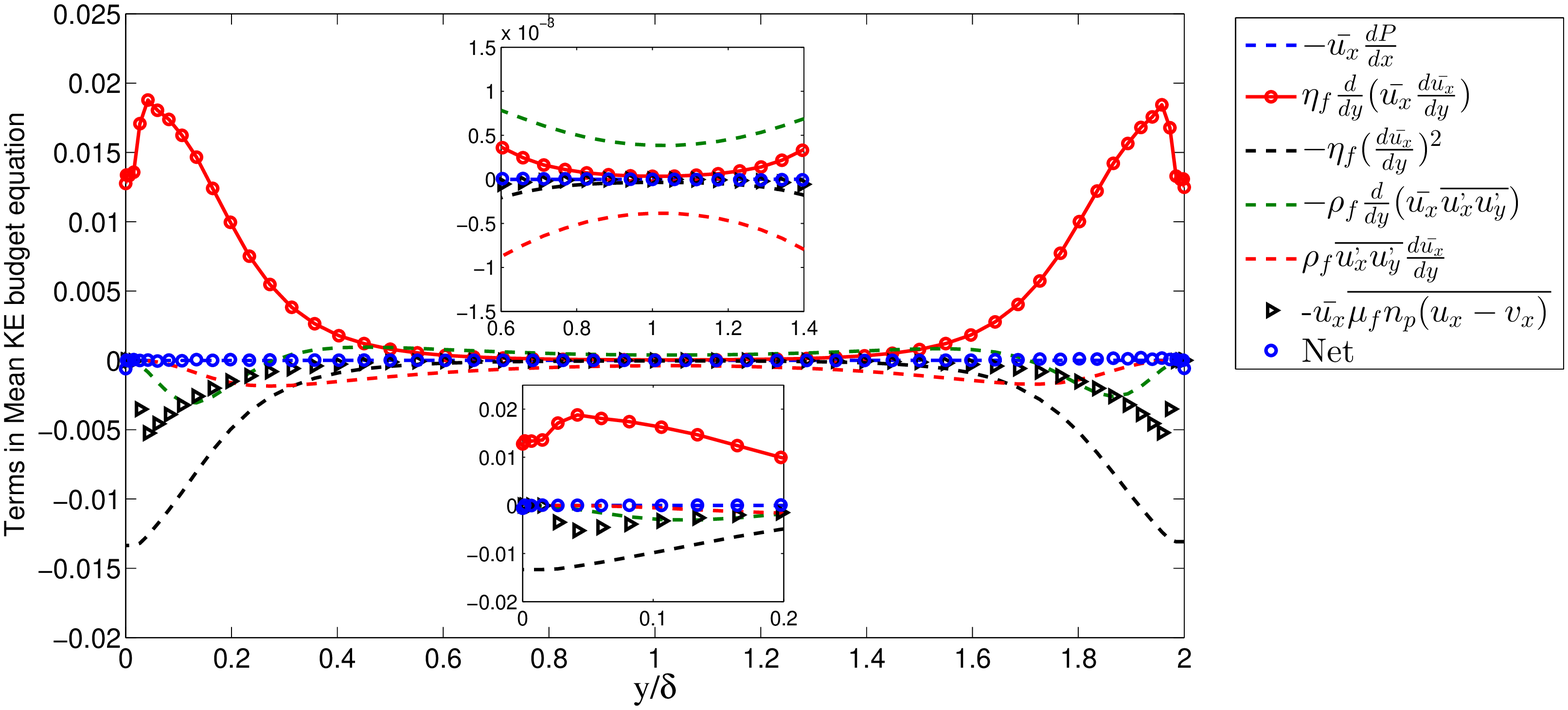}}
 	\caption*{(d)}
 	\end{minipage}\hfill
 	\begin{minipage}{0.5\textwidth}
 		{\includegraphics[width=1.0\linewidth]{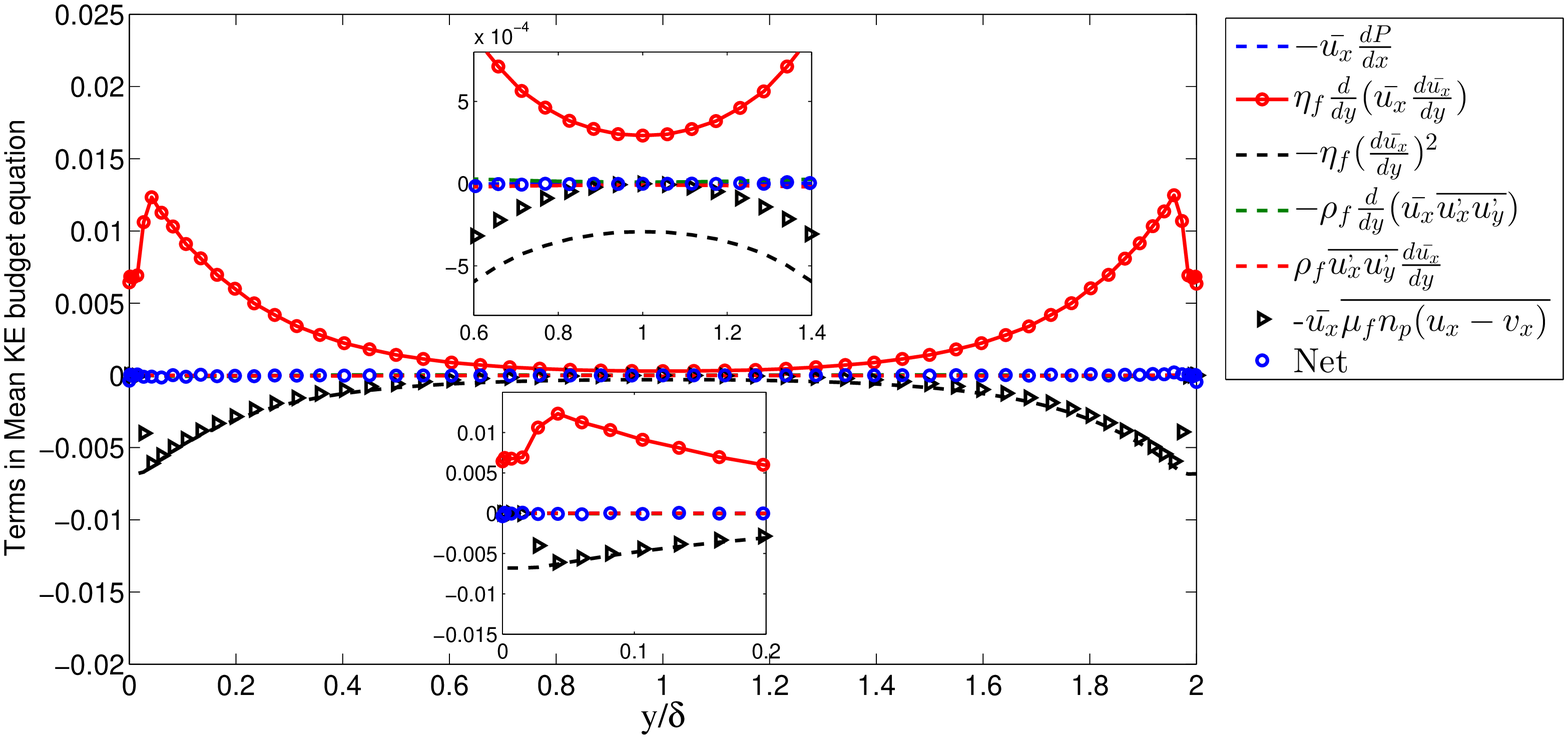}}
 	\caption*{(e)}
 	\end{minipage}\hfill
 	\begin{minipage}{0.5\textwidth}
 		{\includegraphics[width=1.0\linewidth]{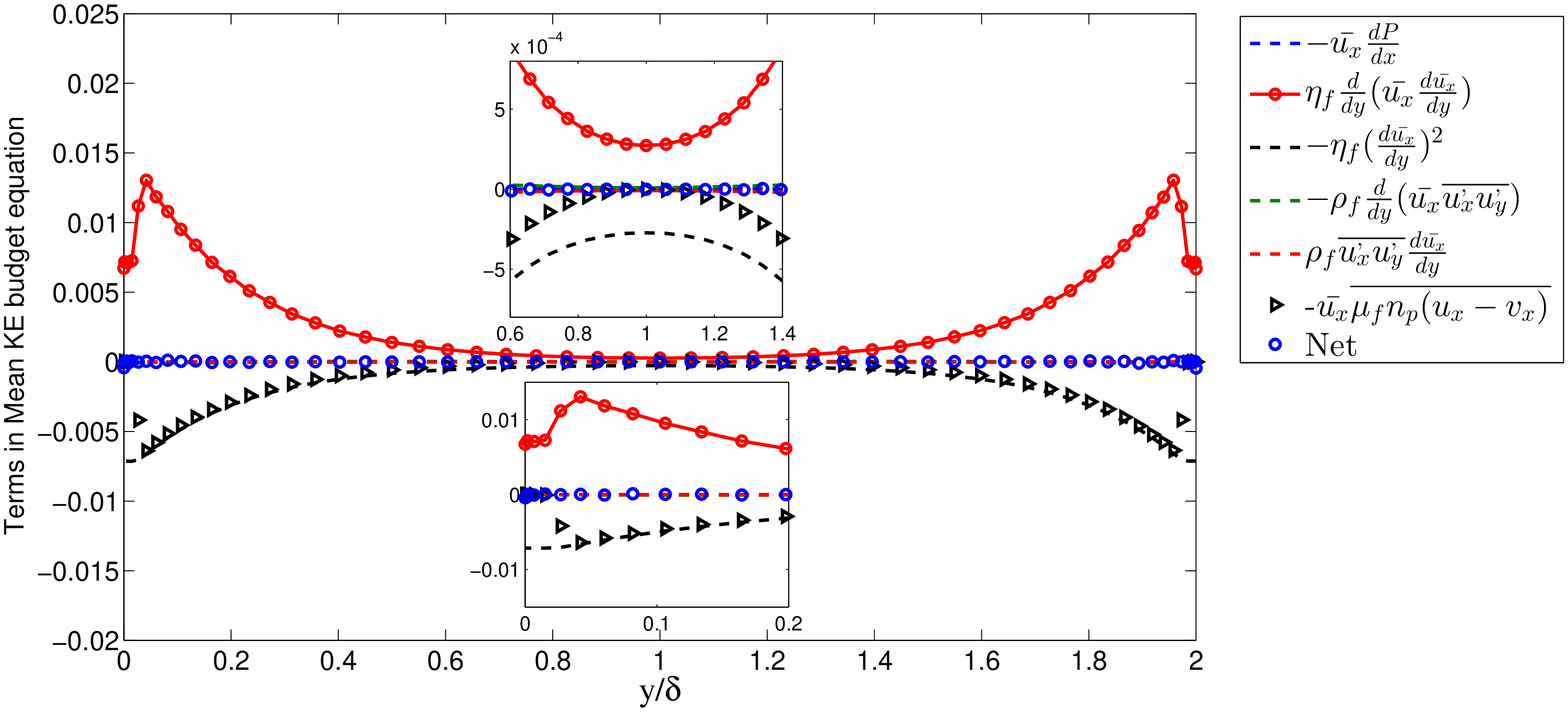}}
 	\caption*{(f)}
 	\end{minipage}\hfill
 	\caption{Fluid phase mean kinetic energy budget at various volume fractions (a) $\phi=1.75\times10^{-4}$, (b) $\phi=3.5\times10^{-4}$, (c) $\phi=7\times10^{-4}$, (d) $\phi=7.875\times10^{-4}$, (e) $\phi=8.3125\times10^{-4}$ and (f) $\phi=8.75\times10^{-4}$}	
 	\label{fig:ke_mean}	
 \end{figure*}

\begin{figure*}[!]
	\includegraphics[width=1.0\linewidth]{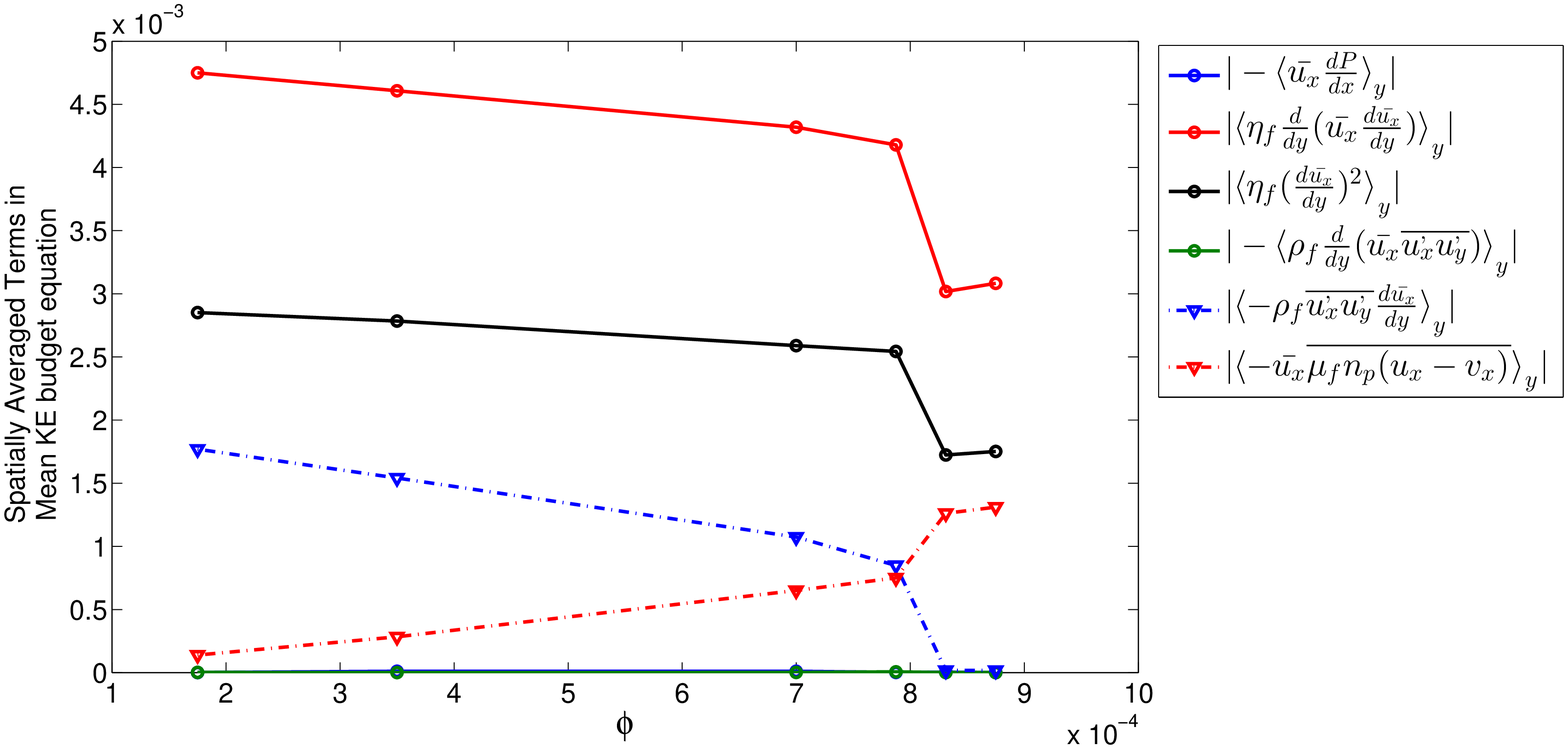}
		\caption{Magnitude of spatially averaged terms of fluid phase mean K.E. as a function of $\phi$}
	\label{fig:av_mean_ke}
\end{figure*}

   . The mean fluid K.E. budget equation is expressed as:
    \begin{align}
   & \underbrace{\eta_f\frac{d}{dy}\left(\bar{u_x}\frac{d\bar{u_x}}{d y}\right)}_{1}+\underbrace{\frac{d}{dy}(\bar{u_x}\tau_{xy}^{Rf})}_{2}-\underbrace{\eta_f\left(\frac{d\bar{u_x}}{dy}\right)^2}_{3}\nonumber\\ &-\underbrace{\tau_{xy}^{Rf}\frac{d\bar{u_x}}{dy}}_{4}-\underbrace{\bar{u_x}\overline{\mu_fn_p(u_x-v_x)}}_{5}=0
    \end{align}
    \label{eq:mean_ke}
The terms in the mean kinetic energy budget equation are as follows:   
   \begin{itemize}
   	\item 1: Transport of mean K.E. due to fluid viscous stress
   	\item 2: Transport of mean K.E. due to Reynolds stress
   	\item 3: Viscous dissipitaion of mean K.E. due to mean shear
   	\item 4: Decrease in mean K.E. \textcolor{black}{associated with the} shear production of turbulence
   	\item 5: Dissipation of mean K.E. due to mean particle drag
   \end{itemize}
The effect of particle volume fraction on fluid phase mean kinetic energy budget is shown in figure \ref{fig:ke_mean} along with the absolute value of each of the spatially averaged terms in figure \ref{fig:av_mean_ke}.
For the terms of mean K.E. budget equation, qualitatively different kinds of behaviours are observed for volume fraction below and above $\phi=7.875\times10^{-4}$. Below $\phi=7.875\times10^{-4}$ , in figures \ref{fig:ke_mean} (a) to (d), it is observed that the terms dominating the mean fluid K.E. budget changes from the near-wall region to the central region of the Couette-flow. At the near-wall region, for all the four volume fractions, mean K.E. due to fluid viscous stress arising out of the mean velocity gradient (term 1) acts as the source of mean K.E., whereas the viscous dissipation remains (term 3) the leading sink term. Transport of mean K.E. due to Reynolds stress (term 2) acts as an important sink in the near-wall region at lower volume fractions (insets of fig \ref{fig:ke_mean} (a) and (b)) and decreases with the increase in volume fraction. Conversely the loss of mean fluid K.E. due to particle reverse drag (term 5) increases with \textcolor{black}{increase in} volume fraction and acts as an significant sink of the mean K.E. in the near-wall region as shown in the insets of figure \ref{fig:ke_mean} (c) and (d). However, at the central region of the Couette-flow decrease of mean K.E. \textcolor{black}{used for the} shear production of turbulence (term 4) becomes the dominant sink term whereas the transport of mean K.E. due to Reynolds Stress (term 2) term acts as the leading source term of the mean K.E. The magnitude of the dominant source and sink terms of the mean K.E. decreases by one order. The values of the dominant source and sink terms of this region, decreases with increasing volume fraction.  
Figure \ref{fig:ke_mean} (e) and (f) shows that above volume fraction $\phi=7.875\times10^{-4}$, the same terms remain as the dominant source and sink terms be it in the the-wall region or in the central region of the Couette-flow, although a decrease of two orders of magnitude is observed at the central region. Transport of mean K.E. due to fluid viscous stresses generated due to mean fluid velocity gradient (term 2) acts as the principal source of mean K.E. across the cross-stream direction. The primary sink terms are observed to be the viscous dissipation term (term 4) and the decrease of mean K.E. due to particle reverse drag (term 6) term. Loss of mean K.E. due to shear production of turbulence (term 5) and the transport of mean K.E. due to fluid Reynolds Stress term (term 3) become negligible in this regime of volume loading.      
\begin{figure*}[!]
	\includegraphics[width=1.0\linewidth]{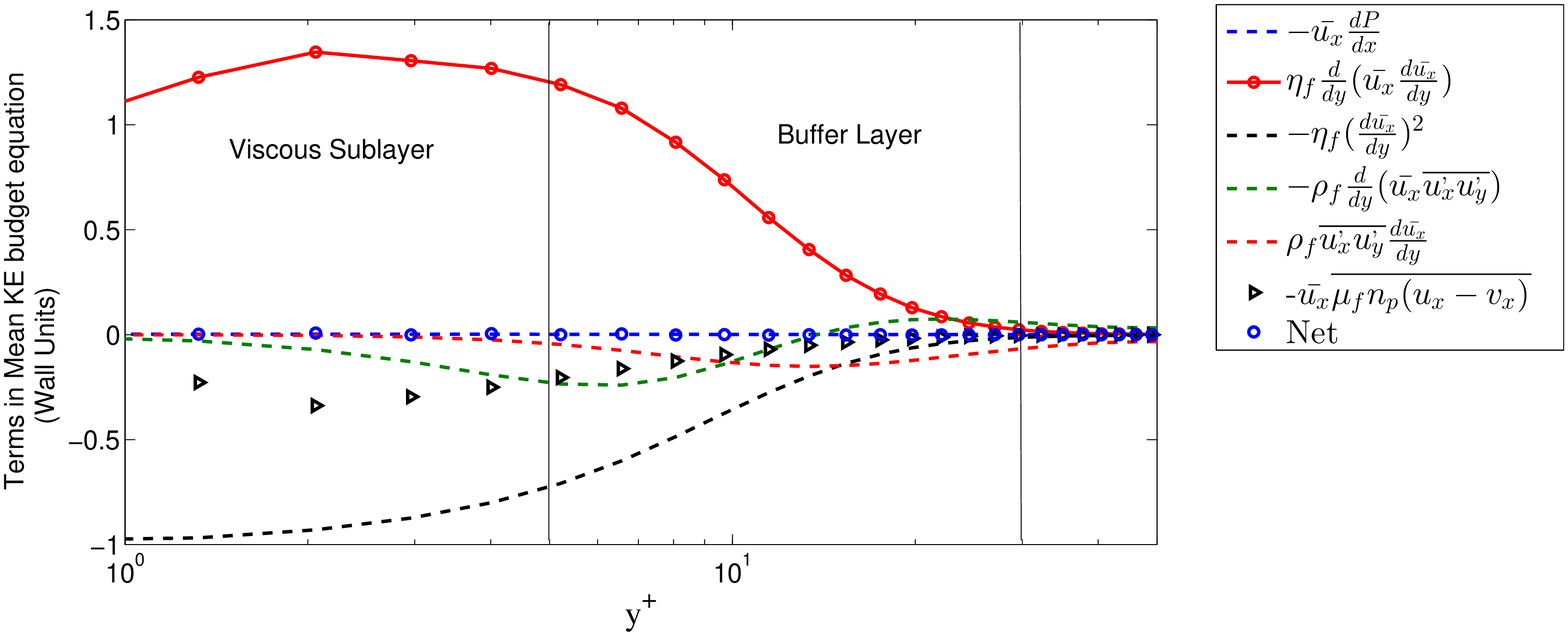}
		\caption{ Terms of fluid phase mean K.E. as a function of cross-stream position in wall-scaling for $\phi=7\times10^{-4}$}
	\label{fig:wall_mean_ke}
\end{figure*}
\\The variation of \textcolor{black}{different components of} mean K.E. budget terms \textcolor{black}{along the} cross-stream distance in wall-units are \textcolor{black}{presented} for lower $\phi$ values. Figure \ref{fig:wall_mean_ke} shows the behaviour of mean k.e terms of the fluids as a function of $y^+$ at $\phi=7\times10^{-4}$. The transport of mean K.E. due to viscous stress arising out of mean fluid velocity gradient acts as the main source in the viscous sub-layer and drastically decreases in magnitude in the buffer-layer. The viscous dissipation of mean K.E. term, the dominant dissipation term, shows a similar behaviour: highest near wall and drastically decreasing in the buffer layer. The dissipation term due to particle reverse drag acts as the sink in the viscous sub-layer, and reduces to a very small magnitude in the buffer layer. 
\textcolor{black}{Peak production of turbulent K.E. occurs in the buffer layer. A similar amount of energy is lost from the mean flow.}
This occurrence takes place along with the change of transport term of mean K.E. due to Reynolds Stress from being a sink term to source term. 
\\ In order to get an average measure of all the terms across different cross-stream positions of the Couette flow, a spatially averaged term $\langle[\cdot]\rangle_y$ derived by averaging the term $[\cdot]$ over half-width $\delta$ of the Couette flow is defined as: 
\begin{equation}
\label{eq:spatially_averaged}
    \langle  [\cdot] \rangle_y= \frac{1}{\delta}\int_{y=0}^{\delta} [\cdot]dy
\end{equation}
Figure \ref{fig:av_mean_ke} shows the variation of the absolute values of the spatially averaged term of the mean K.E. equation with volume fraction. A sharp change in the magnitudes are observed in between $\phi=7.875\times10^{-4}$ and $\phi=8.3125\times10^{-4}$. In this region, ,the magnitude of spatial averaged values of the dominant terms e.g. the transport term of mean K.E. due to fluid viscous stress generated by mean velocity gradient, the viscous dissipation of mean K.E. and loss of mean K.E. \textcolor{black}{for turbulence production} show a sharp decrease and loss of mean K.E. due to particle reverse drag an abrupt increase, although less sharper. The extent in decrease for the shear production of turbulence \textcolor{black}{is maximum}. This \textcolor{black}{motivates us} to look into the various terms of the fluctuating K.E. budget of the fluid phase in order to understand the sharp attenuation, occurring above $\phi=7.875\times10^{-4}$.        
\subsection{Fluid phase fluctuating kinetic energy budget equation}
\label{sec:fluc_ke_budget}
\begin{figure*}[!]
	\includegraphics[width=1.0\linewidth]{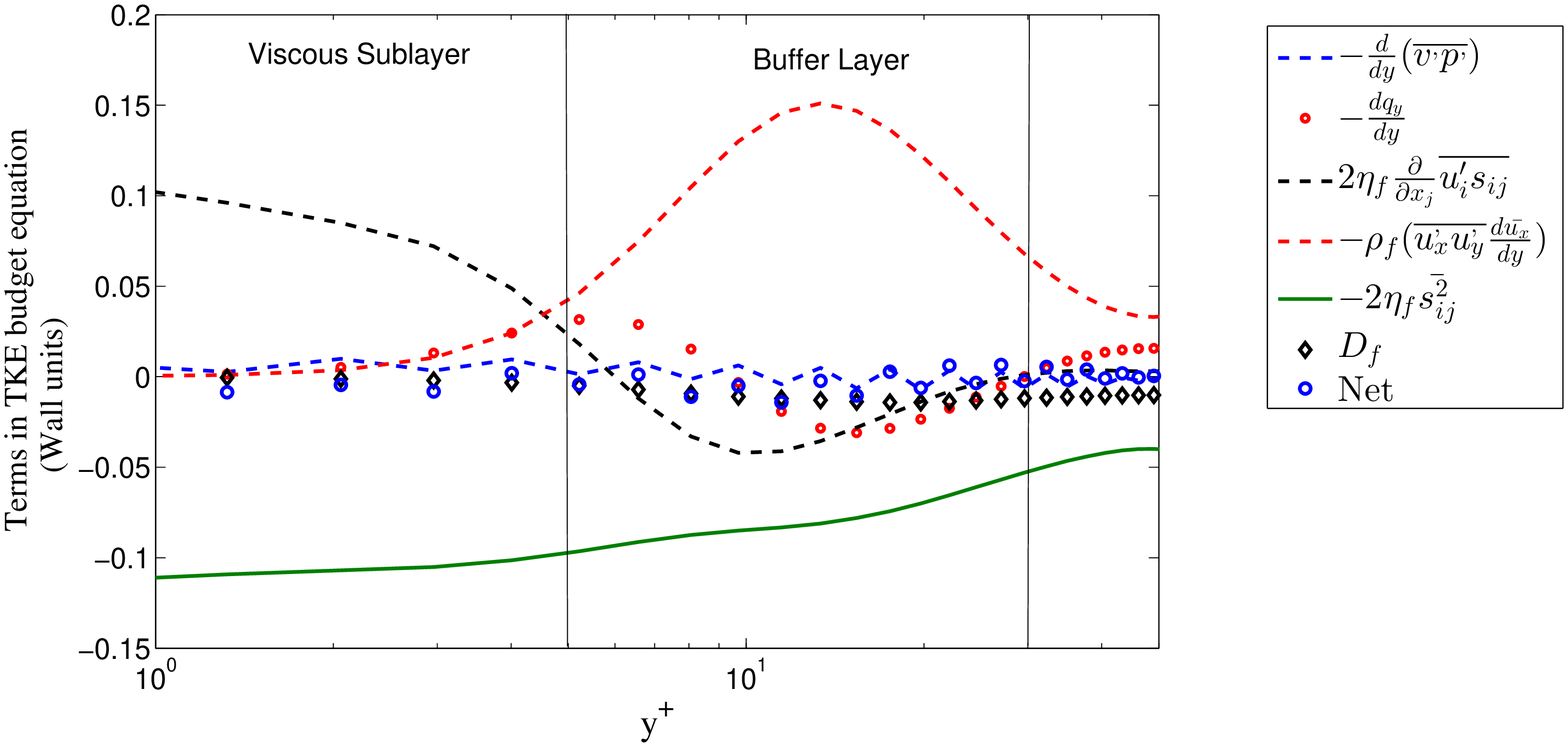}
		\caption{ Terms of fluid phase fluctuating K.E. as a function of cross-stream position in wall-scaling for $\phi=7\times10^{-4}$}
	\label{fig:wall_turb_ke}
\end{figure*}

In the backdrop of wall-bounded turbulent flows, fluctuating kinetic energy fluxes and hence the energy transfer is studied at \textcolor{black}{different wall-normal locations. A detailed analysis of fluctuating K.E. is required in the context of sharp modulation of turbulence fluctuation.} 
The fluctuating kinetic energy budget equation is given by:
 \begin{align}
 	&-\underbrace{\frac{dq_y}{dy}}_{1}-\underbrace{\frac{d}{dy}(\overline{u_y'p'})}_{2}+\underbrace{2\eta_f\frac{\partial}{\partial  x_j}\overline{u_i's_{ij}}}_{3} \nonumber\\&+\underbrace{\tau_{xy}^{Rf}\frac{d\bar{u_x}}{dy}}_{4}-\underbrace{2\eta_f\overline{s_{ij}^2}}_{5}-\underbrace{D_f}_{6}=0
 \end{align}
 \label{fig:turb_ke}
The various terms of the fluctuating turbulent K.E. equation can be given as:
\begin{itemize}
	\item 1: $q_y=\frac{1}{2}\rho_f\overline{u_y'u'^2}$ is the energy flux due to fluid fluctuating velocity
	\item 2: Transport of T.K.E. due to pressure fluctuation 
	\item 3: Transport of T.K.E. due to viscous stresses, where $s_{ij}=\frac{1}{2}\left(\frac{\partial u'_i}{\partial x_j}+\frac{\partial u'_j}{\partial x_i}\right)$ is the fluctuating strain rate tensor
	\item 4: Shear production of turbulence
	\item 5: Viscous Dissipation of fluctuating energy
	\item 6:  $D_f=\overline{\mu_fn_pu_x(u_x-v_x)}+\overline{\mu_fn_pu_y(u_y-v_y)}+\overline{\mu_fn_pu_z(u_z-v_z)}$\\$-\bar{u_x}\overline{\mu_fn_p(u_x-v_x)}$ is the difference between total energy dissipation rate and the mean energy dissipation rate due to the presence of particles
\end{itemize} 
The effect of particle volume fraction on fluid phase fluctuating kinetic energy budget before and after attenuation is given in figure \ref{fig:turb_ke}.\\
\begin{figure*}[!h]
 \begin{minipage}{0.5\textwidth}
 	{\includegraphics[width=1.0\linewidth]{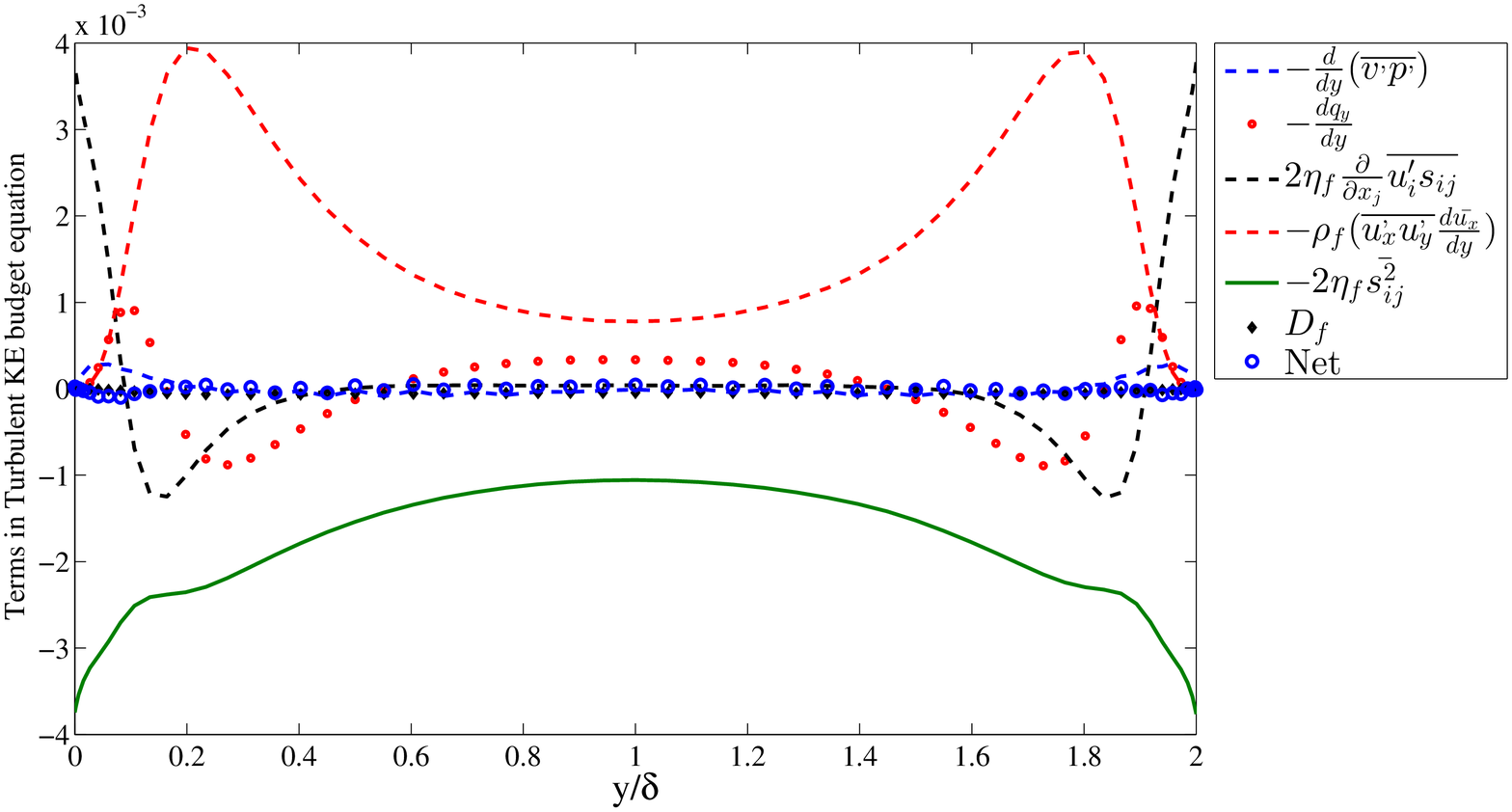}}
 	\caption*{(a)}
 \end{minipage}\hfill
\begin{minipage}{0.5\textwidth}
 	{\includegraphics[width=1.0\linewidth]{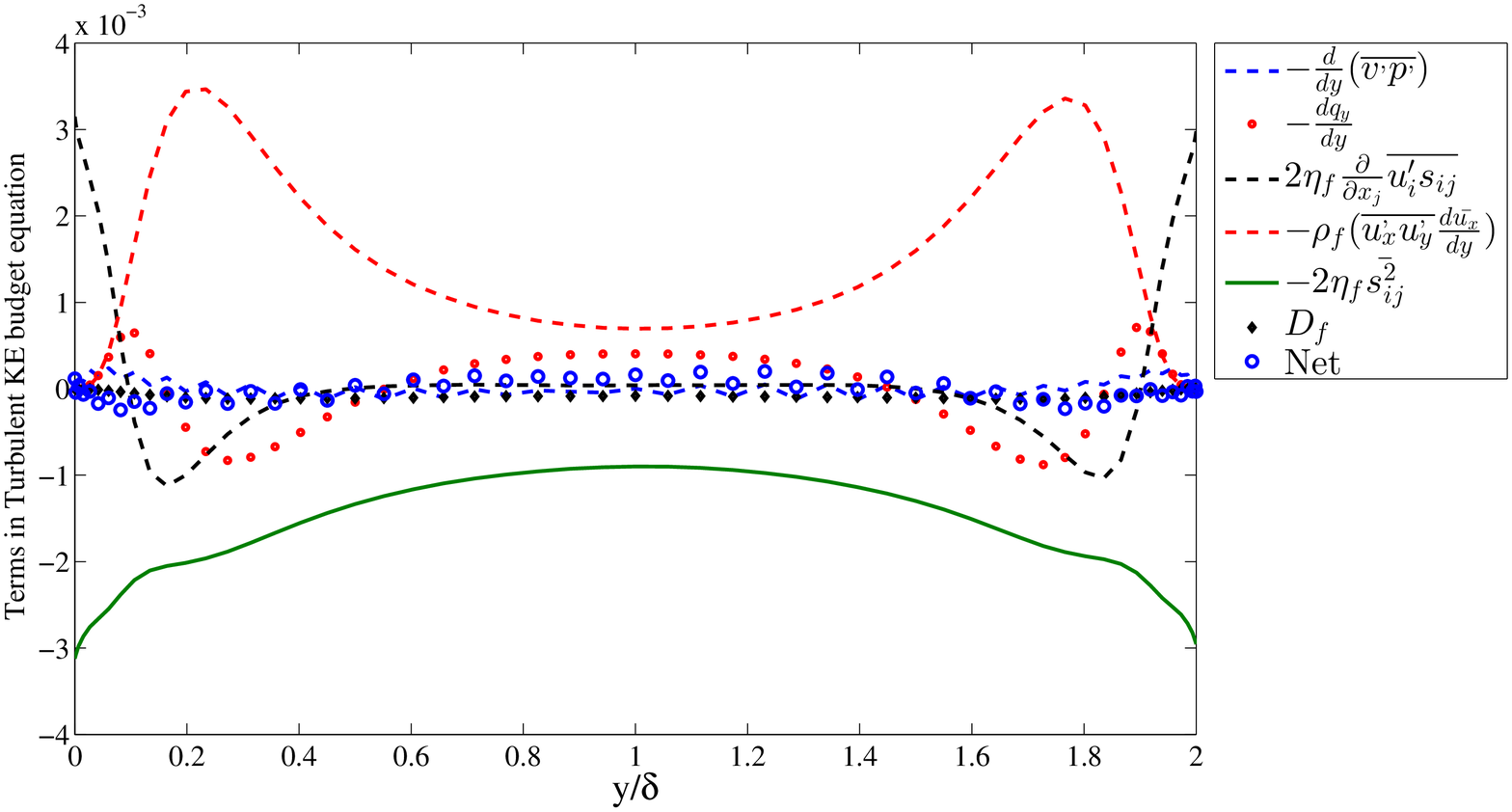}}
 	\caption*{(b)}
 \end{minipage}\hfill
 \begin{minipage}{0.5\textwidth}
 	{\includegraphics[width=1.0\linewidth]{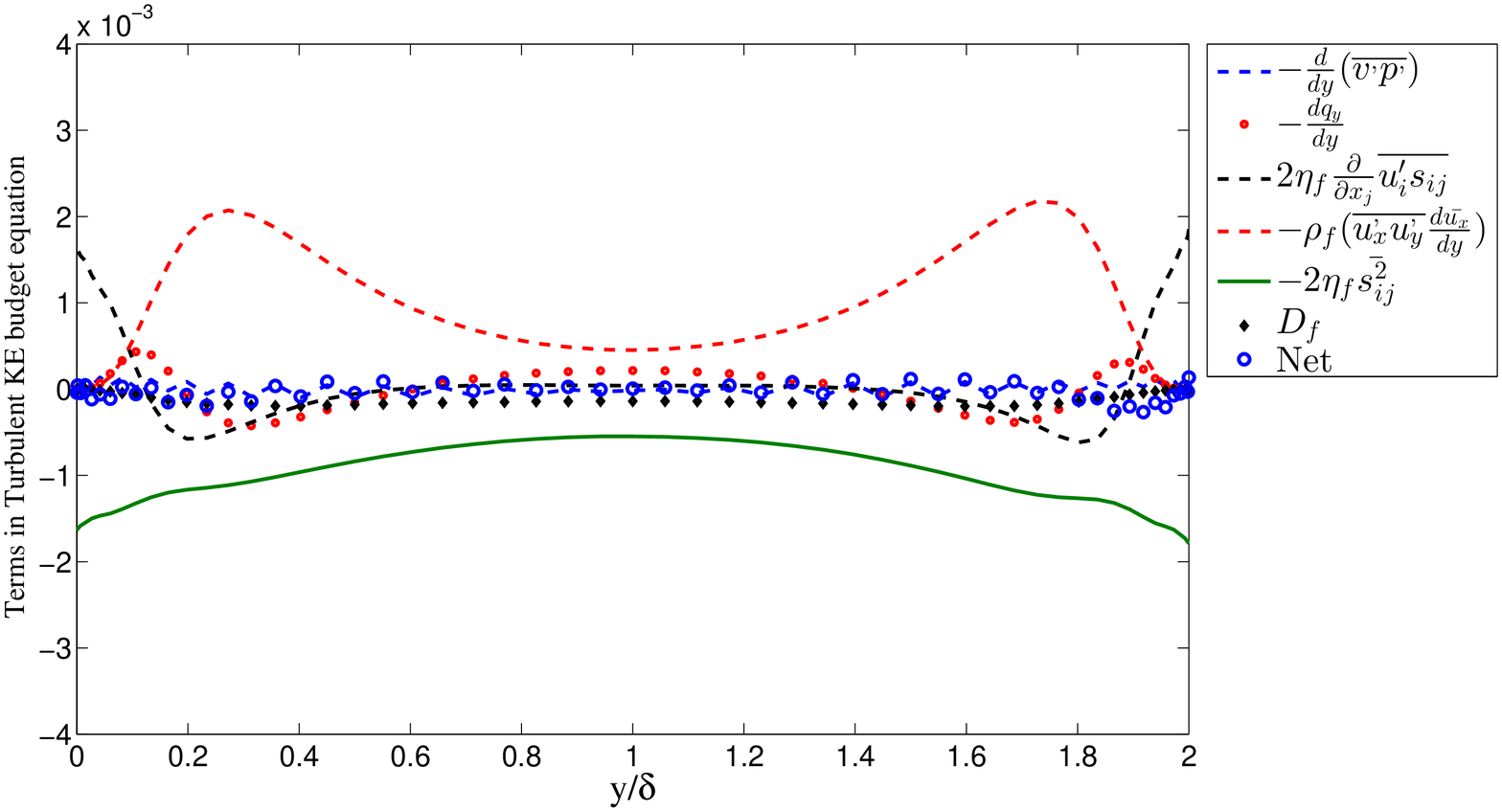}}
 	\caption*{(c)}
 \end{minipage}\hfill
 \begin{minipage}{0.5\textwidth}
 	{\includegraphics[width=1.0\linewidth]{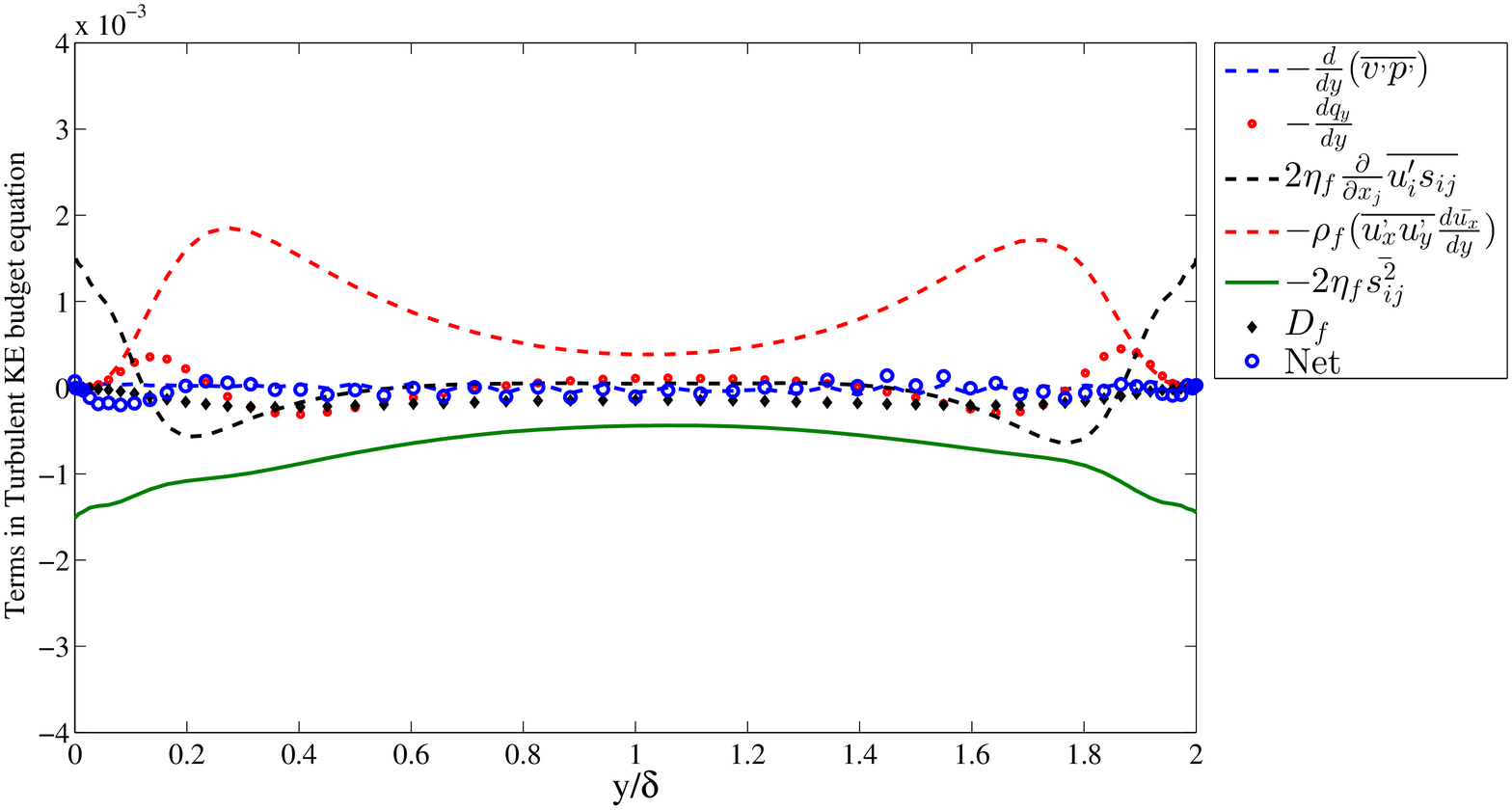}}
 	\caption*{(d)}
\end{minipage}\hfill
 \begin{minipage}{0.5\textwidth}
 	{\includegraphics[width=1.0\linewidth]{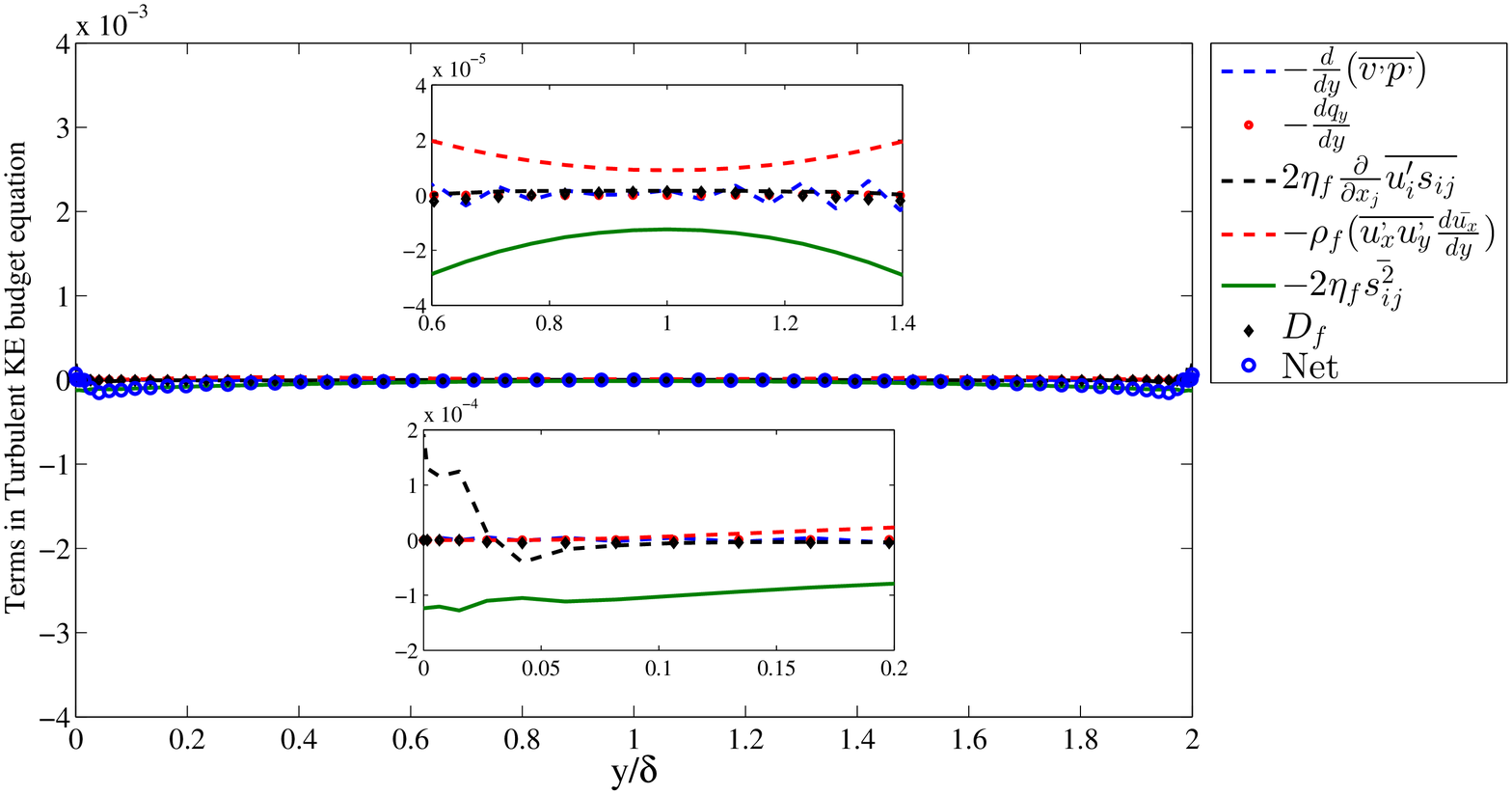}}
	\caption*{(e)}
\end{minipage}\hfill
\begin{minipage}{0.5\textwidth}
	{\includegraphics[width=1.0\linewidth]{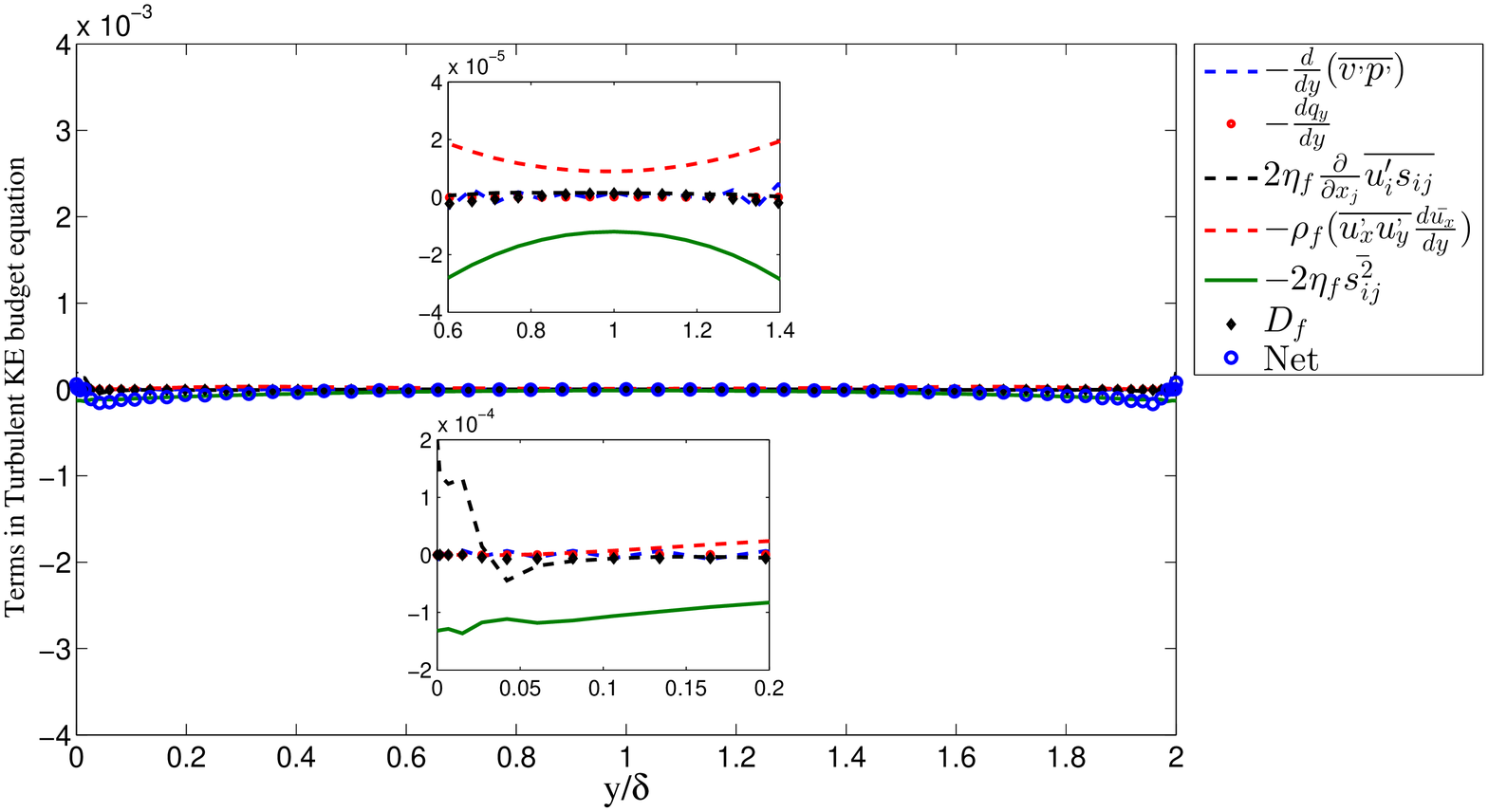}}
	\caption*{(f)}
\end{minipage}\hfill
 	\caption{Fluid phase fluctuating kinetic energy budget at various volume fractions (a) $\phi=1.75\times10^{-4}$, (b) $\phi=3.5\times10^{-4}$, (c) $\phi=7\times10^{-4}$, (d) $\phi=7.875\times10^{-4}$, (e) $\phi=8.3125\times10^{-4}$ and (f) $\phi=8.75\times10^{-4}$}	
 	\label{fig:turb_ke}		       
 \end{figure*}
 It is evident from figures \ref{fig:turb_ke} (a)-(d) that the shear production of turbulence term (term 4) acts as the primary source term of the fluctuating kinetic energy across most of the cross-stream positions except at the walls. This term is generated at the expense of the mean kinetic energy of the fluid as evident from the discussion in the subsection \ref{sec:mean_ke_budget}. The viscous dissipation term (term 5) is the dominant sink term across all the $y/\delta$ positions. However, the magnitude of shear production of turbulence and the viscous dissipation is observed to decrease with increasing volume fraction. Additionally, the transport of fluctuating K.E. due to fluid viscous stresses (term 3) and the transport of fluctuating K.E. due to fluid fluctuating velocity (term 1) are observed to decrease with the decrease of the dominant source and sink term with an increase in volume loading. Fig. \ref{fig:turb_ke} shows a drastic decrease in shear production of turbulence, along with the viscous dissipation term across all the cross-stream positions of the Couette-flow. At the central position these terms remain the dominant source and sink terms respectively. Similarly in the near wall region the principal source and sink terms remain unchanged with respect to figure \ref{fig:turb_ke}. The behaviour of these terms across channel-width is shown in the representative figure \ref{fig:wall_turb_ke}. The transport of fluctuating K.E. due to fluid viscous stresses (term 3) acts as the main source of fluctuating K.E. at the walls due to the higher strain rate in the near wall region.     

 The qualitative behaviours of the terms in fluctuating K.E. budget equation as a function of cross-stream position scaled in wall-units $y^+$ for a volume fraction of $7\times10^{-4}$ is shown in figure \ref{fig:wall_turb_ke} before the attenuation. As a very well-accepted fact , the shear-production of turbulence takes place at in the buffer layer \citep{andrade2018analyzing}. In the viscous sub-layer, due to the higher strain rate, the fluctuating K.E. is imparted to the fluid through fluid viscous stresses. The transport of fluctuating K.E. through fluid velocity fluctuation alters its role of being a source to sink in the buffer-layer and then becomes again a source in the log-law region. The loss in fluid fluctuating K.E. due to particle reverse drag remains very less in magnitude across all the cross-stream distance and volume fractions.       
\begin{figure*}[!h]
	\includegraphics[width=1.0\linewidth]{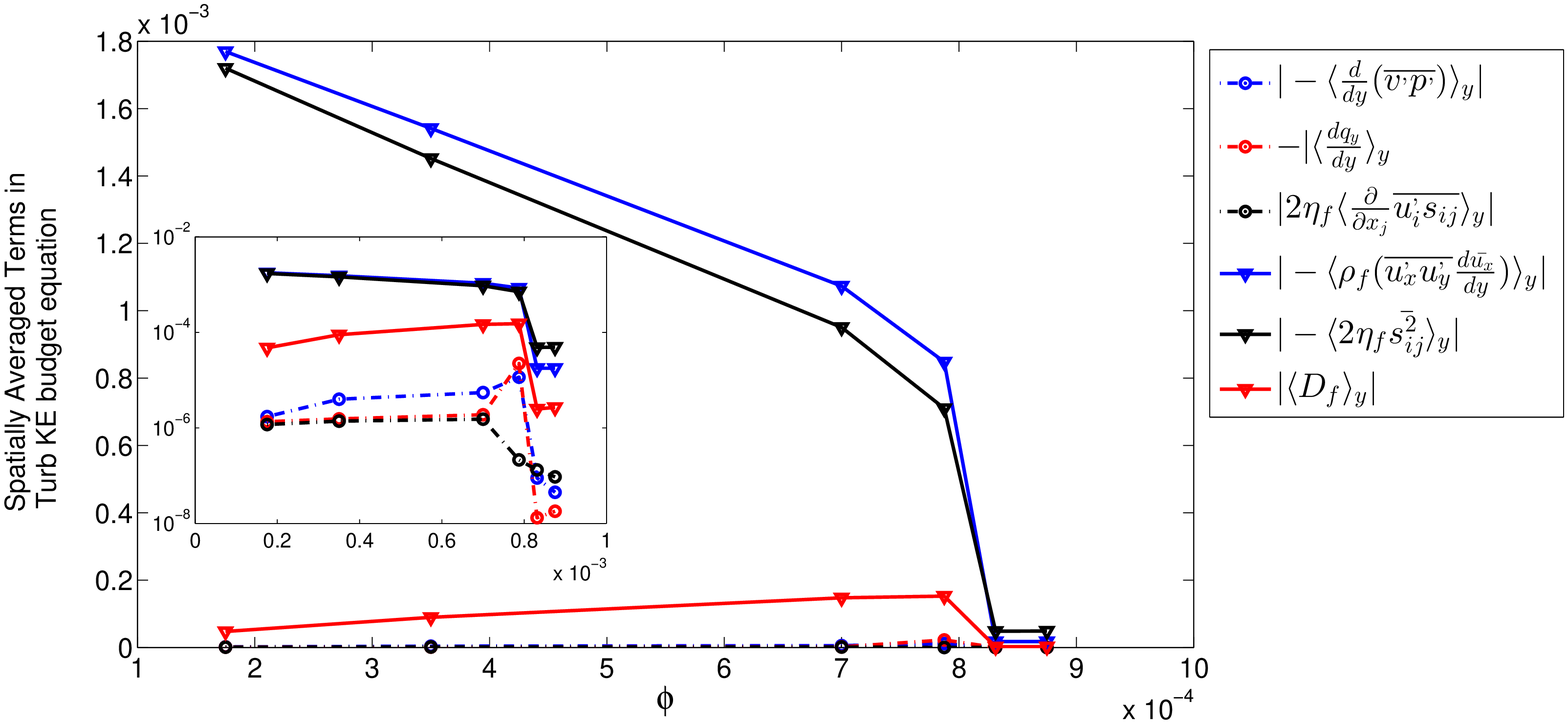}
		\caption{Magnitude of spatially averaged terms of fluid phase fluctuating K.E. as a function of $\phi$}
	\label{fig:av_turb_ke}
\end{figure*}
\\The terms of the fluctuating K.E. budget equation are spatially averaged using equation \ref{eq:spatially_averaged} and the magnitudes are shown  as a function of $\phi$ in figure \ref{fig:av_turb_ke}. From the figure it is evident that a catastrophic decrease in shear-production of turbulence takes place, at a volume fraction of $\phi=7.875\times10^{-4}$. This results in a similar decrease of the viscous dissipation as well. This decrease is observed to be linear for these two dominant terms in the lower volume fractions. The decrease in fluid fluctuating K.E. due to particle drag is almost linear \textcolor{black}{with partial loading} up to a critical volume fraction \textcolor{black}{which is followed} by a sharp decrease.
\\\textcolor{black}{The figure also depicts that dissipation due to particle at the fluctuating scale is more than one order of magnitude smaller than the dissipation due to particle drag at the mean scale.}	
\clearpage

\section{Step-wise Particle Injection Study}
\begin{figure}[!]
	\includegraphics[width=0.65\textwidth]{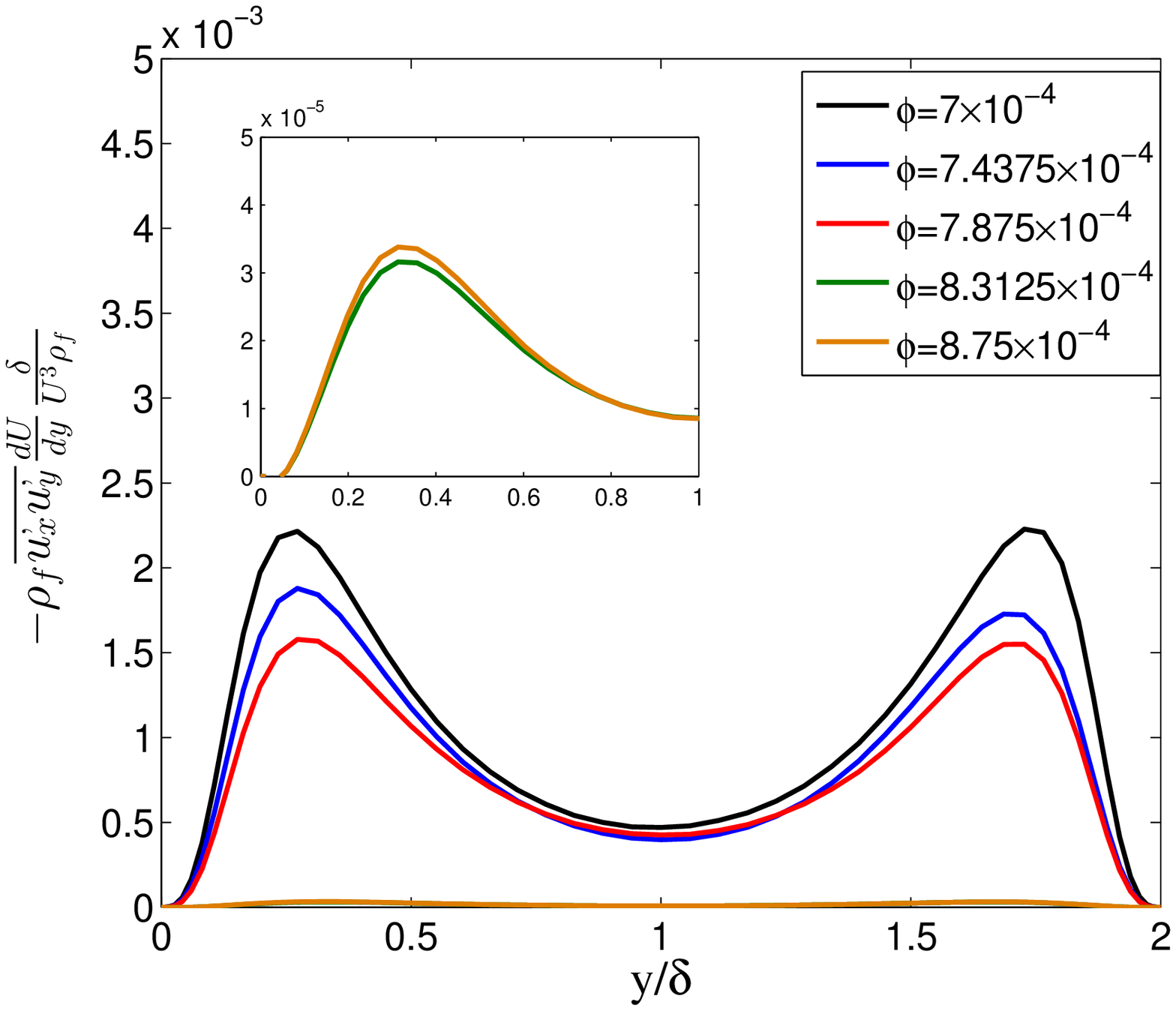}
	\caption{Profile of non-dimensionalized turbulence production term $-\rho_f\overline{u_x^, u_y^,}\frac{dU}{dy}\frac{\delta}{U^3\rho_f}$ at various steps of volume fraction in the step-wise particle injection study} 
	\label{fig:prod_step}
\end{figure}

In order to confirm the fact that it is only the presence of the particles \textcolor{black}{with a particular volume fraction} that drive the drastic discontinuous \textcolor{black}{collapse} in the fluid phase turbulence \textcolor{black}{and the phenomenon is not initial condition dependent}, a step-wise particle injection study is carried out. This study has been initiated from a statistically steady state previously achieved for a turbulent suspension with particle volume fraction $\phi=7\times10^{-4}$. In the first step, a step-injection of 500 of particles, with random initial positions and velocities, is done such that the volume fraction of the suspension increases to $\phi=7.4375\times10^{-4}$  and then the system is allowed to reach statistical steady-state. Except the volume fraction, all the other parameters in this study, are kept unchanged as discussed in \ref{sec:sim_meth}. Further step injections with 500 of particles are carried out sequentially and resulted in volume fractions $7.875\times10^{-4}$, $8.3125\times10^{-4}$, $8.75\times10^{-4}$ in second third and fourth step respectively. 
\begin{figure}[!]
	\includegraphics[width=0.65\textwidth]{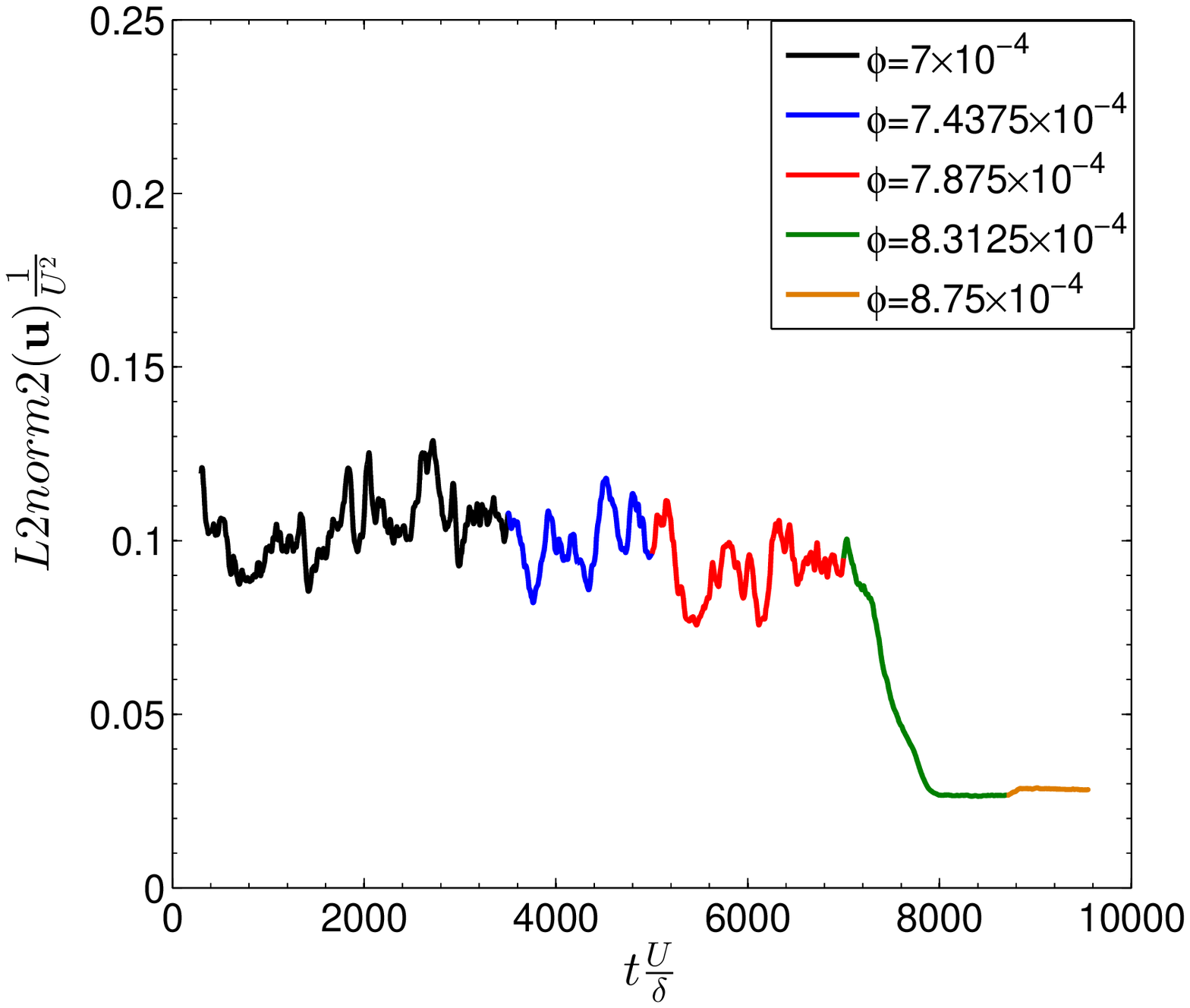}
	\caption{Time evolution of $L2norm2(\mathbf u)$ at various steps of volume fraction in step-wise particle injection study} 
	\label{fig:L2norm_step}
\end{figure}
\\ A few very interesting phenomena are observed and are shown in figures \ref{fig:L2norm_step} and \ref{fig:prod_step}. Figure \ref{fig:L2norm_step} shows the temporal evolution $L2norm2(\mathbf{u})$ which is defined as:
\begin{math}
L2norm2(\mathbf{u})=\int_{0}^{L_x}\int_{0}^{2\delta}\int_{0}^{L_z} (\textbf{u}\cdot\textbf{u}) dxdydz
\end{math}
where $\mathbf{u}$ represents the fluid velocity disturbance field evolved from the initial numerical perturbation (standard method of DNS). It is to be mentioned that the 'zero time' is counted from the start of the simulation of the unladen fluid subjected to the initial numerical perturbation. After the statistical steady value of the $L2norm2(\mathbf u)$ is achieved, 8000 particles are injected to initiate the simulation of two-way coupled particle-laden sheared turbulent flow.
From the figure it is evident that the addition of particles over $\phi_{cr}=7.875\times10^{-4}$ \textcolor{black}{reduces} the fluid disturbance field.  
\begin{figure*}[!]
	\includegraphics[width=1.0\textwidth]{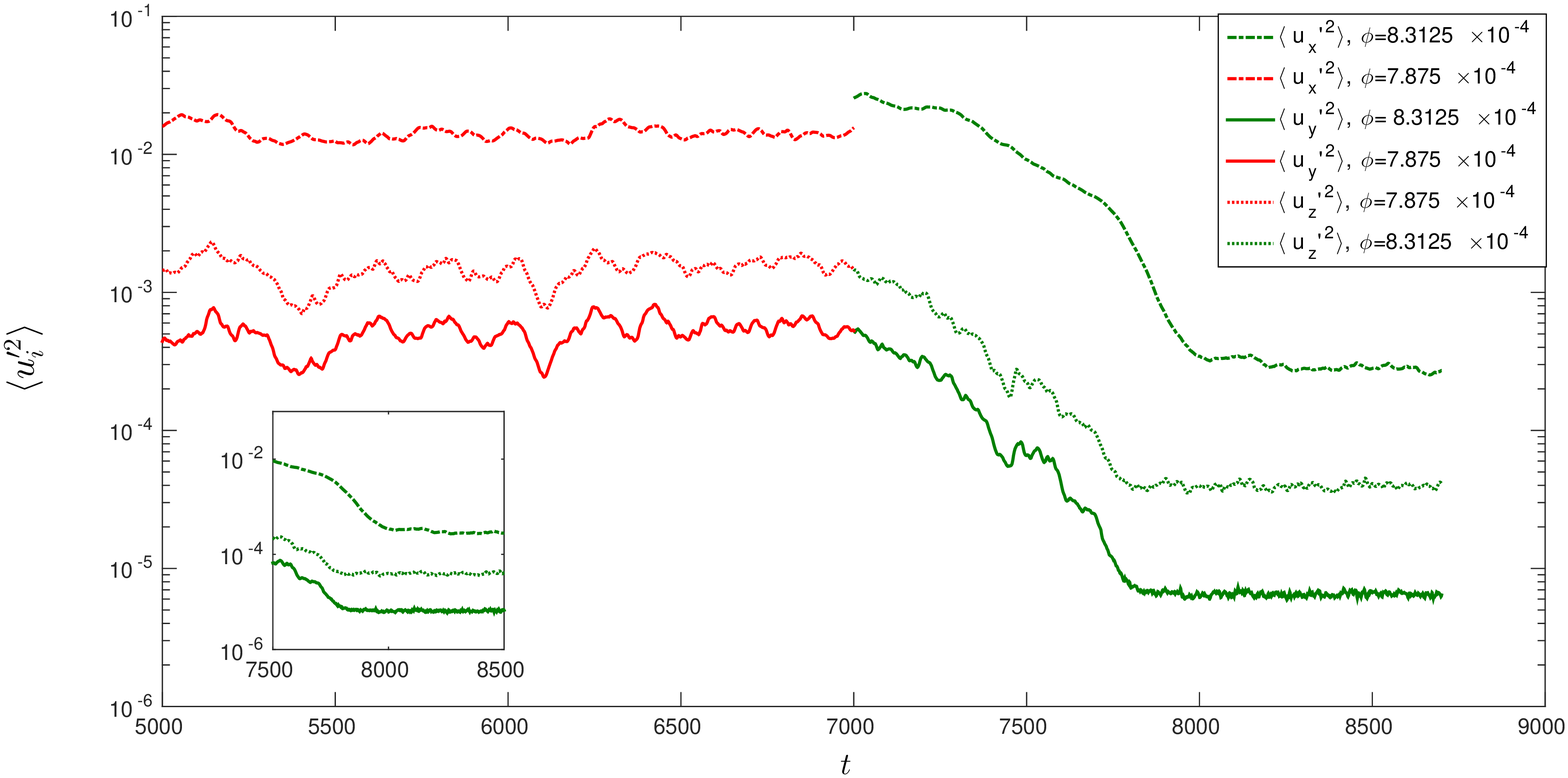}
	\caption{Time evolution of mean square fluid velocity $\langle{u_i^{,2}}\rangle$ about the point of critical volume fraction} 
	\label{fig:uu_ms}
\end{figure*}
Figure \ref{fig:prod_step} shows that \textcolor{black}{that there is a} drastic decrease of non-dimentionalized shear-production of turbulence term due to further injection of the particles over $\phi_{cr}=7.875\times10^{-4}$. This decrease is almost two-orders of magnitude. Hence it is quite evident that in step-wise injection study the critical volume fraction remains unchanged. The temporal evolution of the \textcolor{black}{second moments of different components of fluid velocity fluctuations} is studied about the critical volume fraction, for the second and third particle injection step. The decrease for individual mean square fluid velocity component is observed to decrease drastically (between one and two orders of magnitude), qualitatively similar to that of the \textcolor{black}{decrease in} $L2norm2(\mathbf u)$ and the shear production of turbulence term. The cross-stream velocity fluctuation is observed to be decreased before the streamwise velocity fluctuation. This carries a signature that decrease in $\overline{u_y^{,2}}$ brings about a drastic decrease in Reynolds stress which in turn affects the turbulence production term.      
\section{Step-wise Particle Removal Study}
After the third step of particle injection, the discontinuous decrease in the turbulence is observed at $\phi=8.3125X10^{-4}$. Consequently it becomes relevant to check the effect of step-wise particle removal from the system on the temporal evolution of a few crucial turbulent statistics. \textcolor{black}{Such a strategy helps us to identify whether there exists any hysteresis in the phenomenon of turbulence collapse}. A step removal of 500 particles is carried out which lowers the volume fraction from $\phi=8.3125X10^{-4}$ to $\phi=7.875X10^{-4}$. Another step-removal run is carried out along with the injection of a numerical  disturbance field that is traditionally used in DNS \textcolor{black}{to transform laminar base state to a turbulent profile}. 
\begin{figure}[!]
\centering
	\includegraphics[width=0.65\textwidth]{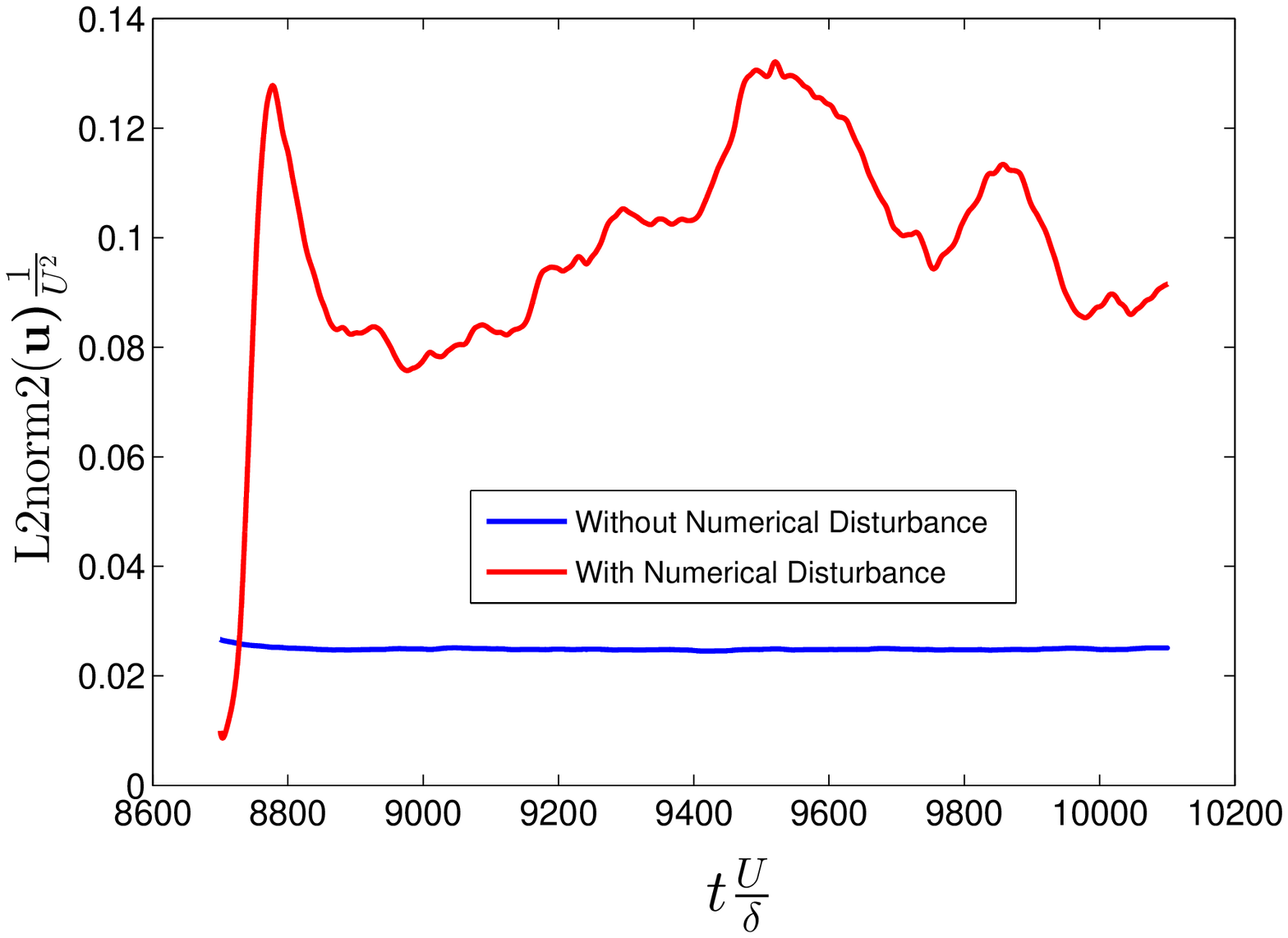}
	\caption{Time Evolution of $L2norm2(\mathbf u)$ in the step-wise particle removal study in presence and in absence of additional noise} 
	\label{fig:L2norm_step_removal}
\end{figure}
\begin{figure}[!]
\centering
	\includegraphics[width=0.65\textwidth]{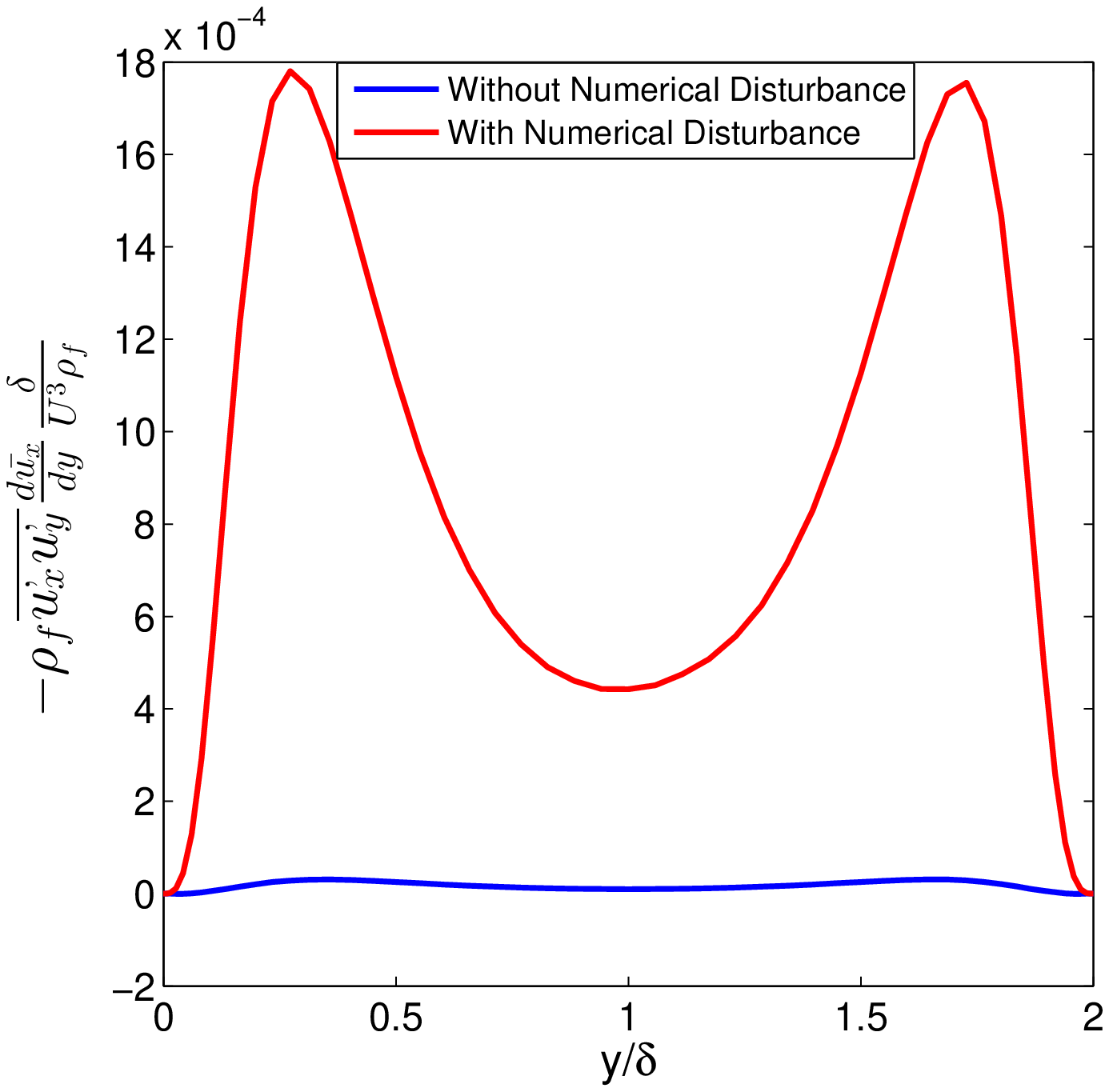}
	\caption{Profile of non-dimensionalized turbulence production term $-\rho_f\overline{u_x^, u_y^,}\frac{dU}{dy}\frac{\delta}{U^3\rho_f}$ in the step-wise particle removal study in presence and in absence of additional noise} 
	\label{fig:prodstep_removal}
\end{figure}
Figure \ref{fig:L2norm_step_removal} and figure \ref{fig:prodstep_removal} reveals the difference between the two runs. It is observed that the removal of the particles is not sufficient to bring back the magnitude of L2norm or the shear production of turbulence term to the values in the pre-attenuation phase. Numerical disturbance is required to restore the turbulence disrupted at $\phi_{cr}$. \textcolor{black}{Such an observation also indicates that when particle volume fraction is greater than $\phi_{cr}$, the fluid flow becomes laminarized.}  
\\ The step-wise particle injection and removal study together show that it is the presence of particles that is exclusively behind the discontinuous transition \textcolor{black}{and no hysteresis is observed.}
This modulation in turbulence \textcolor{black}{leads to a laminar state} and the turbulent state could not be restored by step-removal of particles only without the application of disturbance in the flow-field. 
\section{Key Observations on Modifications in Fluid Phase Dynamics}
In summary, a discontinuous decrease of fluid turbulence intensity, mean square velocity and Reynolds stress was observed beyond a volume fraction of $\phi\sim 7.875\times10^{-4}$ for $St\sim367$ and $Re=750$. The discontinuous decrease happens along with discontinuous modification in the mean fluid velocity profile and mean fluid velocity gradient statistics. In the mean momentum budget, the momentum transport due to viscous stress term drastically reduces above the $\phi_{cr}$.
Transport of mean K.E. terms due to fluid viscous stress and viscous dissipation of mean K.E. show a sharp drop due to the transition. The drastic reduction of shear production of turbulence along with the viscous dissipation of turbulent K.E. are two important phenomenon occurring during the discontinuous transition. The step-wise particle injection and removal studies revealed that it is the presence of particles which is majorly behind this discontinuous transition.
\section{Turbulence Modulation and Patterns in Streamwise Velocity and Vorticity Fluctuations}
\label{sec:contours}
In the previous sections, the discontinuous transition of turbulence is mainly studied through statistics and energy budgets. It is also relevant to study the streamwise velocity and vorticity contours, namely the primal \textcolor{black}{characteristics} of the \textcolor{black}{large scale motions}. In this section, the effect of \textcolor{black}{increasing particle volume loading} on contours of streamwise velocity fluctuations and vorticity is shown and is compared with the unladen fluid. All the contour plots are shown at a single instant of time after statistically stationary state is achieved. 
\textcolor{black}{It is worth to mention that no significant difference is observed even if we compare two such instantaneous frames captured with a time difference of 10 non-dimensional units. For particle laden cases, the position of the particles are denoted by the black dots.}

\begin{figure*}[!h]
\centering
\includegraphics[width=1.0\linewidth]{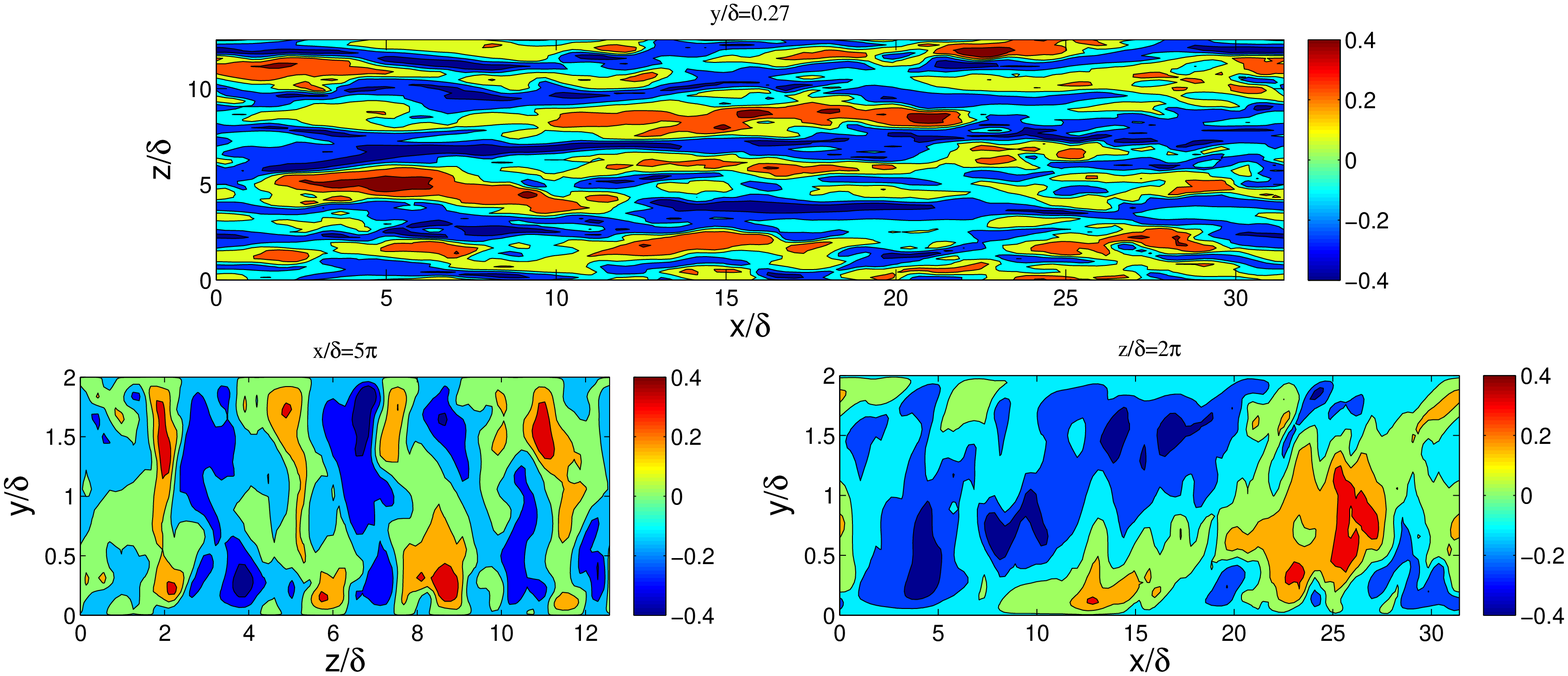}
	\caption{Contours of streamwise velocity fluctuations in unladen fluid phase showed in three different planes as shown in the figures} 
	\label{fig:unladen_vel_contour}
\end{figure*}
\begin{figure*}[!h]
\centering
\includegraphics[width=1.0\linewidth]{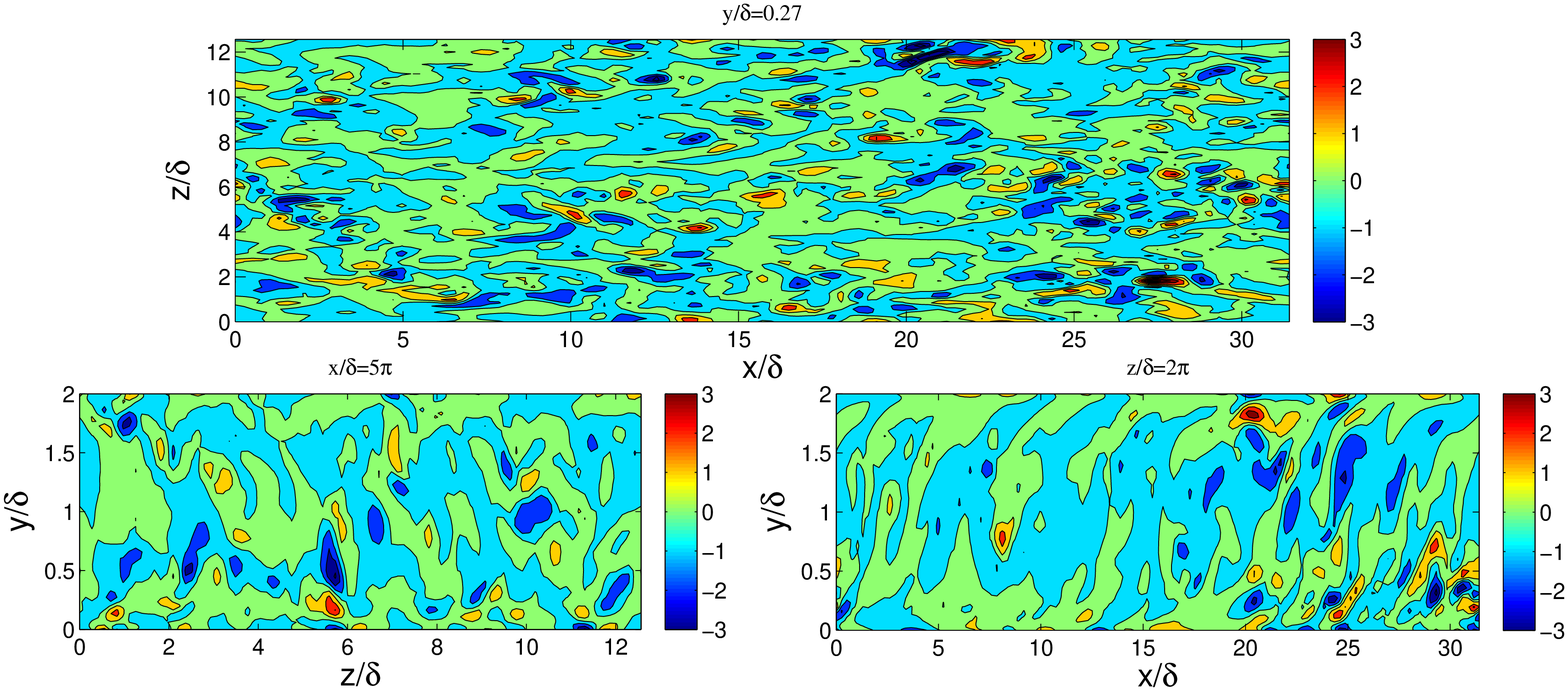}
	\caption{Contours of streamwise vorticity fluctuations in unladen fluid phase showed in three different planes as shown in the figures} 
	\label{fig:unladen_vort_contour}
\end{figure*}
\begin{figure*}[!h]
\centering
\includegraphics[width=1.0\linewidth]{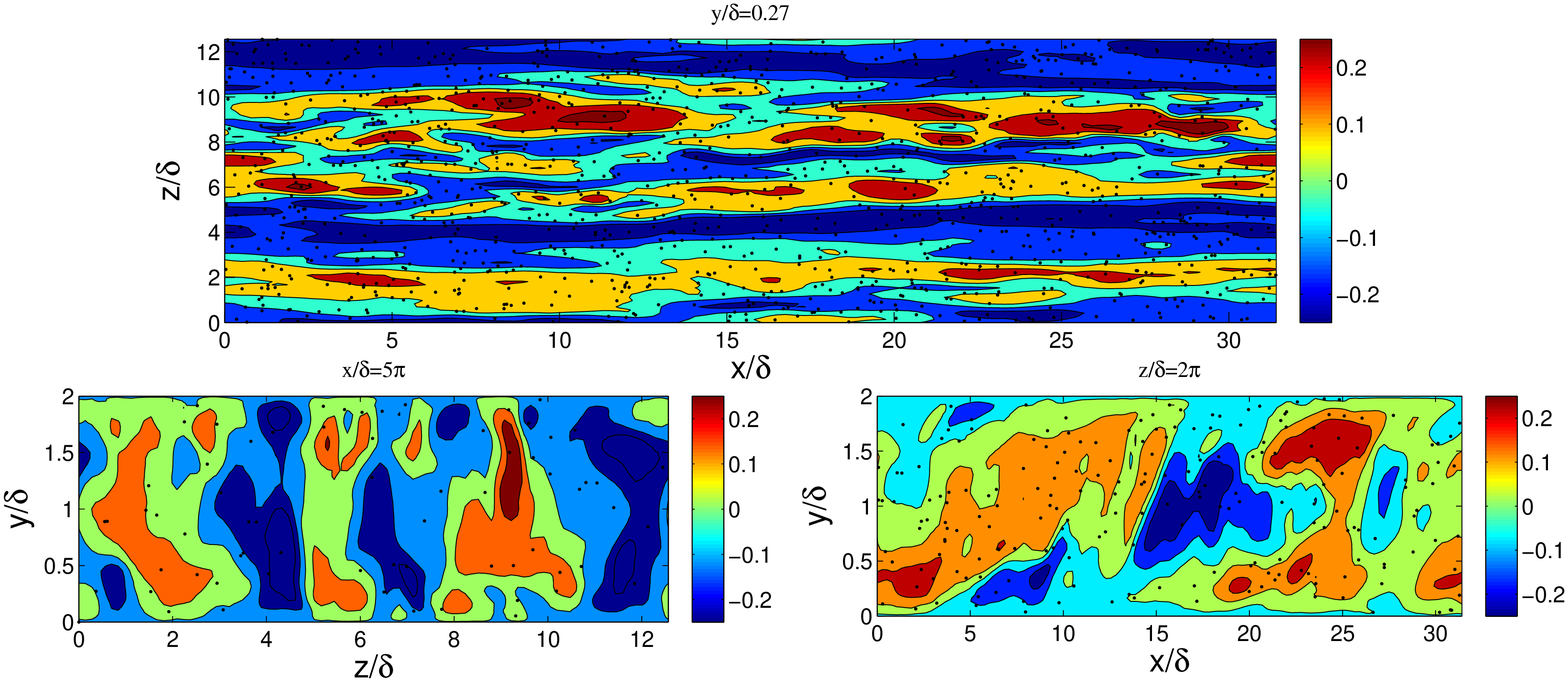}
	\caption{Contours of streamwise velocity fluctuations of fluid phase laden with particles of Volume Fraction $\phi_{cr}=7.875X10^{-4}$ showed in three different planes as shown in the figures} 
	\label{fig:ideal_vel_contour_b4_tr}
\end{figure*}
\begin{figure*}[!h]
\centering
\includegraphics[width=1.0\linewidth]{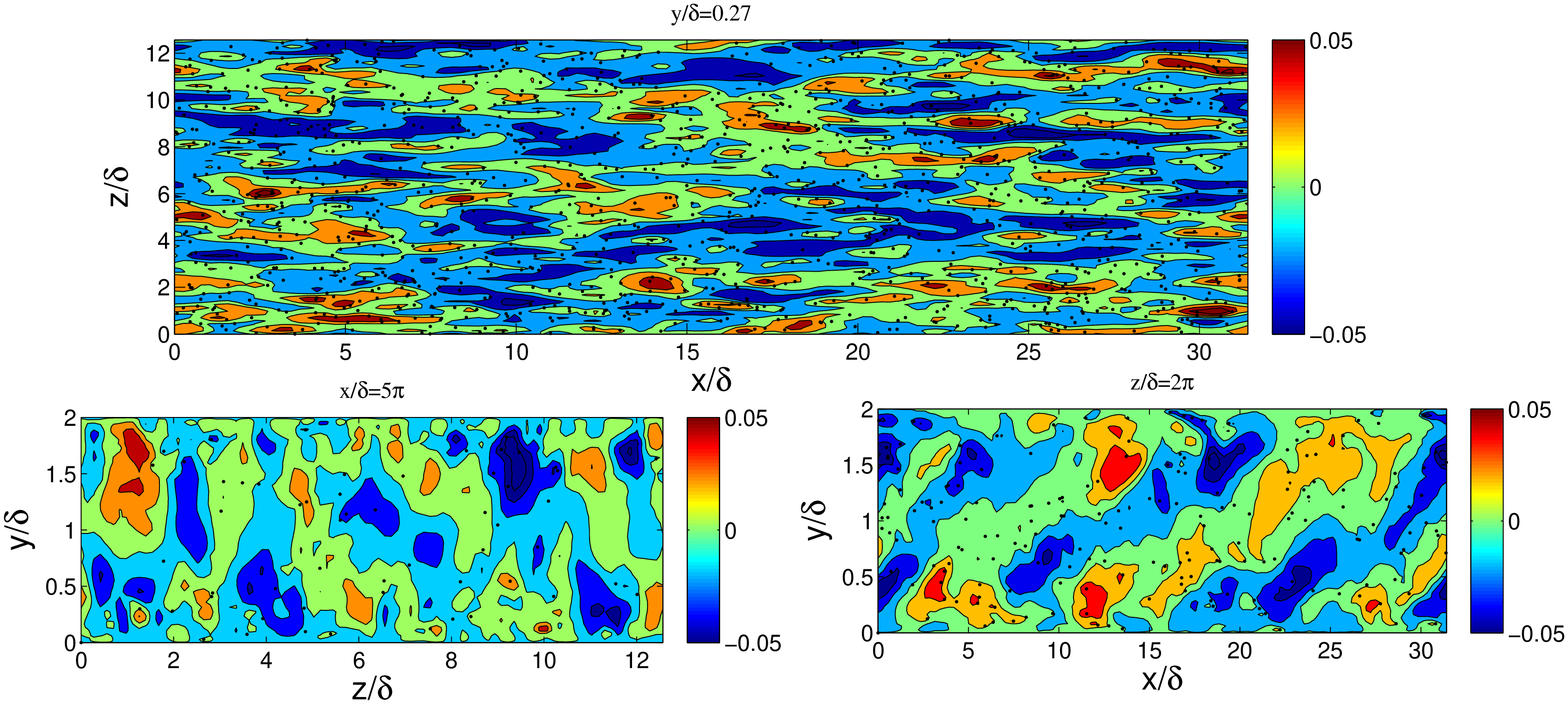}
	\caption{Contours of streamwise velocity fluctuations of fluid phase laden with particles of volume fraction $\phi=8.3125X10^{-4}$ showed in three different planes as shown in the figures} 
	\label{fig:ideal_vel_contour_aftr_tr}
\end{figure*}
The streamwise velocity fluctuation contours \textcolor{black}{shown in fig \ref{fig:unladen_vel_contour},} demonstrate the streaky structures in the streamwise-spanwise (x-z) plane near the wall $y/\delta=0.27$. The generation, evolution and nature of the streaks in the perspective of turbulent coherent structures are  well studied in last few decades for turbulent unladen fluid in turbulent boundary layer flow, \citep*{kline1967structure,praturi1978stereoscopic,head1981new} homogeneous isotropic turbulent flow \citep*{sekimoto2016direct, dong2017coherent} and turbulent Couette flow \cite{papavassiliou1997interpretation,tsukahara2006dns,hwang2010amplification,pirozzoli2011large}. In Particle-laden turbulent flows, the streaky structures are important to capture  particle-fluid interaction and hence the turbulence modification \citep*{hamilton1995regeneration, richter2013momentum, richter2014modification, richter2015turbulence}.
Figure \ref{fig:unladen_vel_contour} captures the streamwise velocity fluctuation of the unladen fluid projected in three different 2-D planes. The contours of the x-z plane, captured at the buffer layer, shows the characteristic low-speed velocity streaks spanning the entire simulation box. In the projection in the y-z plane, captured at the middle of the spanwise direction, the low-speed contours are seen to be more probable to occur near wall. It is argued that the existence of low-speed velocity zones are in between two counter-rotating tails, present occur near the wall, of the hair-pin vortex \citep*{Hinze, head1981new, dennis2015coherent}. \textcolor{black}{The phenomenon of ejection of low-speed streaks away from the wall and the sweeps of high-velocity streaks towards the wall at the buffer region create the flow-patterns away from the walls \citep{kline1967structure, Hinze, dennis2015coherent}}. Hence the streamwise velocity contours provide a suggestive (qualitative) rather than definitive understanding of the \textcolor{black}{large scale (coherent)} structures. The contour in the x-y plane shows low-speed region present in an oblique plane with the mean-flow direction. This may bear the signatures of large hair-pin vortices which typically orient themselves at certain angles with the mean-flow \citep{Hinze, head1981new, dennis2015coherent}. \textcolor{black}{However it is not very much clear to us at this point}. Figures \ref{fig:ideal_vel_contour_b4_tr} and \ref{fig:ideal_vel_contour_aftr_tr} show the streamwise velocity contours for turbulent fluid laden with particles before and after \textcolor{black}{the turbulence collapse}. The concentration of heavy inertial particles, as expected, do not show any correlation \textcolor{black}{of particle concentration} with streamwise velocity fluctuation field. 
\textcolor{black}{Figures \ref{fig:unladen_vel_contour} and \ref{fig:ideal_vel_contour_b4_tr} show that contours for streamwise fluctuations in wall parallel planes are very much similar except for particle laden flow ($\phi>\phi_{cr}$, Fig. \ref{fig:ideal_vel_contour_aftr_tr}), few of the smaller scales have been dumped. A similar observation has also been reported by \citet{zhao2010turbulence} in a particle laden channel flow.}
However, (fig: \ref{fig:ideal_vel_contour_aftr_tr}) shows very different qualitative pictures of the streamwise velocity contours. It is observed that the long streaky structures vanish and the intensity of the velocity fluctuations decreased in all the three planes by \textcolor{black}{approximately} one-order of magnitude. Hence discontinuous disruption in turbulence brings about drastic change in streamwise velocity fluctuation contours, qualitatively.
\begin{figure*}[!h]
\centering
\includegraphics[width=1.0\linewidth]{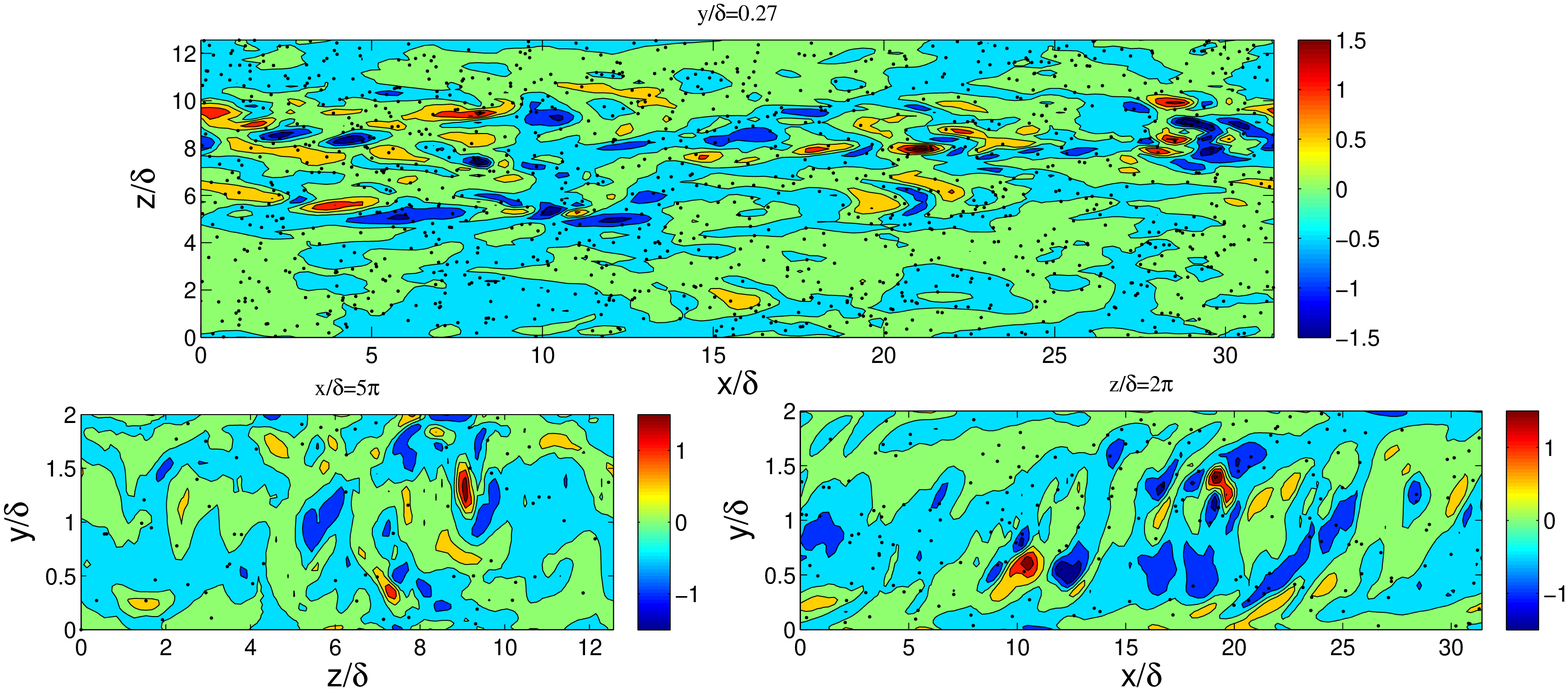}
	\caption{Contours of streamwise vorticity fluctuations of fluid phase laden with particles of Volume Fraction $\phi_{cr}=7.875X10^{-4}$ showed in three different planes as shown in the figures} 
	\label{fig:ideal_vort_contour_b4_tr}
\end{figure*}
\begin{figure*}[!h]
\centering
\includegraphics[width=1.0\linewidth]{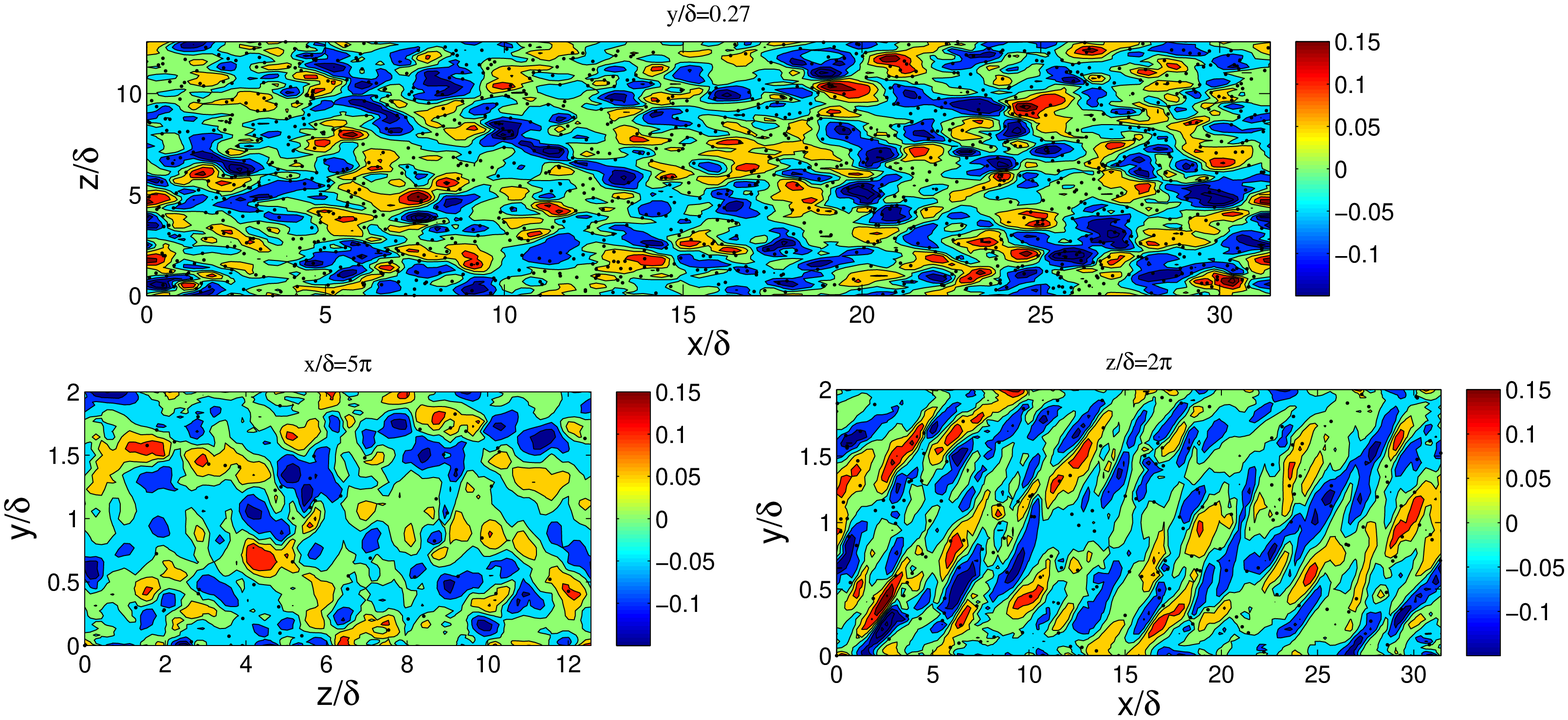}
	\caption{Contours of streamwise vorticity fluctuations of fluid phase laden with particles of Volume Fraction $\phi=8.3125X10^{-4}$ showed in three different planes as shown in the figures} 
	\label{fig:ideal_vort_contour_aftr_tr}
\end{figure*}
The streamwise vorticity contours are captured in the figure\ref{fig:unladen_vort_contour} for unladen fluid and in figures \ref{fig:ideal_vort_contour_b4_tr} and \ref{fig:ideal_vort_contour_aftr_tr} for particle-laden turbulent suspension. 
These high vorticity zones roughly can be seen to exist in opposite signs (blue patches and red/yellow patches) arranged near to each-other. In the context of trains of hair-pin vortices, the streamwise counter rotating high-magnitude fluid zones occur at the tail region of the larger stable structures. This is more prominent in x-z plane where high magnitude vorticity zones are observed to exist in patches. The low-magnitude rotating zones (green and sky-blue) are observed to be of \textcolor{black}{larger sizes in all the three planes of projections}. The vorticity contours in x-y plane (\textcolor{black}{Fig. \ref{fig:unladen_vort_contour}, \ref{fig:ideal_vort_contour_b4_tr} and \ref{fig:ideal_vort_contour_aftr_tr}}) arrange themselves in an oblique plane inclined \textcolor{black}{to} the mean-flow direction, qualitatively very similar to what has observed for the streamwise velocity fluctuation contours. The concentration of heavy inertial particles, not showing any correlation with streamwise vorticity field is similar to the behaviour observed in the streamwise velocity field as well. Discontinuous modulation in turbulence lowers the magnitude of the streamwise vorticity by about one order as observed in figure \ref{fig:ideal_vort_contour_aftr_tr} and hence the larger low-rotating vortical contours are not observed. Thus the qualitative study of the contours of streamwise velocity and vorticity, shows some elementary signatures of breaking-down of coherent structures due to discontinuous turbulence disruption caused by high-inertial particles.\textcolor{black}{The fluctuations in streamwise velocity and vorticity may be generated due to particle phase fluctuation has been discussed later.} 
\section{Role of Inter-particle Collisions on Turbulence Modulation}
The discontinuous disruption of turbulence observed in the particle-laden turbulent shear flow is found out to be a result of drastic reduction of the shear production of turbulence \textcolor{black}{due to increase in} particle number density in the system. The contour plots reveal the break down of the streamwise velocity streaks after \textcolor{black}{complete collapse of the turbulence}. All of these observations reveal the interaction between particle and fluid phase thorough reverse feed-back force to be the crucial factor for the \textcolor{black}{turbulence attenuation. But}, it is not quite intuitive to comment on whether the presence of inter-particle collisions or their nature would bring about any change in the nature of the turbulence modification. In this section two different cases are studied \textcolor{black}{turbulence modulation when inter-particle and wall-particle collisions are inelastic} and for the case where the inter-particle collisions are switched \textcolor{black}{but particle-wall collision are activated.}   
\subsection{Effect of inelastic collisions on turbulence modulation}
\label{sec:inelastic_collision_2way}
The effect of inelastic collisions on turbulence is studied by keeping the co-efficient of restitution ($e$) to 0.9 for inter-particle and wall-particle collisions. Fluid phase velocity statistics namely the mean velocity and the second moments \textcolor{black}{of the fluctuating velocity} contours of streamwise velocity fluctuations and vorticity \textcolor{black}{fluctuations} of the fluid phase.
\subsubsection{\textbf{Fluid velocity Statistics}}
    \begin{figure}[h!]
        	\includegraphics[width=0.65\linewidth]{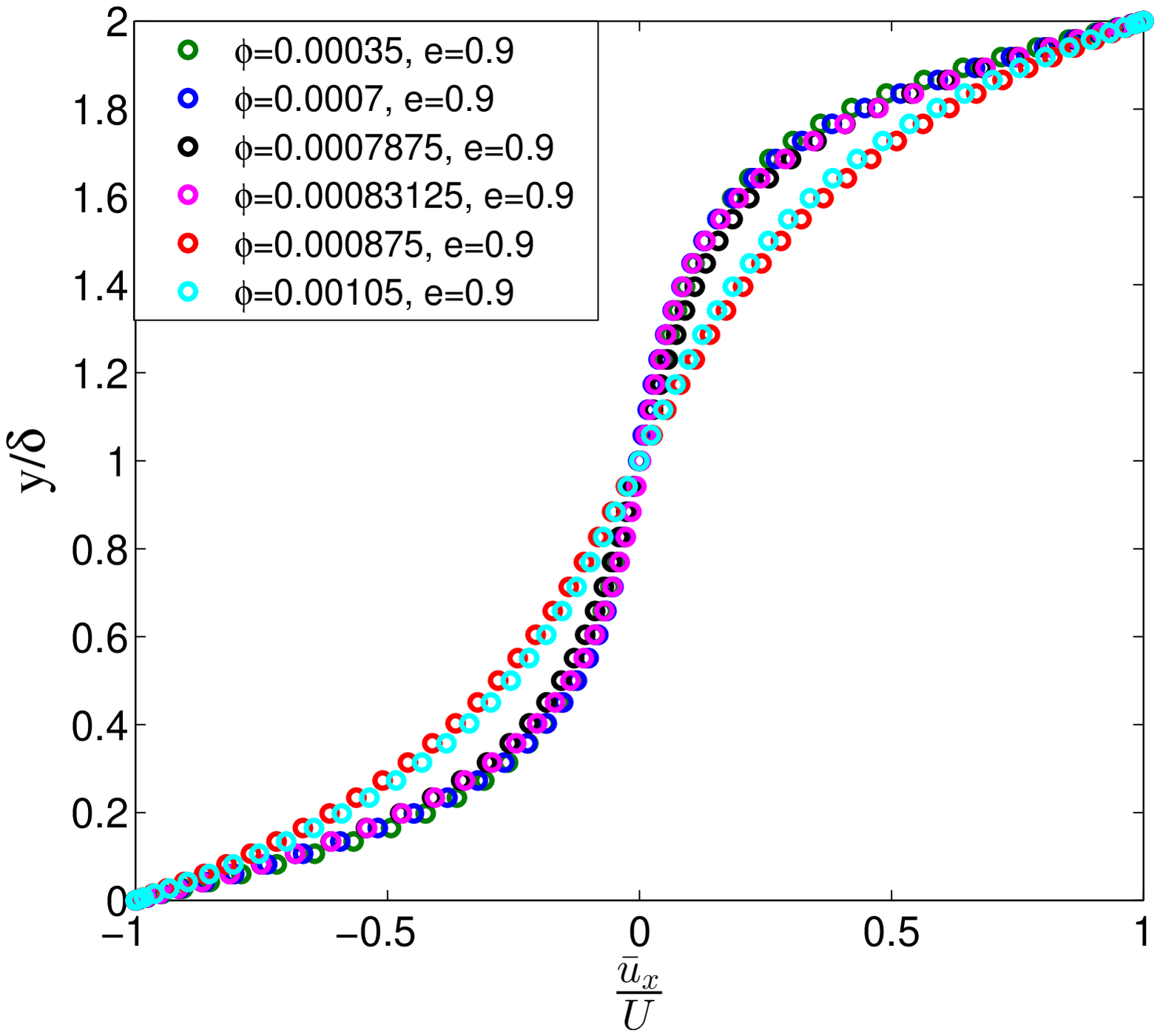}
    	\caption*{(a)}
  	\includegraphics[width=0.65\linewidth]{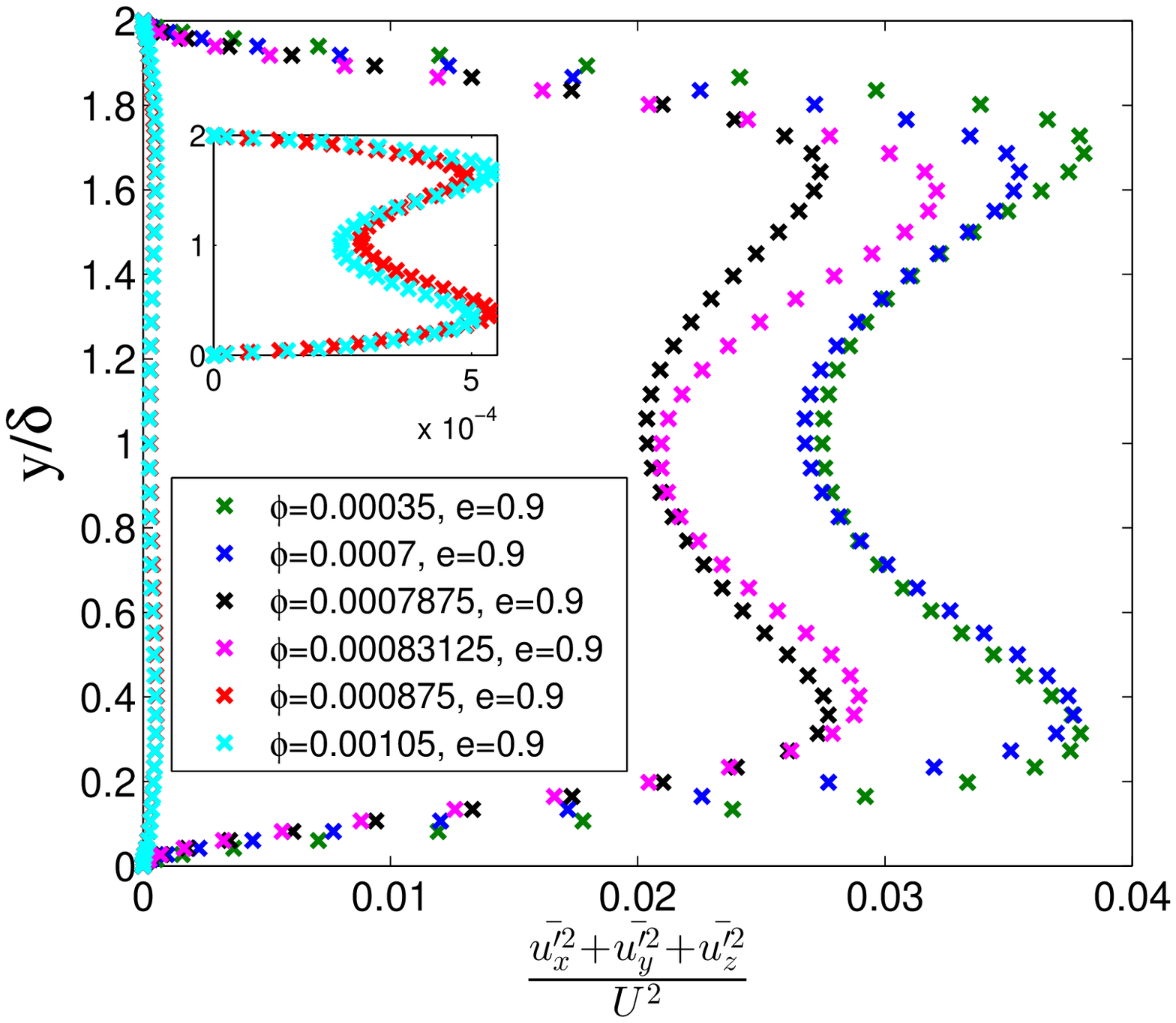}
  	\caption*{(b)}
  	\caption{Effect of particle volume fraction on (a) fluid mean velocity and (b) turbulent kinetic energy in presence of inelastic collisions $e=0.9$}
  	\label{fig:inelastic_mean_tke}
    \end{figure}
\begin{figure*}[h!]
    \begin{minipage}{0.45\textwidth}
        	\includegraphics[width=1.0\linewidth]{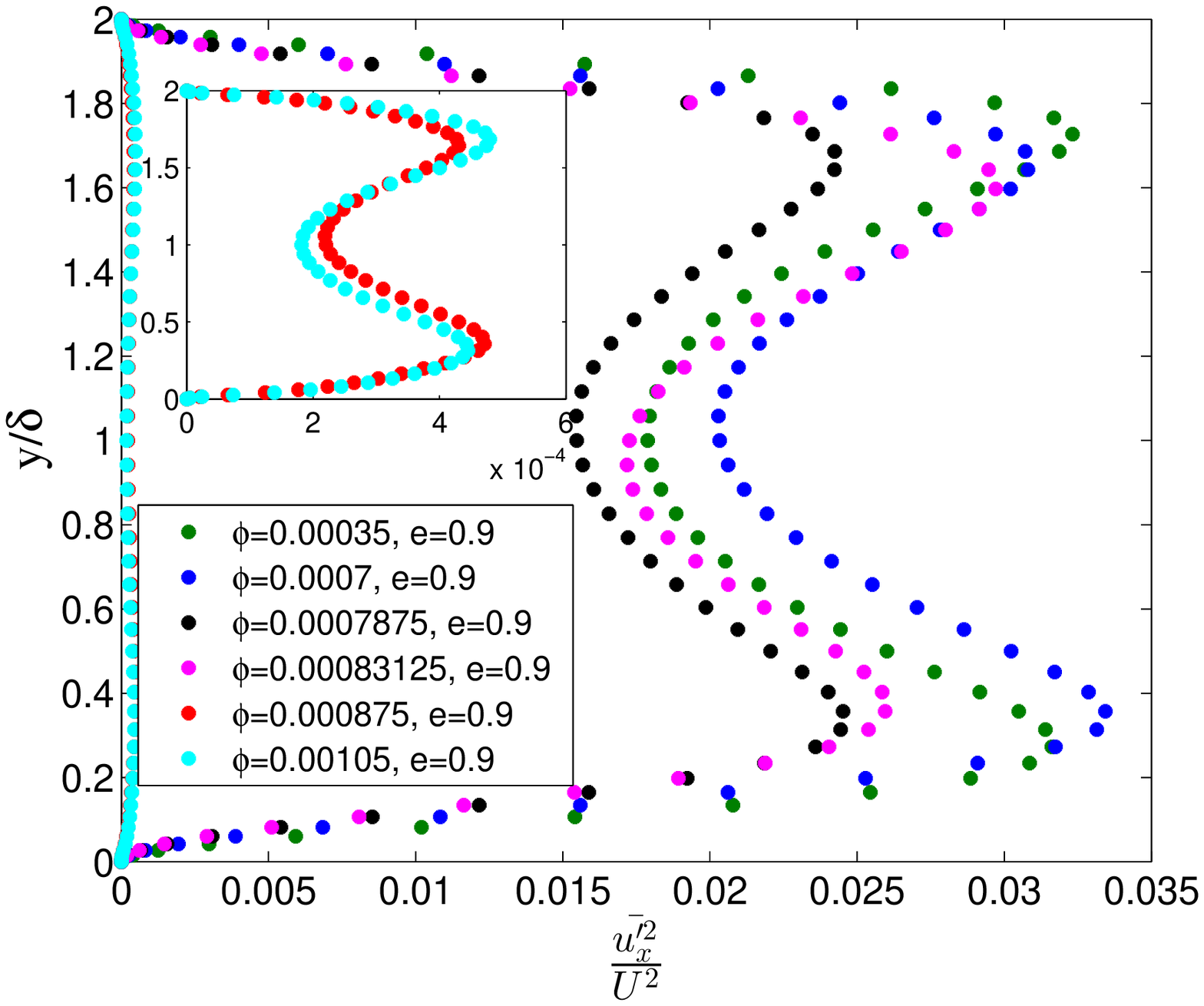}
    	\caption*{(a)}
    \end{minipage}
    \begin{minipage}{0.45\textwidth}
  	\includegraphics[width=1.0\linewidth]{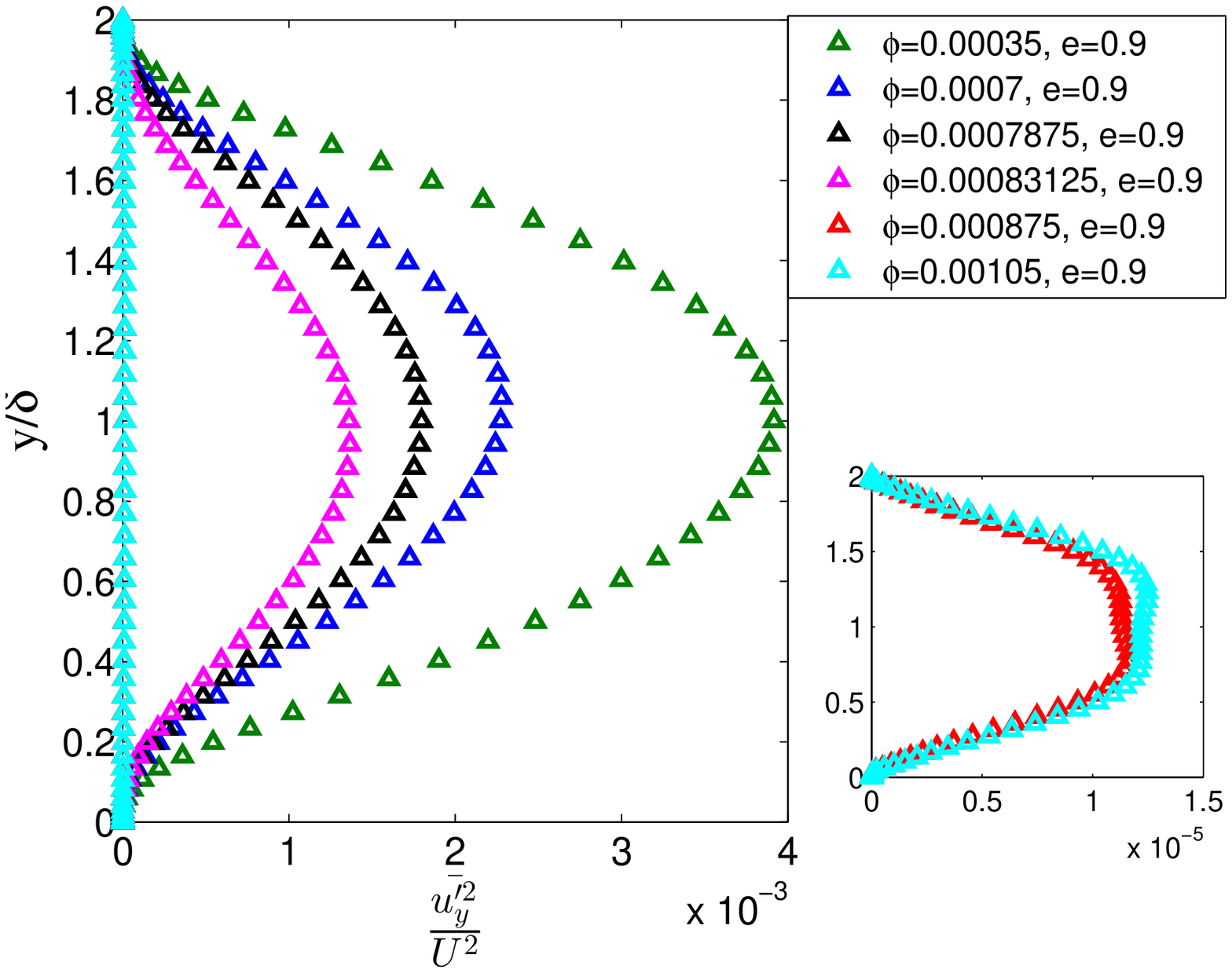}
  	\caption*{(b)}
  	\end{minipage}
  	\begin{minipage}{0.45\textwidth}
  	\includegraphics[width=1.0\linewidth]{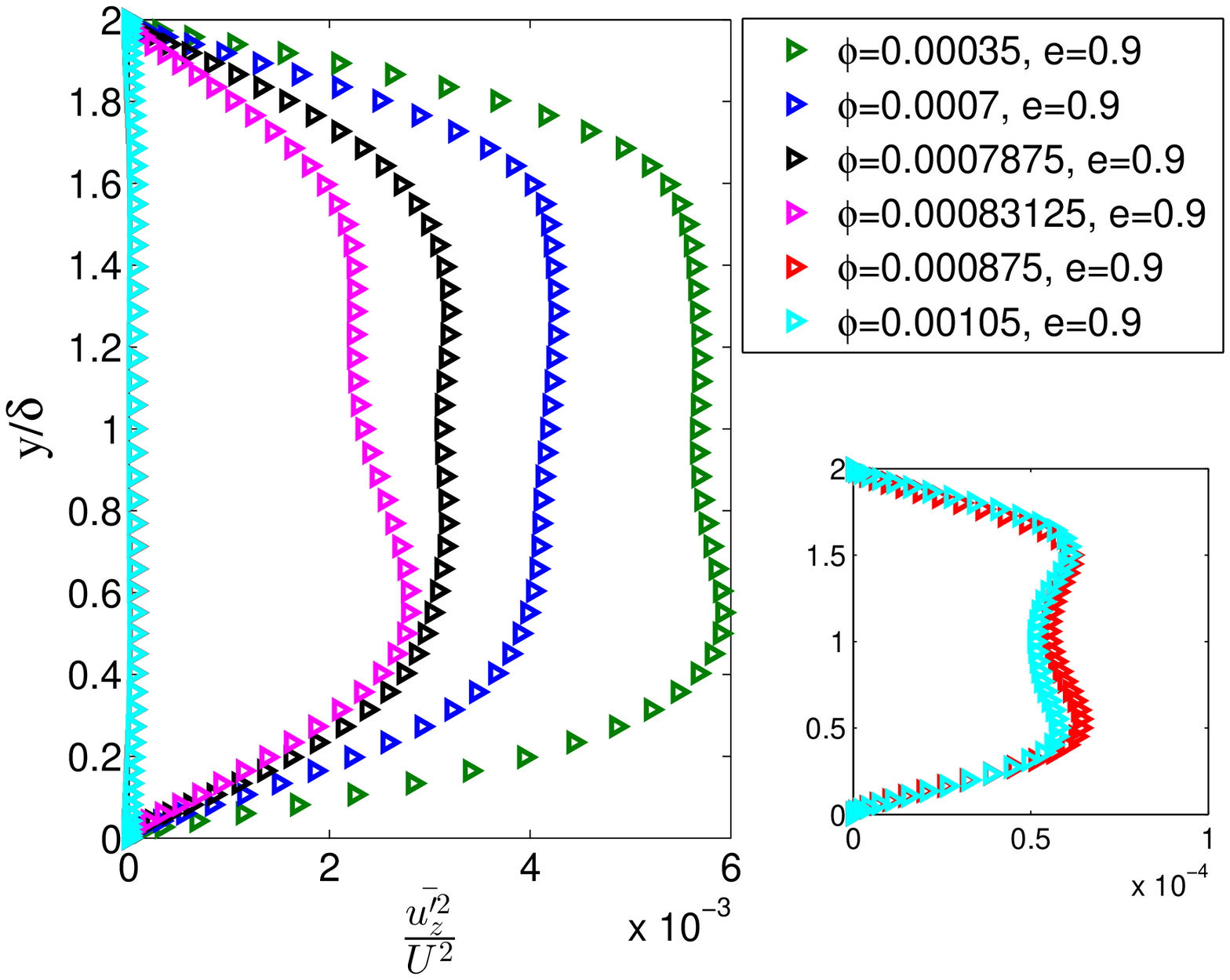}
  	\caption*{(c)}
  	\end{minipage}
  	\begin{minipage}{0.45\textwidth}
  	\includegraphics[width=1.0\linewidth]{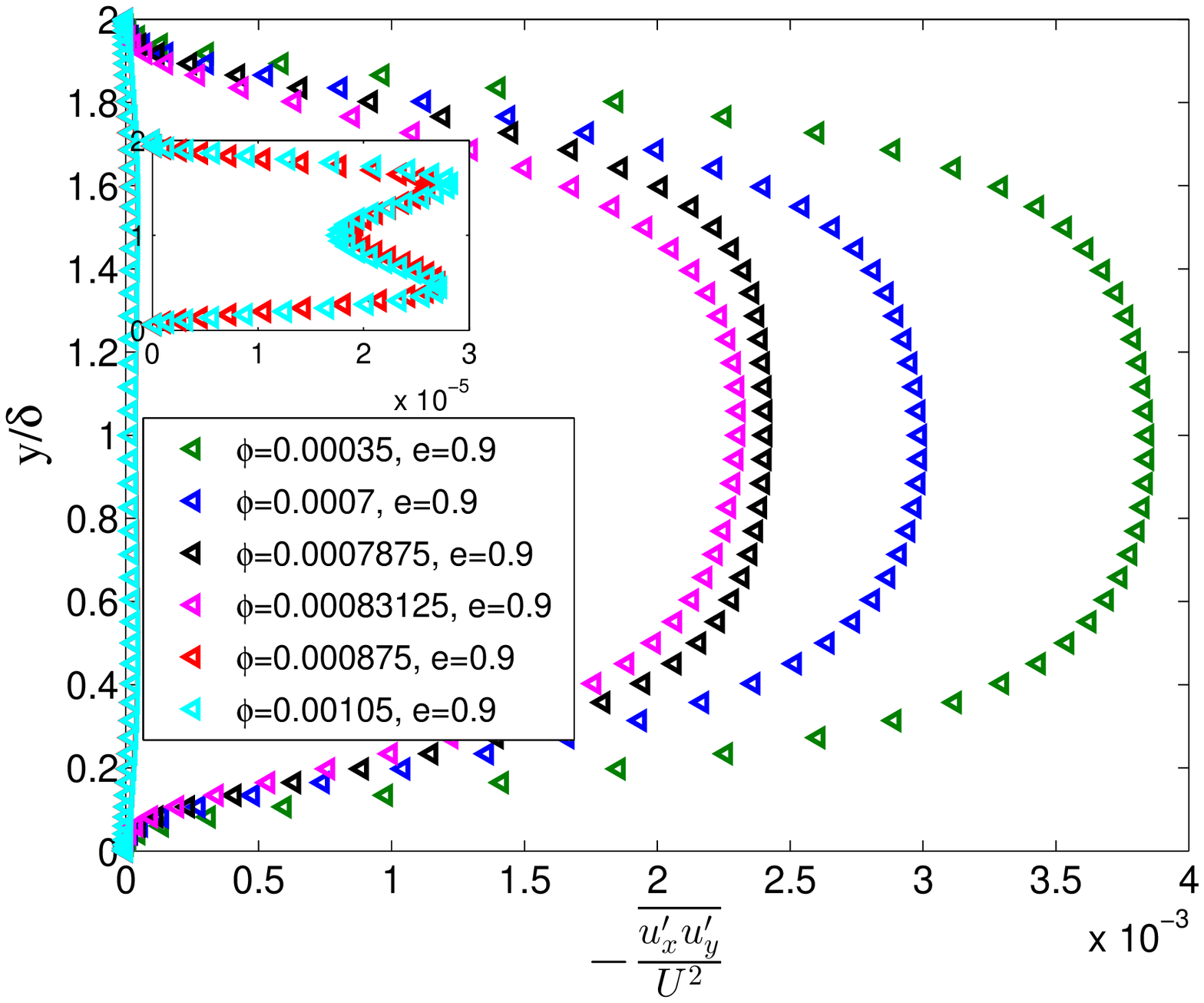}
  	\caption*{(d)}
  	\end{minipage}
  	\caption{Effect of particle volume fraction on fluid phase velocity fluctuations (a) $\overline{u_x^{,2}}$, (b) $\overline{u_y^{,2}}$ (c) $\overline{u_z^{,2}}$ and (d) $-\overline{u_x^,u_y^,}$ in presence of inelastic collisions $e=0.9$}
  	\label{fig:inelastic_uu_ms}
    \end{figure*}

  Variation in the mean fluid velocity and fluid turbulent K.E. with volume fraction is shown in figure \ref{fig:inelastic_mean_tke}. The mean fluid velocity follows two different mean velocity profiles. \textcolor{black}{Mean velocity profiles of the} fluid phase for suspensions with volume fractions greater than $8.3125\times10^{-4}$ \textcolor{black}{are} comparatively flatter \textcolor{black}{as shown in} figure \ref{fig:inelastic_mean_tke}(a). In this regime the turbulent kinetic energy is decreased by two-orders of magnitudes figure \ref{fig:inelastic_mean_tke}(b).
  \\ The second moments of the fluid velocity is shown in figure \ref{fig:inelastic_uu_ms}. It is evident that for volume fraction greater than $8.3125\times10^{-4}$, the magnitudes for all the second moments decrease drastically (about two-orders of magnitudes). The magnitudes of cross-stream and spanwise mean square velocities and the fluid Reynolds stress decrease monotonically with increase in volume loading till a critical volume fraction $\phi_{cr}=8.3125\times10^{-4}$. The drastic decrease in fluid velocity fluctuations and hence the turbulent kinetic energy above critical volume fraction is also observed in presence of elastic inter-particle collisions as discussed in section \ref{sec:Fluid_phase_stats}.  
  Between elastic and slightly inelastic collisional cases, the differences are observed in the decreased magnitudes of the fluid velocity fluctuations and a marginally increased critical volume fractions due to inelastic collisions. 
\subsubsection{\textbf{Fluid-phase streamwise Velocity and Vorticity Contours and Modulation in Turbulence}}

\begin{figure*}[h!]
\centering
\includegraphics[width=1.0\linewidth]{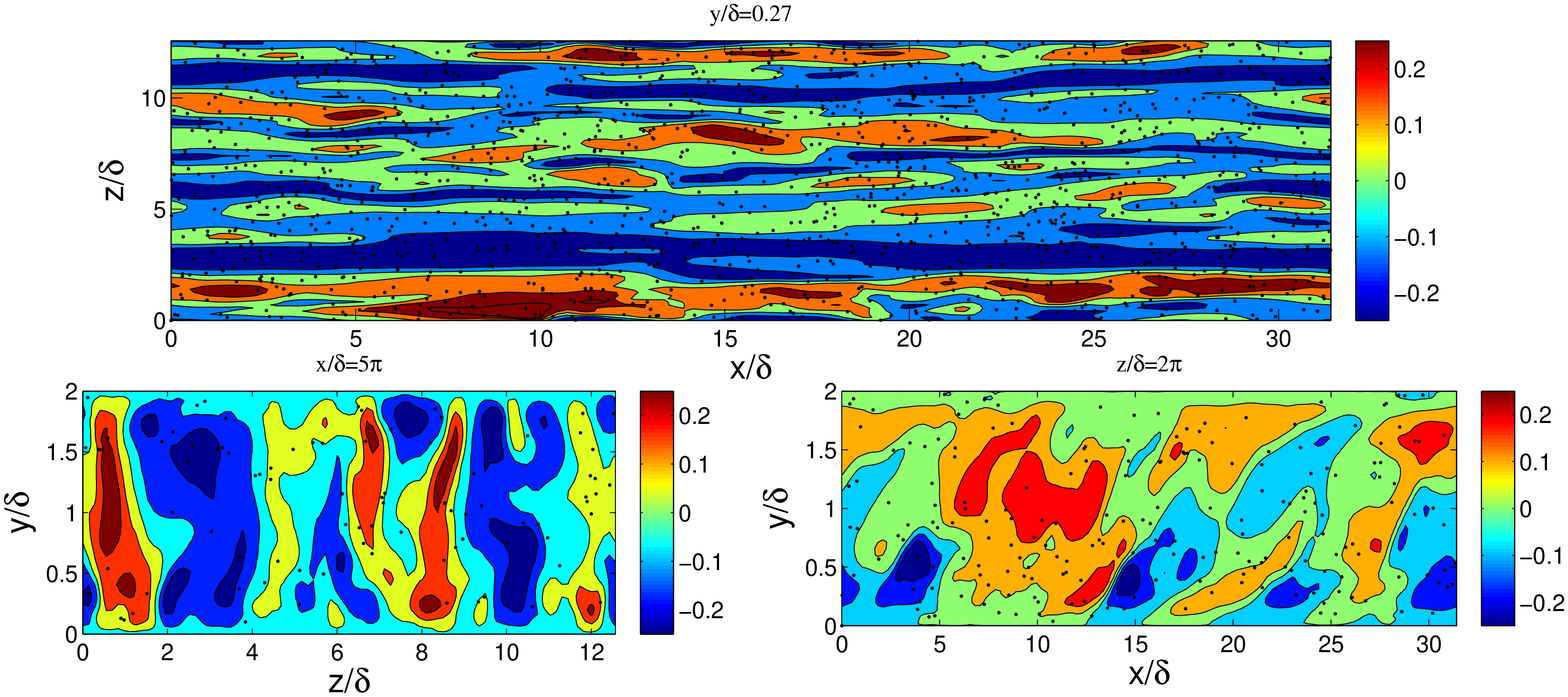}
	\caption{Contours of streamwise velocity fluctuations of fluid phase laden with particles of Volume Fraction $\phi_{cr}=8.3125X10^{-4}$ showed in three different planes as shown in the figures in presence of inelastic ($e=0.9$) inter-particle Collisions} 
	\label{fig:inelastic_vel_contour_b4_tr}
\end{figure*}

\begin{figure*}[!h]
\centering
\includegraphics[width=1.0\linewidth]{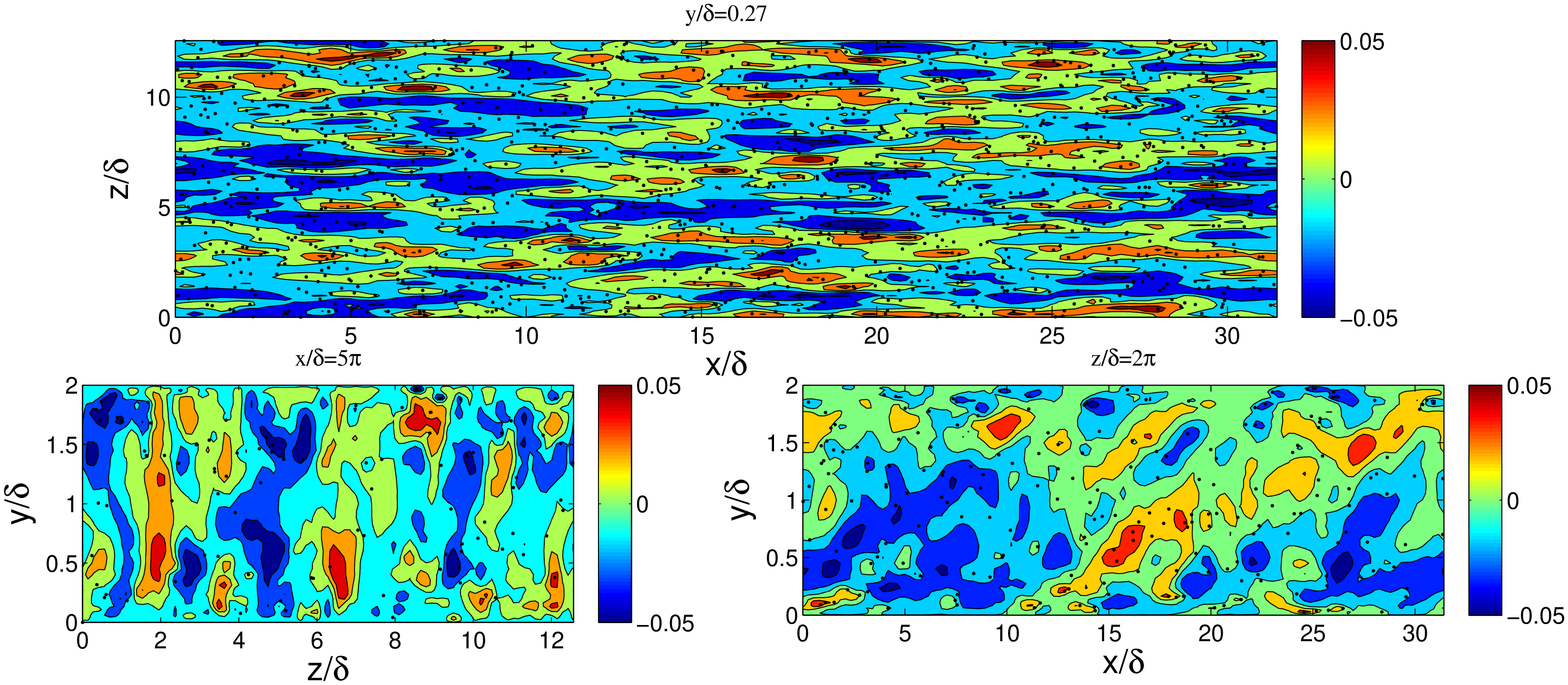}
	\caption{Contours of streamwise velocity fluctuations of fluid phase laden with particles of Volume Fraction $\phi=8.75X10^{-4}$ showed in three different planes as shown in the figures in presence of inelastic ($e=0.9$) inter-particle collisions} 
	\label{fig:inelastic_vel_contour_after_tr}
\end{figure*}

The analysis of the contours of streamwise velocity and vorticity is shown following \ref{sec:contours}.
Figure \ref{fig:inelastic_vel_contour_b4_tr} and \ref{fig:inelastic_vel_contour_after_tr} shows the difference in the fluid streamwise velocity fluctuation fields before and after the discontinuous transition. It is evident that the discontinuous transition reduces the strength of the velocity fluctuation field roughly by one-order of magnitude and breaks down the elongated streamwise velocity streaks, qualitatively similar to what has been observed for perfectly elastic collisions. 
\begin{figure*}[!h]
\centering
\includegraphics[width=1.0\linewidth]{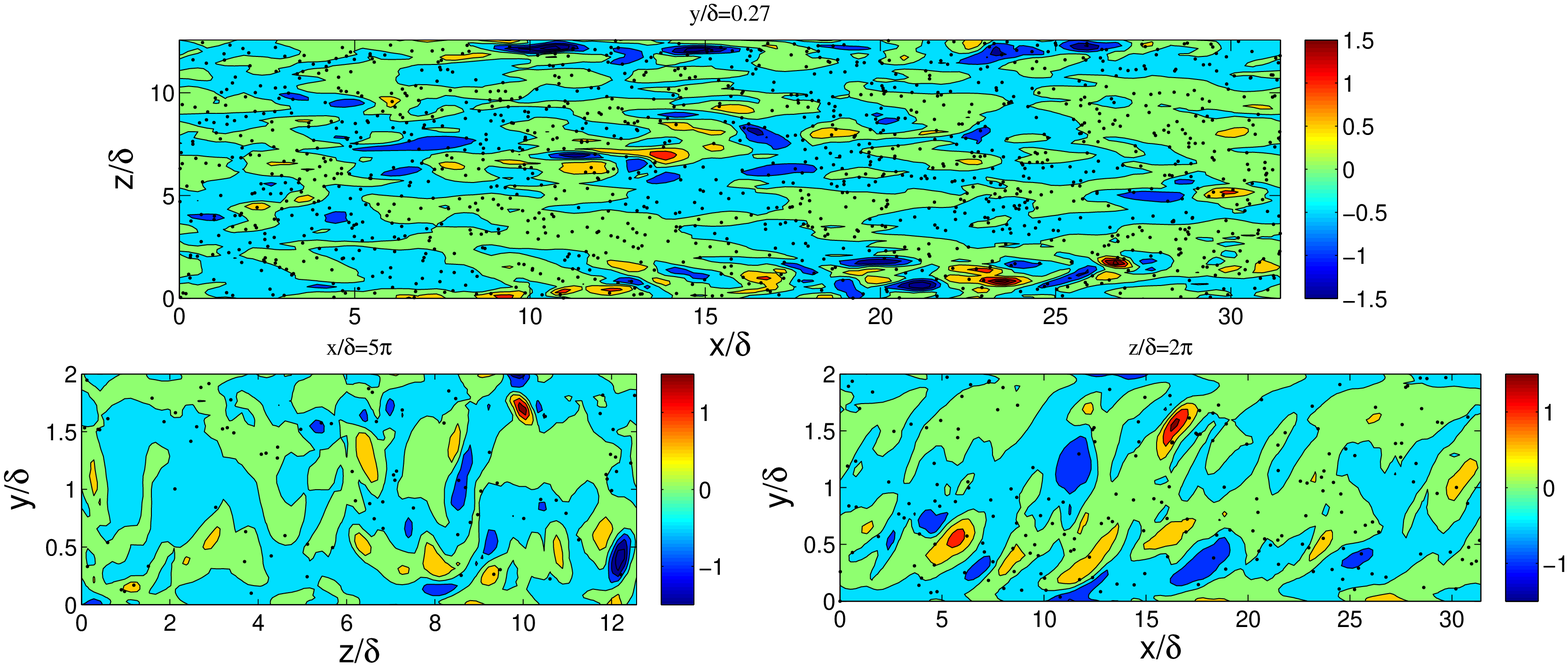}
	\caption{Contours of streamwise vorticity fluctuations of fluid phase laden with particles of Volume Fraction $\phi_{cr}=8.3125X10^{-4}$ showed in three different planes as shown in the figures in Presence of inelastic ($e=0.9$) inter-particle Collisions} 
	\label{fig:inelastic_vort_contour_b4_tr}
\end{figure*}
\begin{figure*}[!h]
\centering
\includegraphics[width=1.0\linewidth]{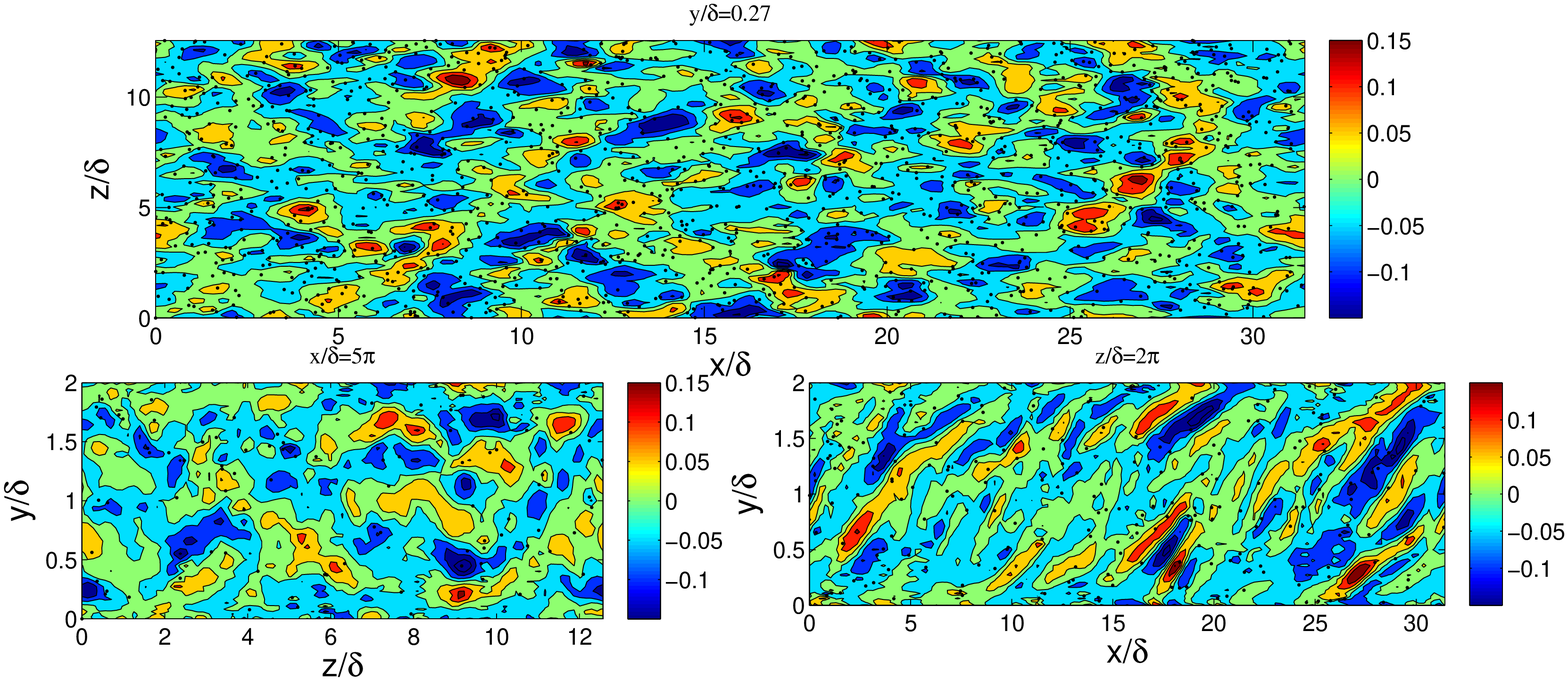}
	\caption{Contours of streamwise vorticity fluctuations of fluid phase laden with particles of volume fraction $\phi=8.75X10^{-4}$ showed in three different planes as shown in the figures in presence of inelastic ($e=0.9$) inter-particle collisions} 
	\label{fig:inelastic_vort_contour_aftr_tr}
\end{figure*}
The fluid streamwise vorticity fields before and after the discontinuous transition are captured in figures \ref{fig:inelastic_vort_contour_b4_tr} and \ref{fig:inelastic_vort_contour_aftr_tr}. Similar to the ideal collisional case, the streamwise vorticity field strength is decreased by one-order of magnitude along with the shortening of low-rotating vorticity zones.
From the above discussion it is evident that the nature of the \textcolor{black}{modulation} in turbulence occurring due to the presence of inertial particles undergoing inelastic collisions with each other and with the walls is drastic and discontinuous very similar to the case of elastically colliding particles. However, inelastic collisions are found to marginally increase the critical volume loading. The reason of this is explored in section \ref{sec:slip_prod_trans}. 
\subsection{Turbulence modulation in the absence of inter-particle collisions}
\label{sec:zero_coll}
In the previous subsection, the effect of slight inelasticity in \textcolor{black}{the} inter-particle collision is found to increase the critical volume fraction for the discontinuous turbulence disruption. The effect of inter-particle collisions could be best studied if the collisional effect is made to be isolated from the particle fluid interaction. 
In this case particles are allowed to move freely through each other without any mechanical contact but undergoes elastic collisional rebound from the walls. Similar to the previous cases, the fluid phase velocity statistics, namely the mean velocity and second moments of the \textcolor{black}{velocity fluctuations} are studied along with the contours of streamwise velocity and vorticity \textcolor{black}{fluctuations} of the fluid phase. 
\subsubsection{\textbf{Fluid velocity Statistics}}
 \begin{figure}[h!]
        	\includegraphics[width=0.7\linewidth]{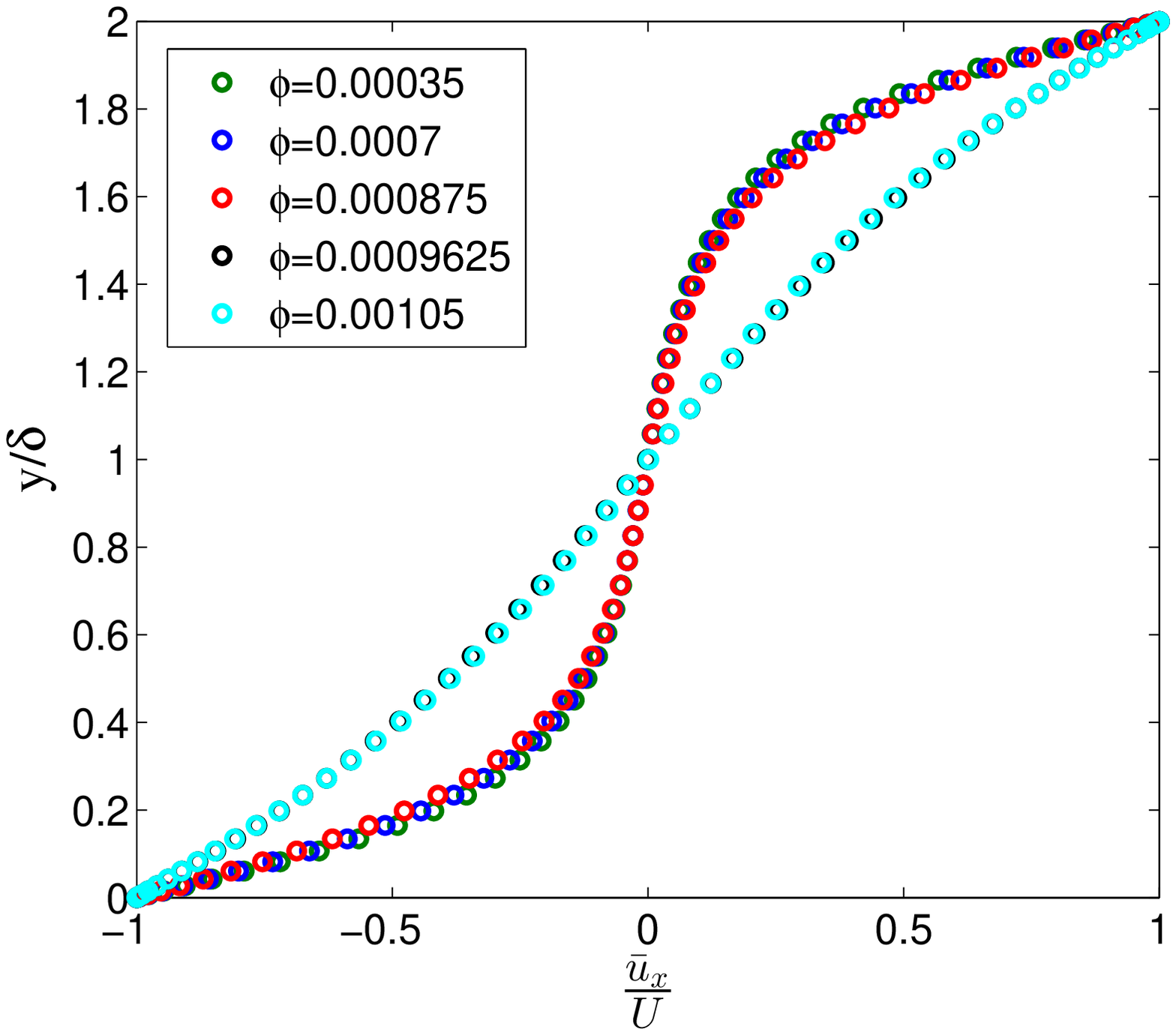}
    	\caption*{(a)}
  	\includegraphics[width=0.7\linewidth]{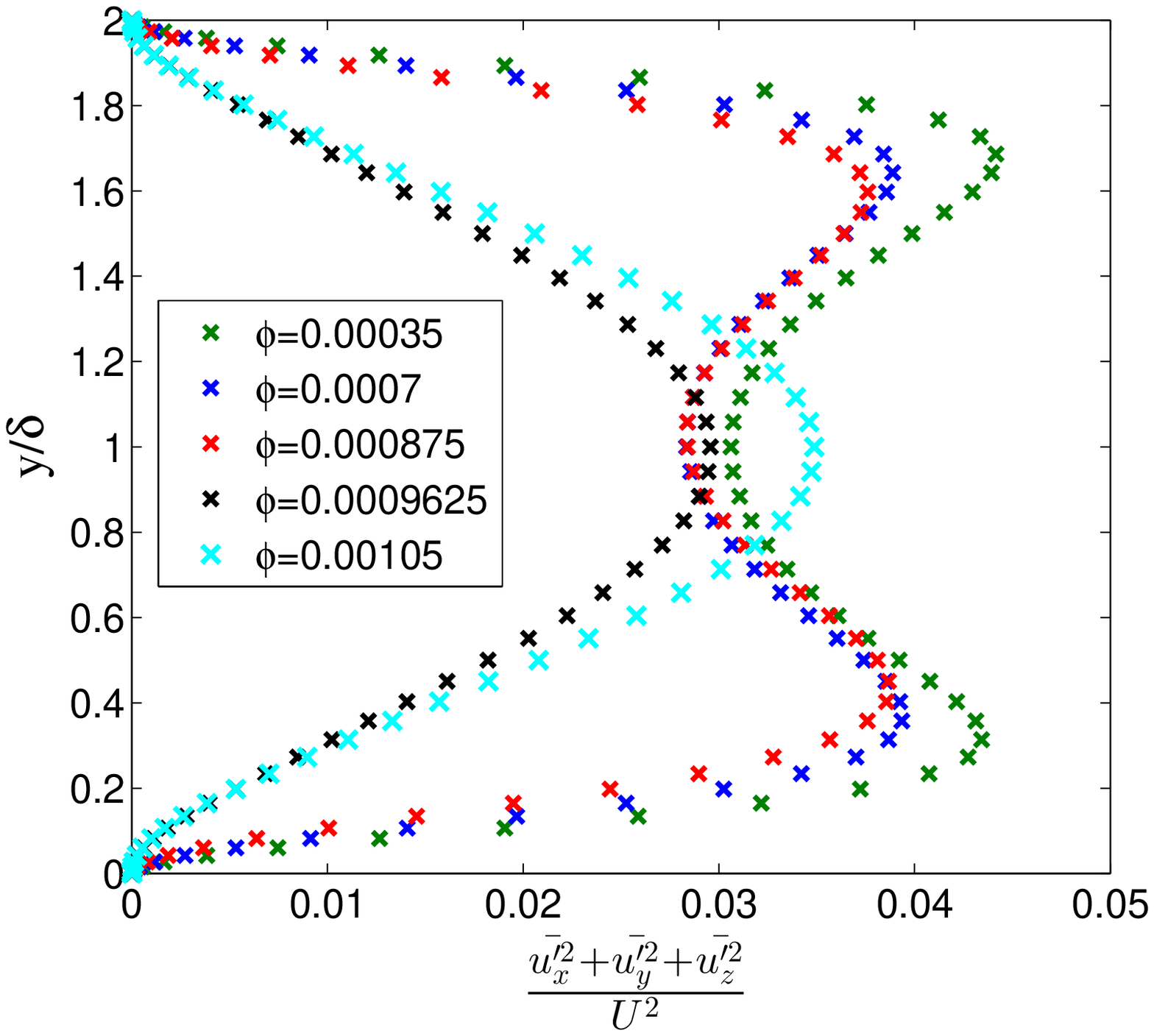}
  	\caption*{(b)}
  	\caption{Effect of particle volume fraction on (a) fluid mean velocity and (b) turbulent kinetic energy in absence of inter-particle collisions}
  	\label{fig:zero_coll_mean_tke}
    \end{figure}
\begin{figure*}[h!]
   \begin{minipage}{0.45\textwidth}
        	\includegraphics[width=1.0\linewidth]{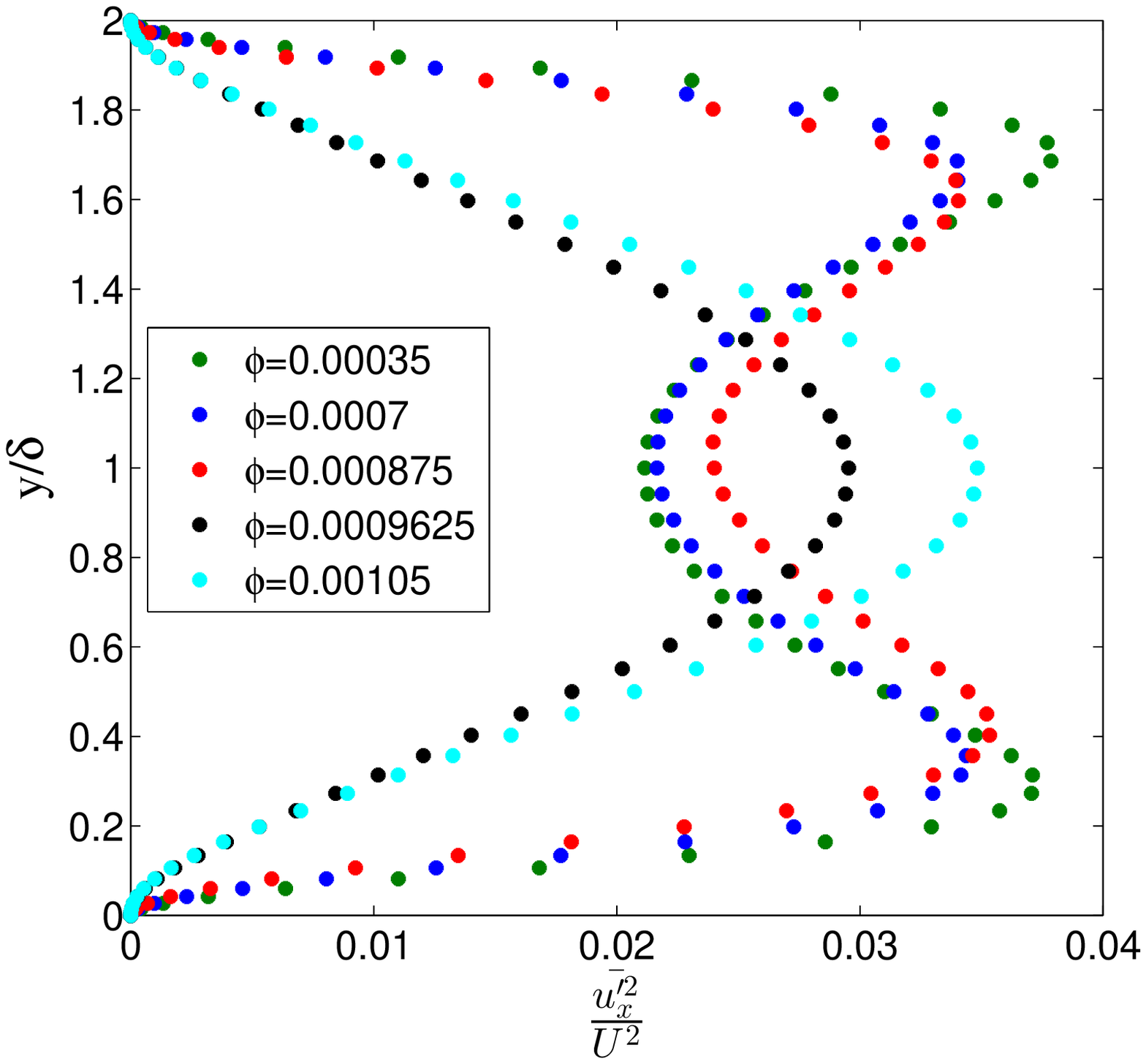}
    	\caption*{(a)}
    \end{minipage}
    \begin{minipage}{0.45\textwidth}
  	\includegraphics[width=1.0\linewidth]{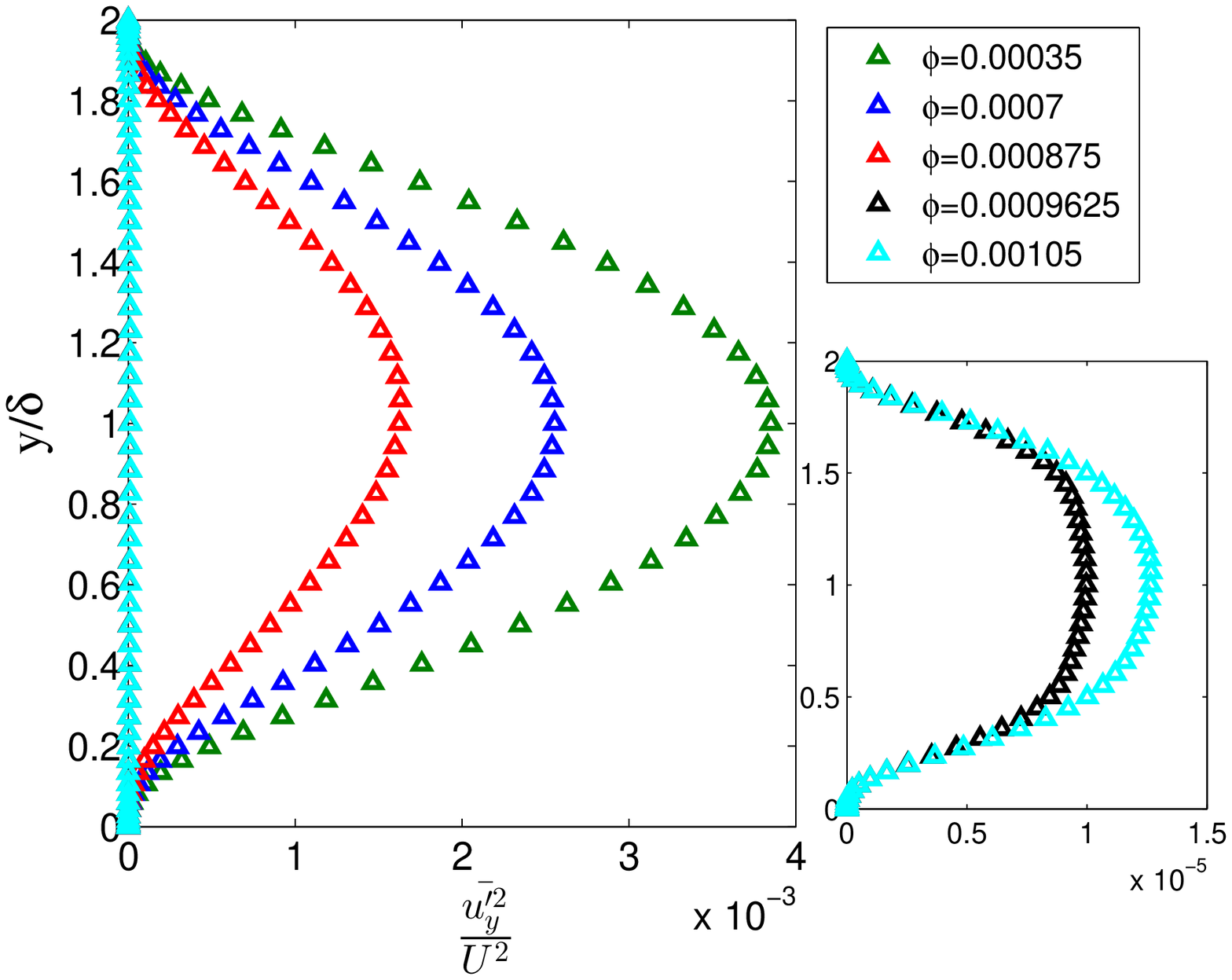}
  	\caption*{(b)}
  	\end{minipage}
  	\begin{minipage}{0.45\textwidth}
  	\includegraphics[width=1.0\linewidth]{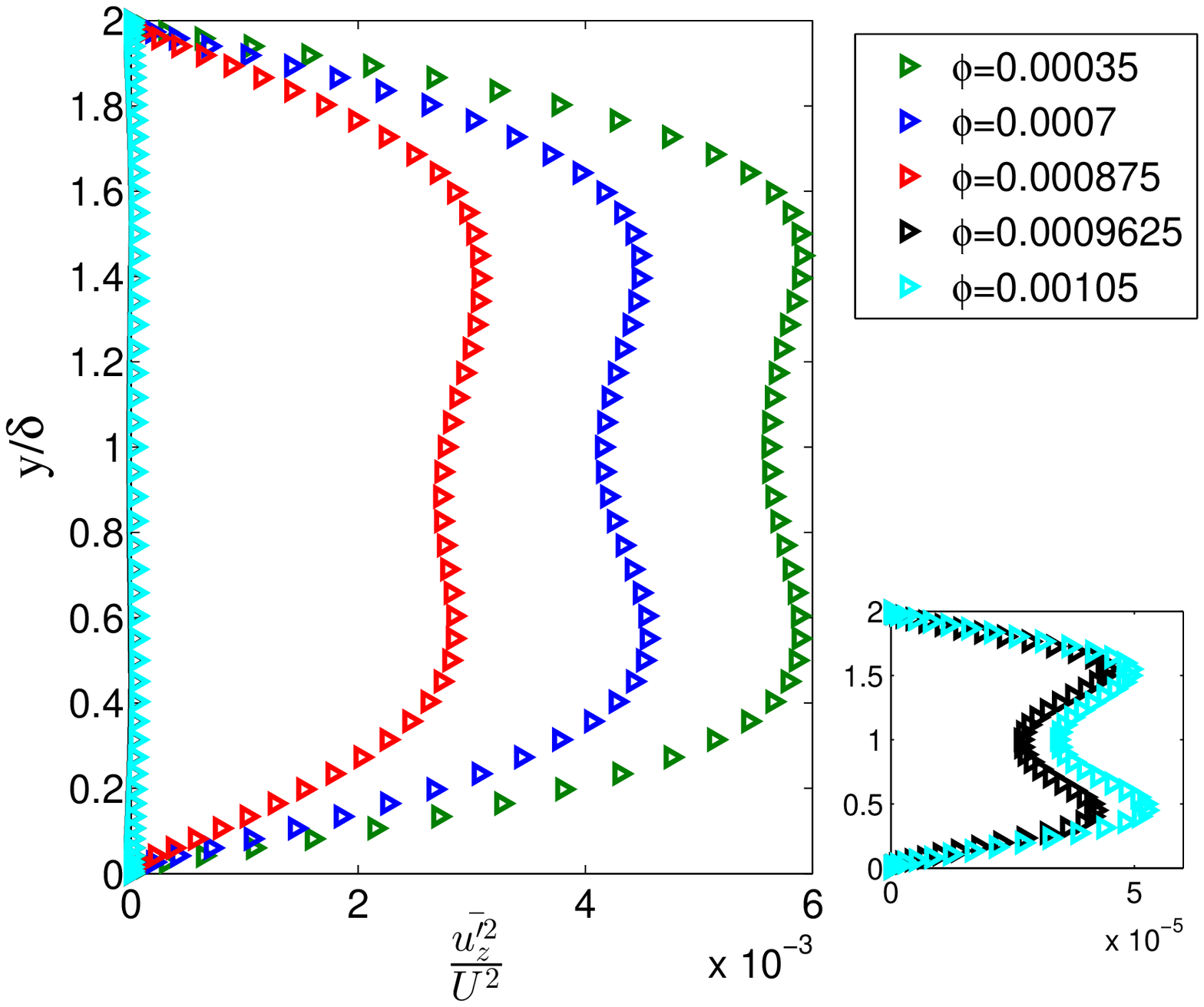}
  	\caption*{(c)}
  	\end{minipage}
  	\begin{minipage}{0.45\textwidth}
  	\includegraphics[width=1.0\linewidth]{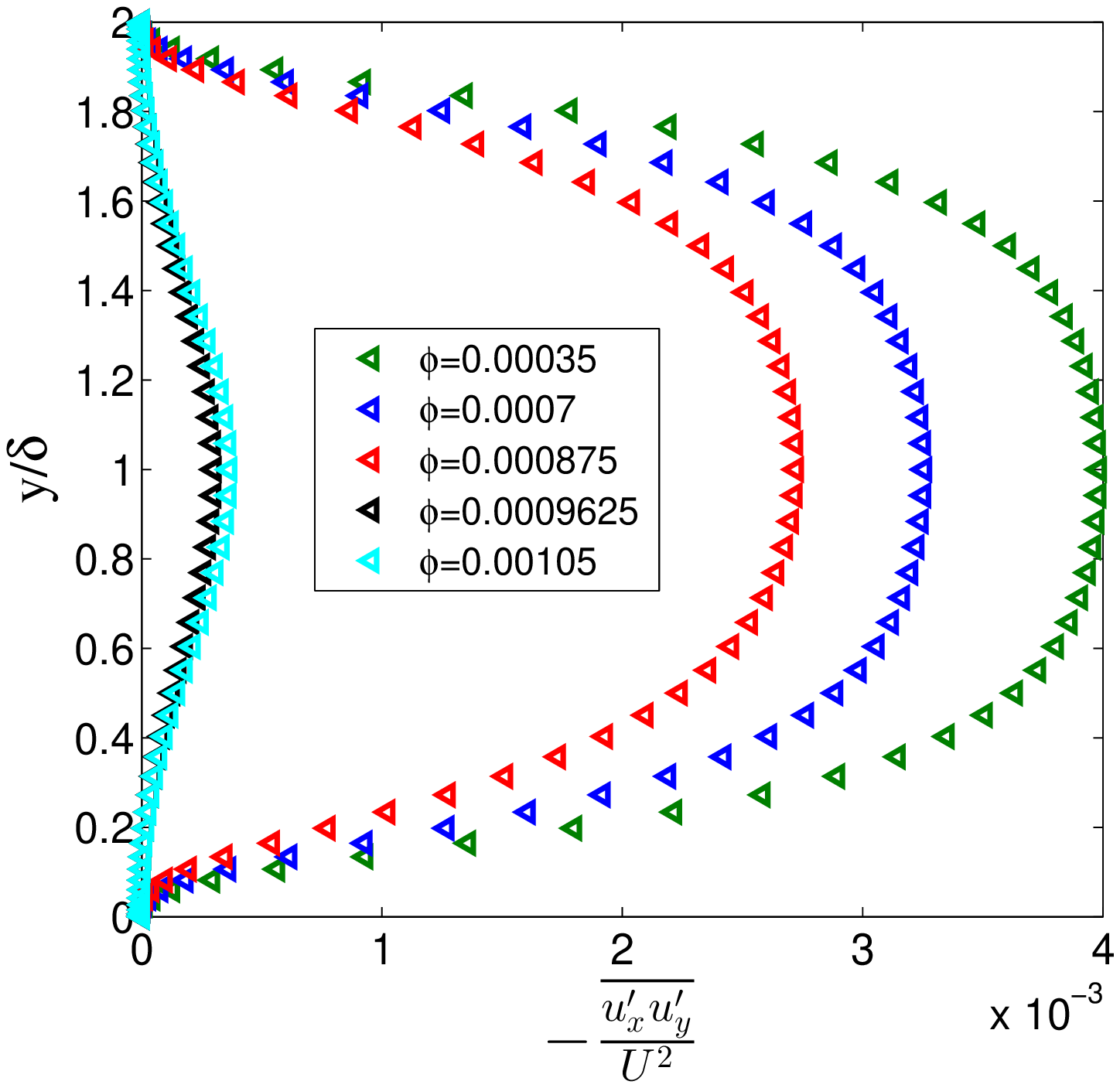}
  	\caption*{(d)}
  	\end{minipage}
  	\caption{Effect of particle volume fraction on fluid phase velocity fluctuations (a) $\overline{u_x^{,2}}$, (b) $\overline{u_y^{,2}}$ (c) $\overline{u_z^{,2}}$ and (d) $-\overline{u_x^,u_y^,}$ in absence of inter-particle collisions}
  	\label{fig:zero_coll_uu_ms}
    \end{figure*}
\begin{figure*}[h!]
    \centering
    \includegraphics[width=0.6\linewidth]{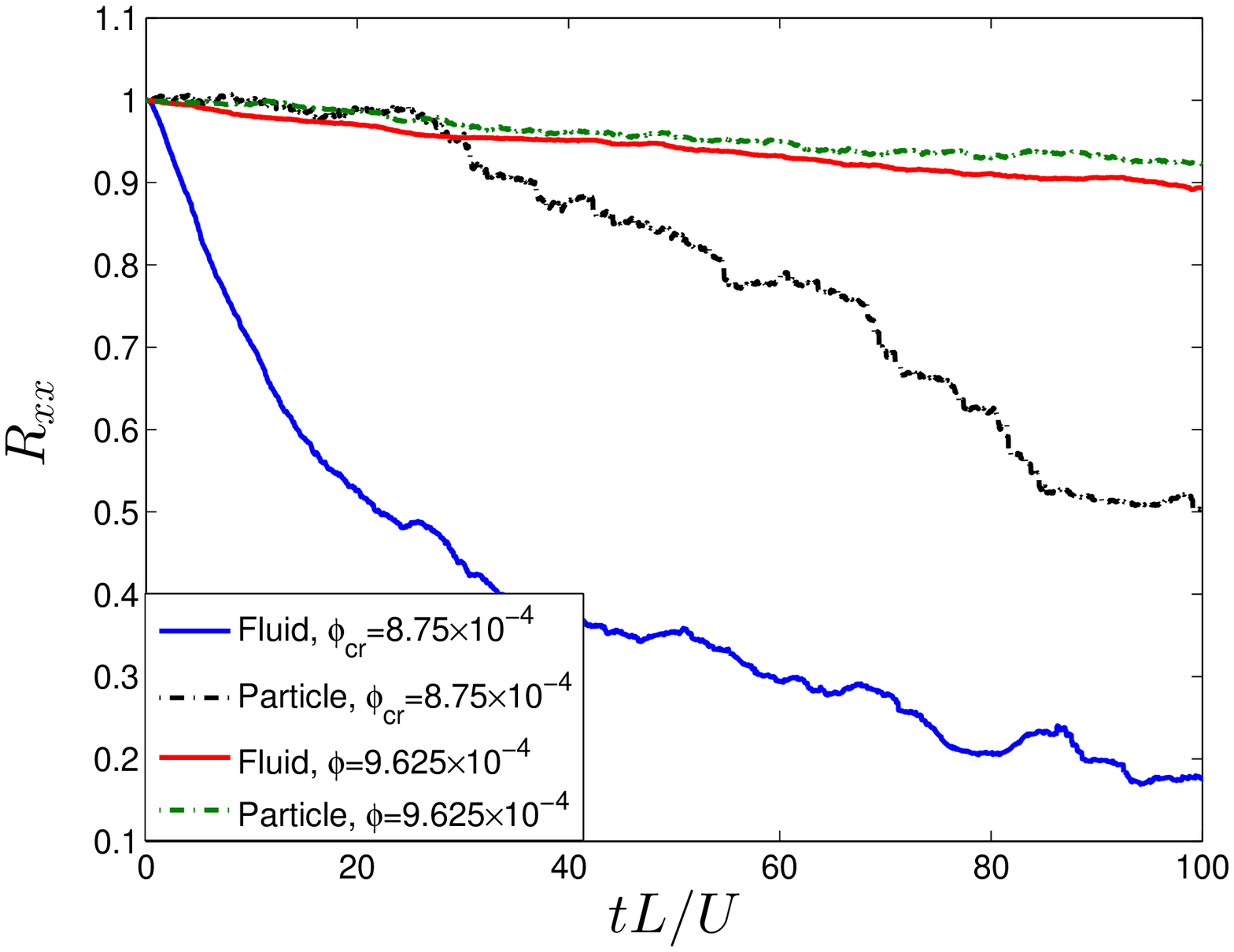}
    \caption{The initial decay of the temporal Lagrangian streamwise velocity auto-correlation $R_{xx}$ of the fluid and the particle phase before and after the transition in turbulence in absence of inter-particle collision computed in the zone $y/\delta:0.6-1.0$}
    \label{fig:PIT}
    \end{figure*}
\begin{figure*}[!h]
			\includegraphics[width=1.0\textwidth,  height=9cm]{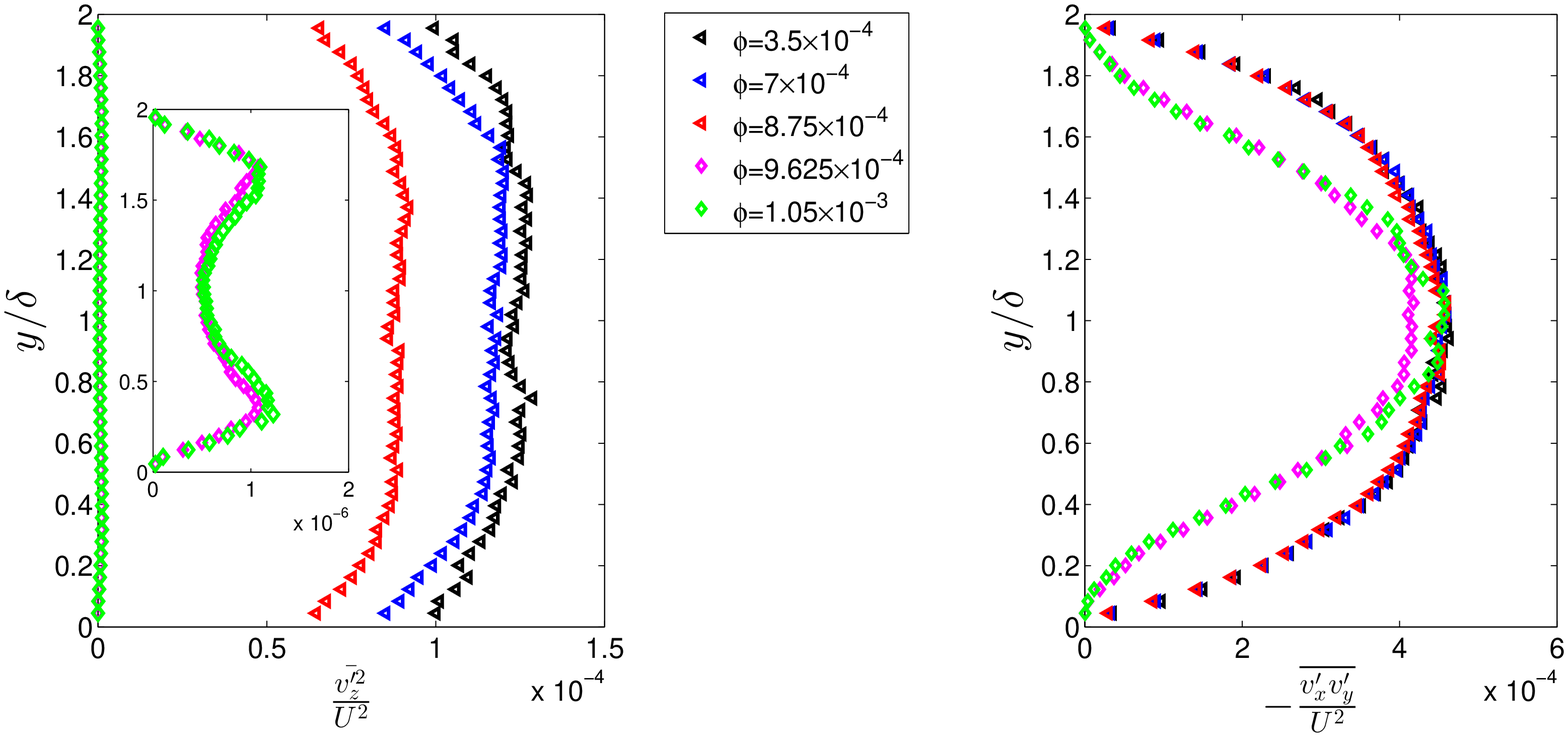}
			\caption{Effect of particle volume fraction on (a) span-wise mean square velocity, (b) particle phase shear stress in the  absence of inter-particle collisions} 
			\label{fig:particle_phase_stat_3_zero_coll}
		\end{figure*}

Figure \ref{fig:zero_coll_mean_tke} shows the variation in mean fluid velocity and fluid turbulent K.E. with volume fraction. The mean fluid velocity follows two different mean velocity profiles. Fluid phase for suspensions with volume fractions greater than $8.75\times10^{-4}$ follow a nearly linear mean velocity profiles similar to laminar Couette-flow profile as shown in figure \ref{fig:zero_coll_mean_tke}(a). Unlike the other cases, the \textcolor{black}{fluctuating} kinetic energy does not show any trend of drastic decrease. Rather the qualitative behaviour changes beyond $\phi=8.75\times10^{-4}$ (Fig.\ref{fig:inelastic_mean_tke}(b)). This is aligned with the change in trend observed for streamwise velocity second moment in figure \ref{fig:zero_coll_uu_ms}(a). It is evident from \ref{fig:zero_coll_uu_ms}(b) and (c) that for volume fraction greater than $8.75\times10^{-4}$, the magnitudes of cross-stream and spanwise second moments decrease drastically (about two-orders of magnitudes). The magnitudes of cross-stream and spanwise mean square velocities and the fluid Reynolds stress decrease monotonically with increase in volume loading till the critical volume fraction $\phi_{cr}=8.75\times10^{-4}$. The magnitude of the fluid Reynolds stress (figure \ref{fig:zero_coll_uu_ms}(d)) is found to be O($10^{-4}$) decrease beyond $\phi_{cr}$. 
    The role of absence of inter-particle collision is most prominent in case of streamwise mean-square fluid velocity. For suspensions with volume fractions higher than $\phi_{cr}$, $\overline{u_x^{,2}}$ is not observed to decrease drastically. Rather, \textcolor{black}{the profile changes its shape}. For unladen fluid, the profile of $\overline{u_x^{,2}}$ shows peaks in the buffer-layer. \textcolor{black}{However, incase of particle laden flows}, in absence of inter-particle collisions, $\overline{u_x^{,2}}$ profile \textcolor{black}{shows} maximum value at the center and monotonically decreases to zero at the wall. The particle phase statistics reported in appendix \ref{sec:Particle Phase Statistics zero_coll} figure  \ref{fig:particle_phase_stat_3_zero_coll}(a) shows that, after the \textcolor{black}{the collapse of fluid turbulence}, the qualitative behaviour of streamwise velocity fluctuation of the particle-phase and the fluid-phase have similar trend and the magnitude of the particle phase is higher. Moreover, the initial temporal decay of Lagrangian velocity correlation reveals a very interesting fact. Due to the higher inertia of the particles, the particle velocity auto-correlation $R_{xx}$ function is expected to decay slowly with respect to that of the fluid phase, as it is observed in figure \ref{fig:PIT} at the critical volume fraction. However, after the transition, the decay of $R_{xx}$ for both the phases is found to be similar which signifies turbulence in fluid phase being induced by the particle phase. \textcolor{black}{Therefore, the present correlation technique may help to make a distinction between the sheared turbulence and the particle driven fluid fluctuation.} 
\subsubsection{\textbf{Fluid-phase streamwise Velocity and Vorticity Contours and Modulation in Turbulence}}
\label{sec:slip_prod_trans}
\begin{figure*}[h!]
\centering
\includegraphics[width=1.0\linewidth]{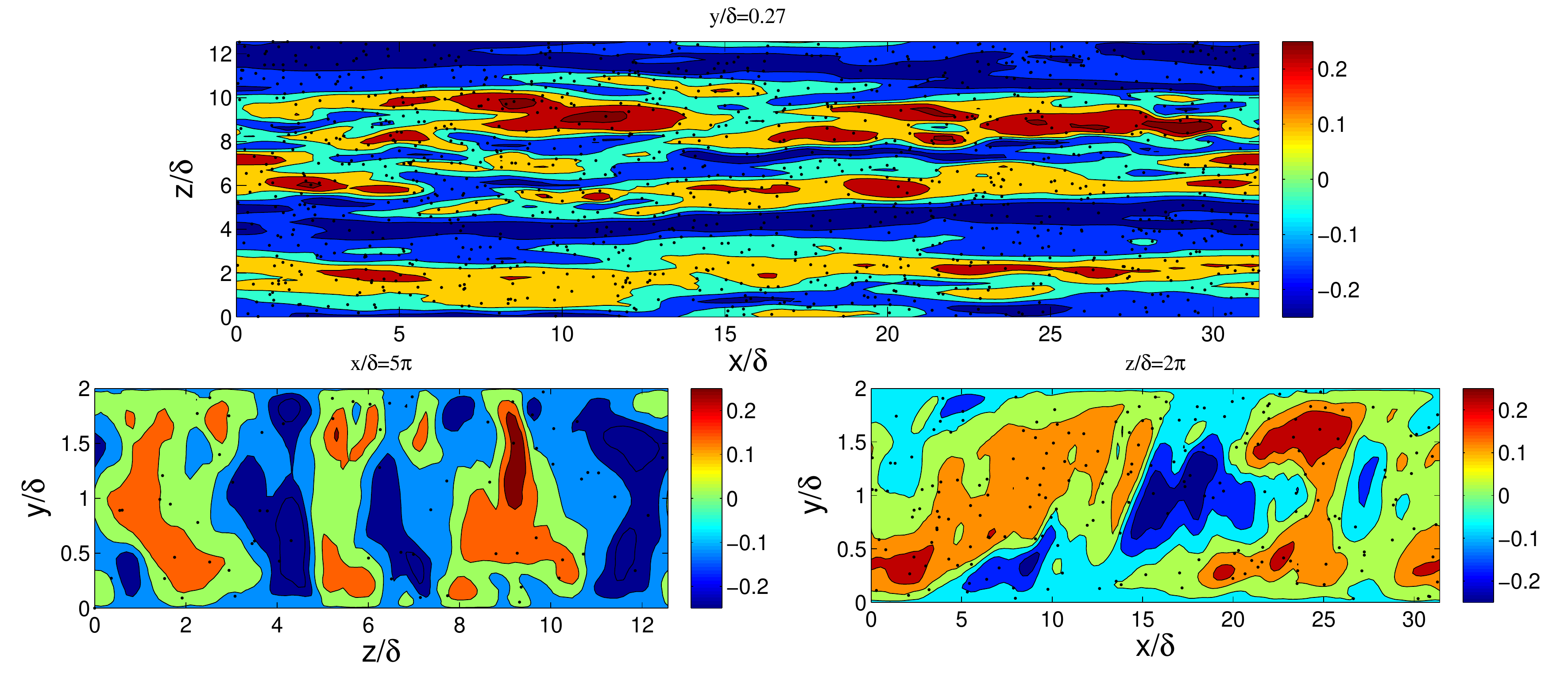}
	\caption{Contours of streamwise velocity fluctuations of fluid phase laden with particles of Volume Fraction $\phi_{cr}=8.75X10^{-4}$ showed in three different planes as shown in the figures in absence of inter-particle collision} 
	\label{fig:no_pp_vel_contour_b4_tr}
\end{figure*}

\begin{figure*}[!h]
\centering
\includegraphics[width=1.0\linewidth]{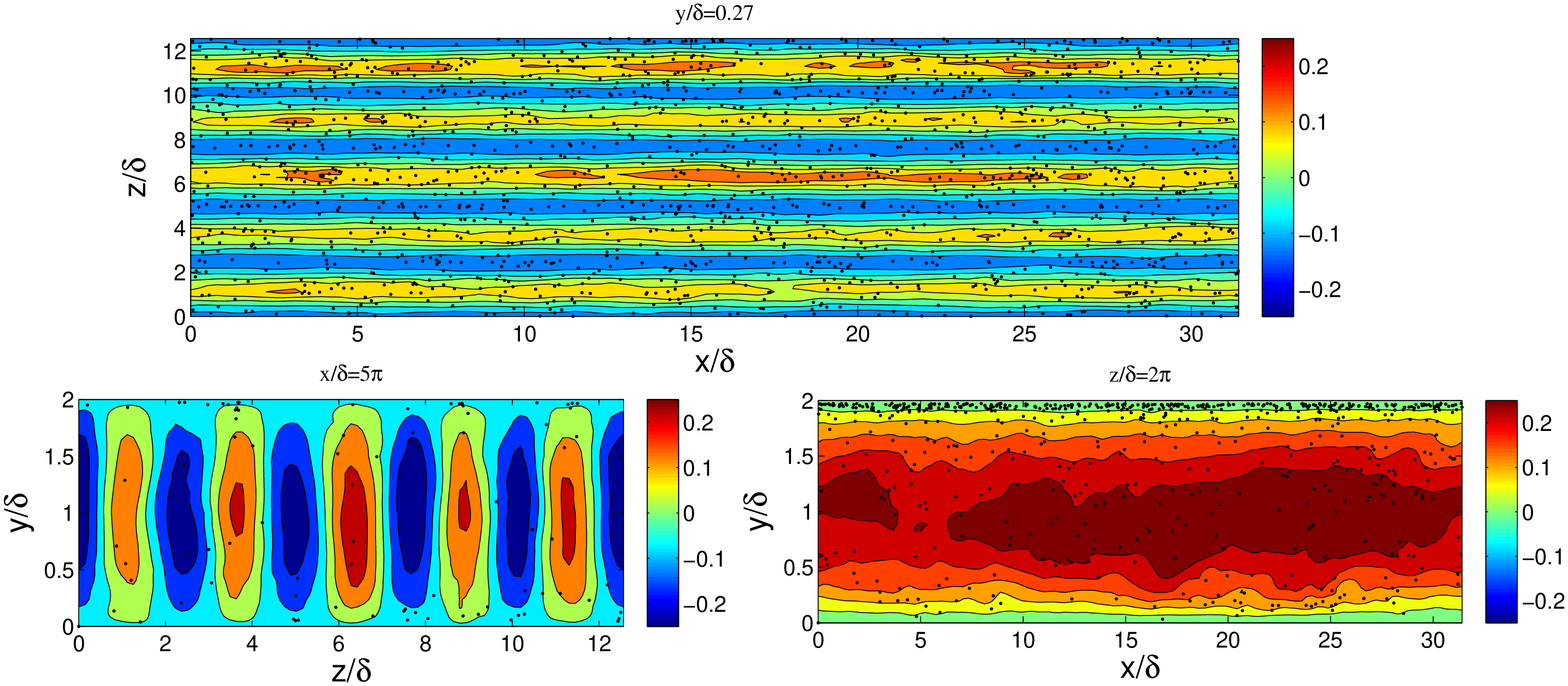}
	\caption{Contours of streamwise velocity fluctuations of fluid phase laden with particles of volume fraction $\phi=9.625X10^{-4}$ showed in three different planes as shown in the figures in absence of inter-particle collision} 
	\label{fig:no_pp_vel_contour_aftr_tr}
\end{figure*}

The contours of fluid phase streamwise velocity fluctuations are \textcolor{black}{shown} in the figures \ref{fig:no_pp_vel_contour_b4_tr} and \ref{fig:no_pp_vel_contour_aftr_tr}. streamwise low-speed velocity streaks, observed in figure \ref{fig:no_pp_vel_contour_b4_tr} before the transition, are qualitatively very similar in nature with the velocity field observed in presence of inter-particle collisions (Figs. \ref{fig:ideal_vel_contour_b4_tr} and
\ref{fig:inelastic_vel_contour_b4_tr}). The transition in turbulence adds some interesting features, not seen in presence of inter-particle collisions, in the streamwise velocity fluctuation field. The magnitude in the streamwise velocity fluctuations are not seen to be decreased unlike the previous cases. Most importantly the velocity field gets arranged in layers of different magnitudes (Fig.\ref{fig:no_pp_vel_contour_aftr_tr}). In x-z plane, contours are found to be arranged parallel to each other spanning the entire box-length. This layer, as seen in y-z plane, do not span the entire height and rather they are arranged in core-shell like structures with the highest fluctuations occurring in central part of the channel. The near-wall low intensity fluctuations are seen in x-z and y-z plane also. Hence in absence of inter-particle collisions, the transition in turbulence bring about sharp change in qualitative behaviour of streamwise velocity fluctuations but not in magnitude. Additionally, it can be inferred that inter-particle collisions play a very important role in breaking up the long streaky structures of streamwise velocity fluctuation field. 

\begin{figure*}[!h]
\centering
\includegraphics[width=1.0\linewidth]{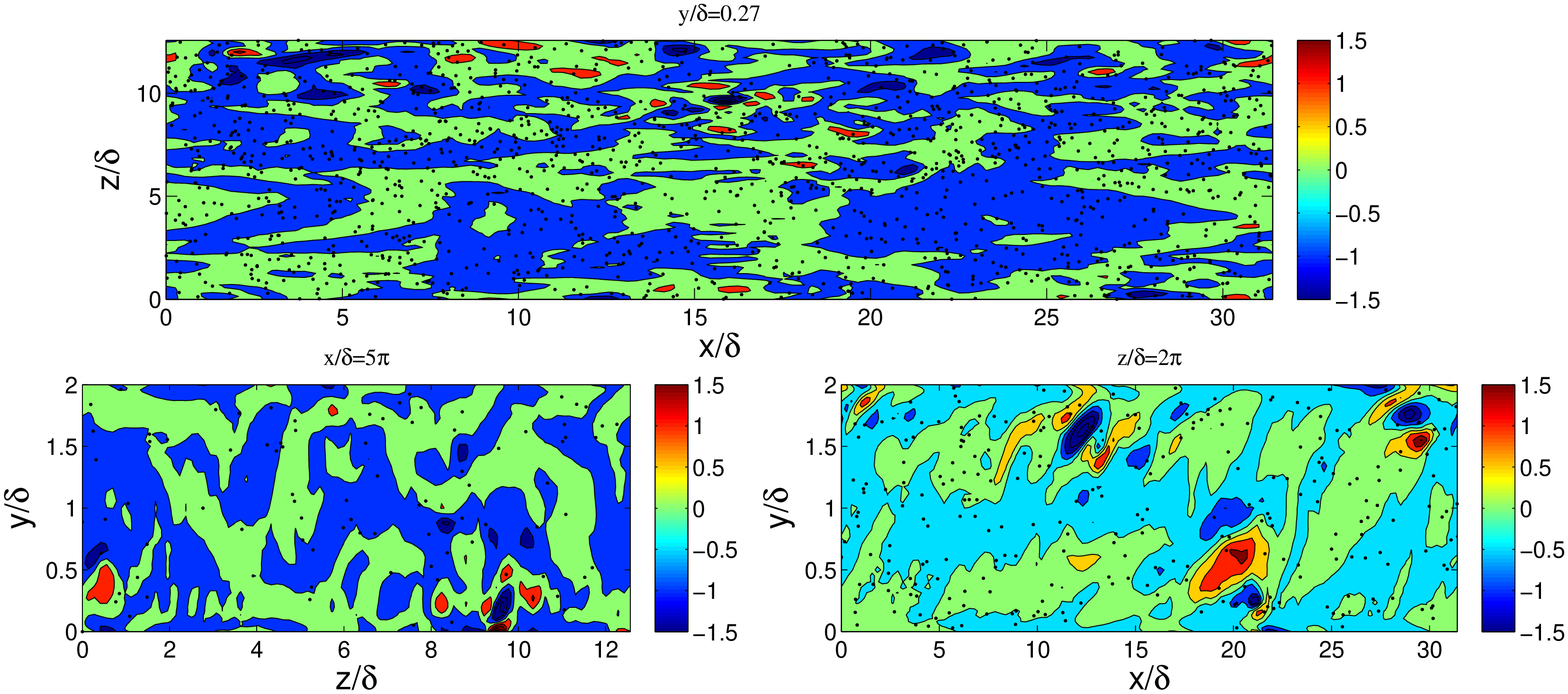}
	\caption{Contours of streamwise vorticity fluctuations of fluid phase laden with particles of volume fraction $\phi_{cr}=8.75X10^{-4}$ showed in three different planes as shown in the figures in absence of inter-particle collision} 
	\label{fig:no_pp_vort_contour_b4_tr}
\end{figure*}
\begin{figure*}[!h]
\centering
\includegraphics[width=1.0\linewidth]{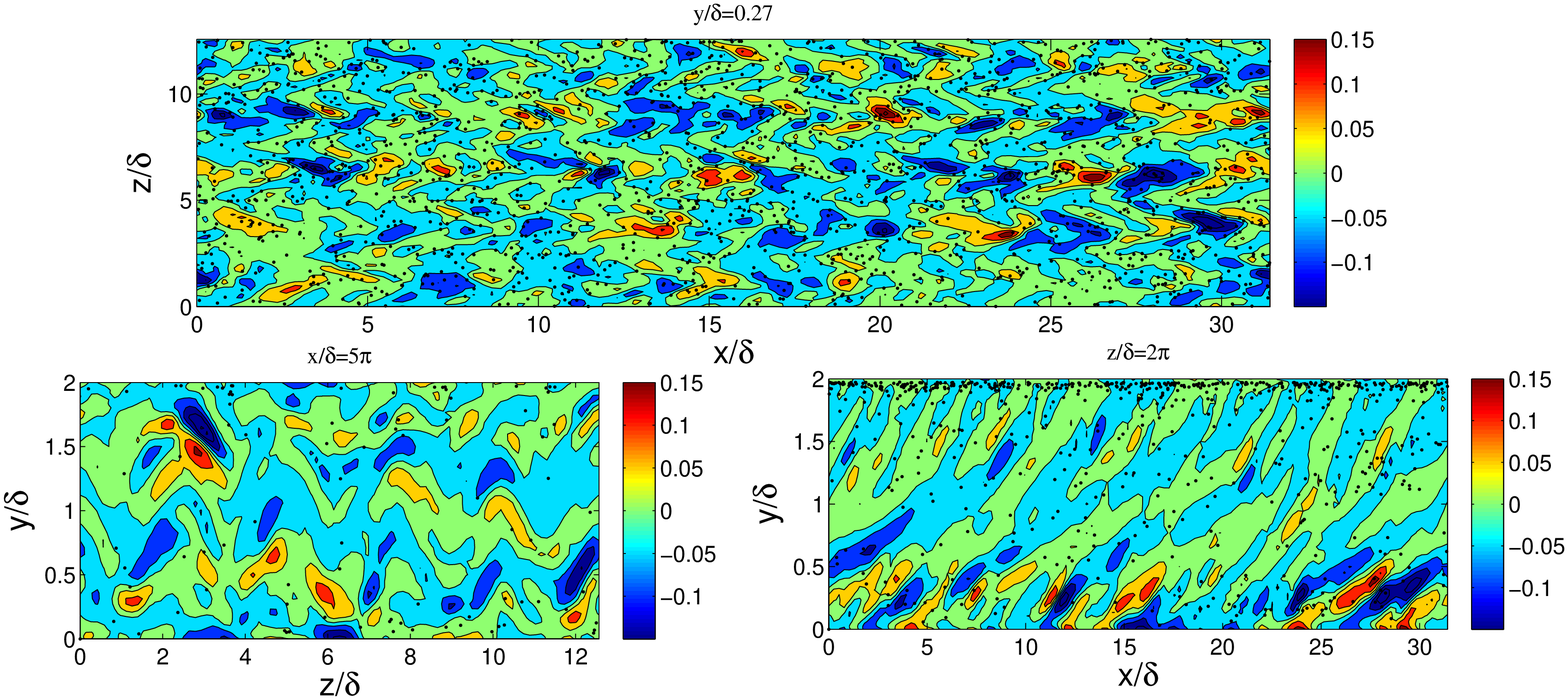}
	\caption{Contours of streamwise vorticity fluctuations of fluid phase laden with particles of volume fraction $\phi=9.625X10^{-4}$ showed in three different planes as shown in the figures in absence of inter-particle collision} 
	\label{fig:no_pp_vort_contour_aftr_tr}
\end{figure*}
The fluid streamwise vorticity fields before and after the discontinuous transition are captured in figures\ref{fig:no_pp_vel_contour_b4_tr} and \ref{fig:no_pp_vel_contour_aftr_tr}.
 The streamwise vorticity field strength is decreased by one-order of magnitude along with the shortening of low-rotating vorticity zones. These traits are qualitatively similar with the observations of streamwise vorticity field in presence of particles as well. The particle concentration field and the spanwise velocity vorticity field is not found to be correlated, although, the particles prefer to concentrate near the wall where streamwise velocity fluctuation is very low. Overall, it is observed that the absence of inter-particle collisions increases the critical volume fraction for the transition in turbulence.

\subsection{\textcolor{black}{Role of inter-particle collision on critical volume fraction}}
\begin{figure}[!h]
\includegraphics[width=0.7\linewidth]{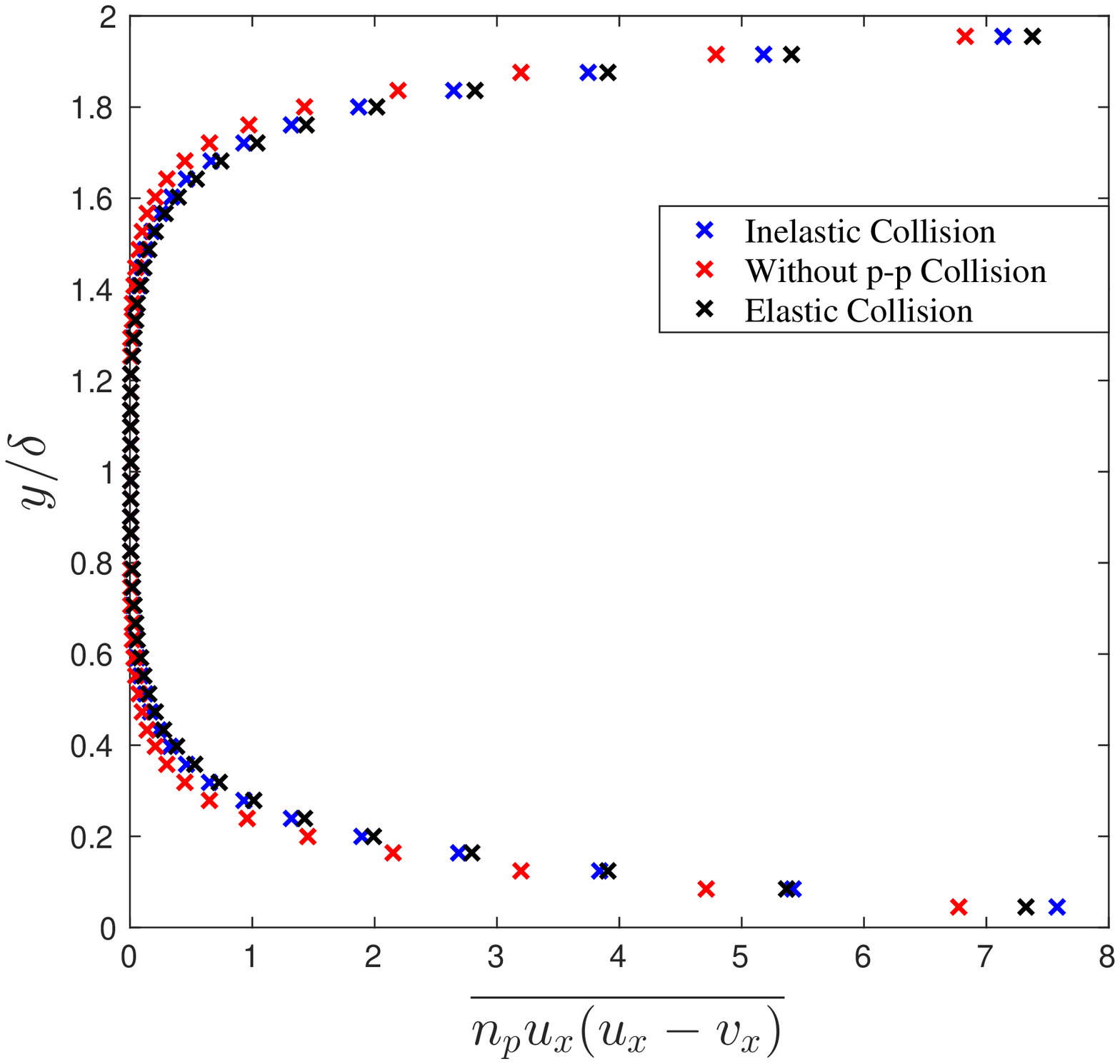}
\caption*{(a)}
\includegraphics[width=0.7\linewidth]{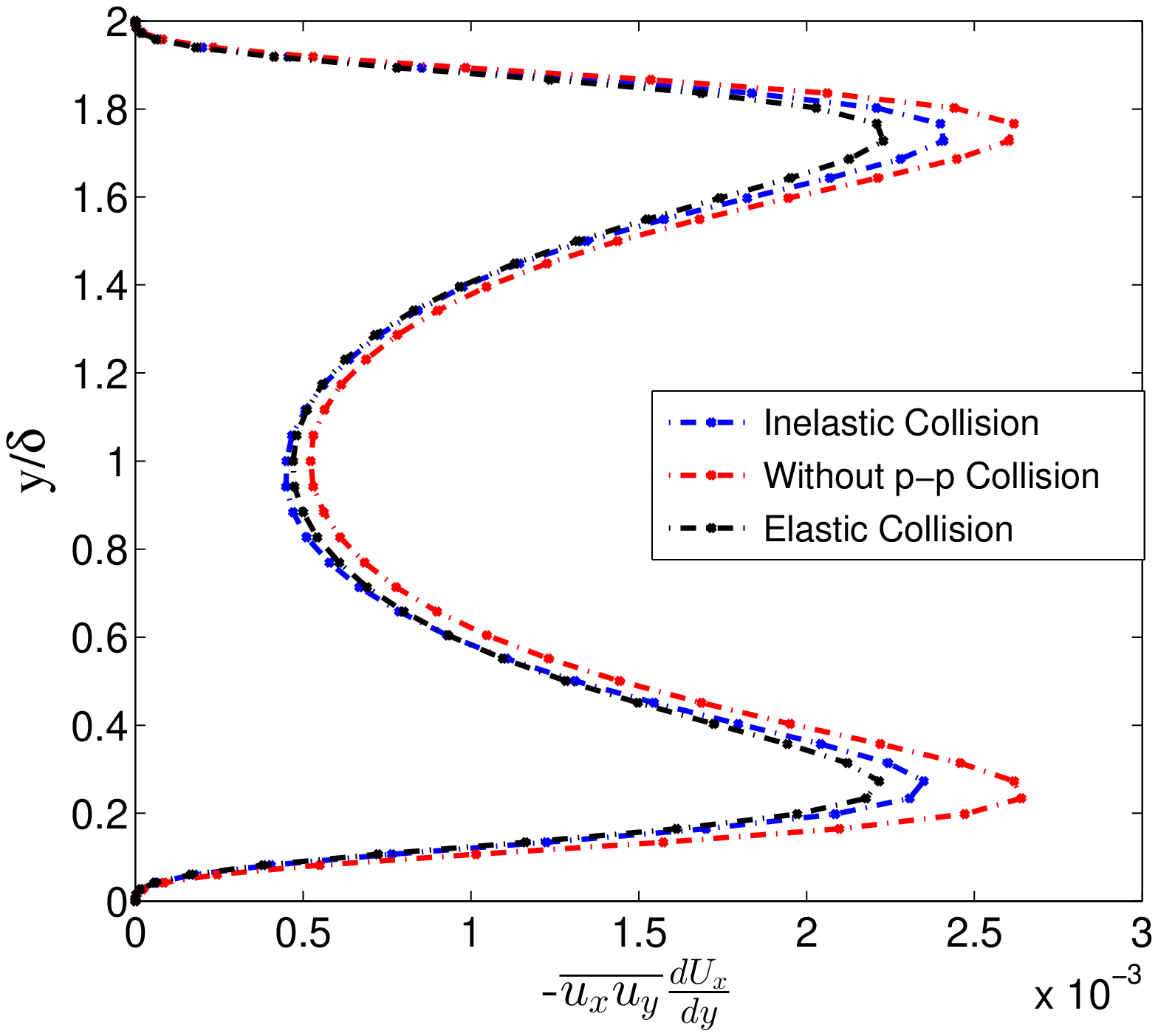}
\caption*{(b)}
	\caption{Effect of nature of collision on (a) $\overline{n_pu_x(u_x-v_x)}$ and (b) turbulence production for volume fraction $7\times10^{-4}$} 
	\label{fig:eff_coll_prod_slip}
\end{figure}

The effect of inter-particle collision is shown in Fig.{\ref{fig:eff_coll_prod_slip}} on two important terms, i.e. (a) $\overline{n_pu_x(u_x-v_x)}$, which is proportional to the mean K.E. dissipation due to particle reverse drag and (b) shear production of turbulence. \textcolor{black}{We have considered a volume fraction which is} less than $\phi_{cr}$ for all of the cases. 
\textcolor{black}{Figure {\ref{fig:eff_coll_prod_slip}} (a) reveals that the mean K.E. dissipation due to particle reverse drag varies marginally with the nature of collision. This dissipation is lowest in absence of inter-particle collision and highest for perfectly elastic collision.} 
\textcolor{black}{The discontinuous \textcolor{black}{decay} in turbulence, discussed previously in this article, happens as a result of simultaneous catastrophic reduction of the shear-production of turbulence. Figure  \ref{fig:eff_coll_prod_slip}(b) shows that  the shear production term is highest in the absence of inter-particle collisions and lowest in presence of perfectly elastic collisions. 
The marginal increase in turbulence production term along with the marginal decrease in the dissipation of mean K.E. due to particle reverse drag is well correlated to the delay in drastic collapse of turbulence or with the increase in critical volume fractions.  This explains why the critical volume fraction of the system without inter-particle collisions is the highest and for the system with ideal elastic inter-particle collisions is the lowest.}
\\\citet*{richter2013momentum} presented a very elegant relation, given below, between particle phase stress $\tau_{particle}$ and the feedback force $F$. Their formulation was based on the derivation of \cite{mito2006effect} where the dispersed phase was modeled using continuum approximation. 
\\\textcolor{black}{ The momentum balance equation for the dispersed phase under continuum approximation can be written by: 
\begin{equation}
\frac{\partial}{\partial t}(c v_i) +  \frac{\partial}{\partial x_j}(c v_i v_j) =-F_i 
\label{eq:mito1}
\end{equation}
where, $c$ represents local instantaneous particle mass, $v_i$ represents particle phase velocity and $F_i$ denotes the reverse force per unit volume.
\\Performing Reynolds averaging procedure on eq. \ref{eq:mito1} and taking the streamwise component : 
\begin{equation}
\frac{\partial}{\partial y}\left(\langle v_x \rangle \langle c v_y' \rangle+\langle c v_x' v_y'\rangle\right) =\langle -F_x \rangle
\label{eq:mito2}
\end{equation}
Here, $v_x'$, $v_y'$ refers to the particle phase velocity fluctuations.
\\Eq. \ref{eq:mito2} can be written through concentration-weighted average form as follows: 
\begin{equation}
\frac{\partial}{\partial y}\left(\langle c \rangle \langle v_x' v_y' \rangle_c\right) =\langle -F_x \rangle
\label{eq:mito3}
\end{equation}
\\ Which upon integration along wall-normal direction yields:
\begin{equation}
\langle c \rangle \langle v_x' v_y' \rangle_c =\int_{0}^{\delta}-\langle F_x(y)\rangle dy =\tau_{particle}
\label{eq:mito4}
\end{equation}
Here  $\langle  \rangle_c$ denotes concentration weighted average and $\tau_{particle}$ being the particle phase stress. }
\begin{figure}[!h]
\includegraphics[width=0.75\linewidth]{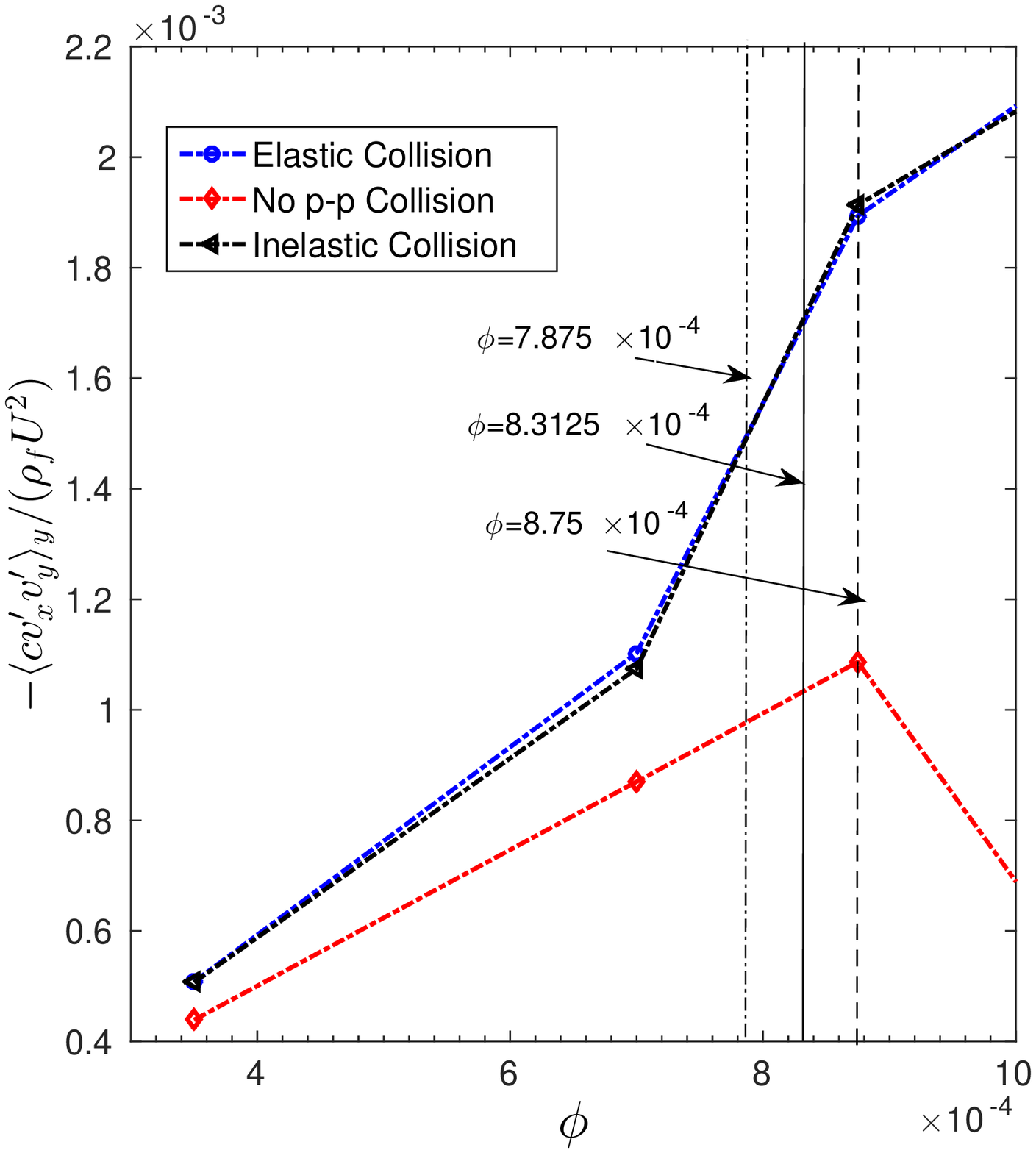}
\caption{Effect of nature of collision on spatial averaged particle stress at various volume fraction}
\label{fig:spat_av_part_re_stress}
\end{figure}
Figure \ref{fig:spat_av_part_re_stress} represents the effect of nature of collision on the variation of spatial average particle stress $\langle c v_x' v_y'\rangle_y$ term as a function of volume fraction $\phi$. Equation \ref{eq:mito4} has shown that the higher particle stress value is due to the higher value of the reverse force integrated over the half width $\delta$. Hence it is observed from the figure that the magnitude of the particle stress and thus the value of $\int_{0}^{\delta}-\langle F_x(y)dy \rangle$ is lowest when the inter-particle collision is switched off. This  causes an increase of $\phi_{cr}$ and delay in turbulence transition in absence of collision. \textcolor{black}{The $\phi_{cr}$ values for the cases are shown by the vertical lines: the dashed line represents $\phi_{cr}$ in absence of particles, the thick line represents the case of inelastic collision and the dashed-dotted line represents the critical volume faction under perfectly elastic collision.} This figure also reveals that above $\phi_{cr}$ the spatial averaged particle stress decreases rapidly. On the contrary, presence of inter-particle collisions maintain the increasing trend of the $\langle c v_x' v_y'\rangle_y$ term with $\phi$ even after turbulence transition.
\clearpage
\section{Conclusion}
This article \textcolor{black}{presents} a detail \textcolor{black}{description} of the fluid phase dynamics of particle-laden turbulent sheared suspension with volume fraction in the range of $\phi=1.75\times10^{-4}$ to $1.05\times10^{-3}$, higher than the cases discussed in our previous work \citep{ghosh2022statistical}. DNS with two-way coupling is used to study the turbulence modulation by the high inertial particles ($St\sim367$). 
\textcolor{black}{Unlike turbulent channel-flow, turbulent Couette flow is driven by the mean shear generated by the differential motion of the walls and hence there is no mean imposed pressure gradient. The total shear stress remains constant across the channel-width. In particle-laden turbulent shear flow the coherent structures, the fluid-particle interaction all differ from channel geometry.}    
Hence the turbulence modulation by high-inertial particles is expected be different. However, the sheared turbulent suspension is found to show a discontinuous \textcolor{black}{collapse} in turbulence, very similar to what has been observed in a vertical channel \citep{muramulla2020disruption}. A detailed analysis of momentum, mean K.E. and turbulent K.E. budget at different volume fractions reveals that the catastrophic decrease in shear production of turbulence plays the major role in the discontinuous transition. Moreover, step-wise particle injection and removal study confirms that the catastrophic decrease is found out to be due to the presence of particles only \textcolor{black}{and does not show any hysteresis}. The transition is also studied through the analysis of streamwise velocity and vorticity fluctuation field. The streaky structures of low-speed streamwise velocity fluctuations are found to be broken down in smaller structure of very small magnitude \textcolor{black}{at higher particle loading}. Particle concentration field is uncorrelated with fluid velocity and vorticity contours as expected due to the very high inertia. 
\\ The next part of the article is \textcolor{black}{focused in} exploring the \textcolor{black}{role} of inter-particle collision on the turbulence transition. In this regard, the ideal elastically colliding particle-laden turbulent suspension is compared with the suspension having inter-particle and wall-particle collisions slightly inelastic. Introduction of slight inelasticity in the inter-particle and wall-particle collision is found to increase the critical volume fraction marginally. The qualitative nature of the streamwise velocity and vorticity fluctuation field remaining similar to the elastic ideal collisional case. The explicit role of the inter-particle collisions is studied by simulating a hypothetical case where only inter-particle collisions are switched off. Here the transition do manifests itself by drastic reduction of streamwise, spanwise mean square velocities and Reynolds stress. On the other hand transition shows only a change in trend of streamwise mean square velocity, similar to that of particle phase instead of unladen fluid phase. This change is found out to be an effect of particle induced fluid turbulence where the timescale of streamwise Lagrangian velocity auto-correlation becomes similar to that of the particle phase. The streamwise velocity fluctuation field shows a unique behaviour, where velocity fluctuations of different magnitude arrange themselves in large parallel structures in x-z plane which gives a core-shell kind of appearance in the y-z plane where the high magnitude zones are present only at the central part. The particle concentration field and the spanwise velocity vorticity field is not found to be correlated, although, the particles prefer to concentrate near the wall where streamwise velocity fluctuation is very low. The absence of inter-particle collisions increases the critical volume fraction for the transition in turbulence as it shows marginally higher shear production of turbulence at the same volume fraction than the cases where the inter-particle collisions are switched on.

\clearpage
	\section{Appendix}
\subsection{Particle phase statistics in absence of inter-particle collisions}
\label{sec:Particle Phase Statistics zero_coll}
\begin{figure*}[!h]
			\includegraphics[width=1.0\textwidth,  height=9cm]{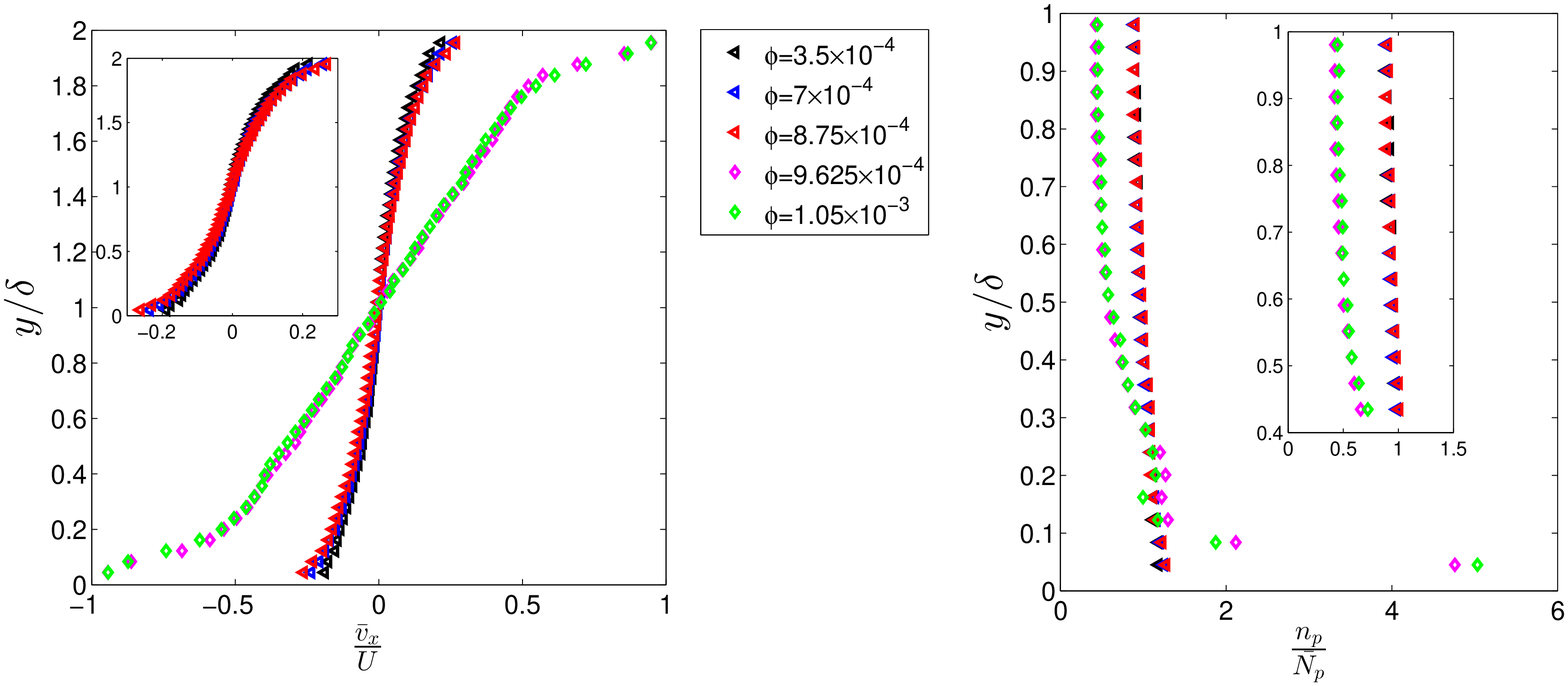}
			\caption{Effect of particle volume fraction on (a) mean velocity,(b) particle number concentration and of the particle phase in absence of inter-particle collisions} 
			\label{fig:particle_phase_stat_1_zero_coll}
		\end{figure*}

\begin{figure*}[!h]
			\includegraphics[width=1.0\textwidth,  height=9cm]{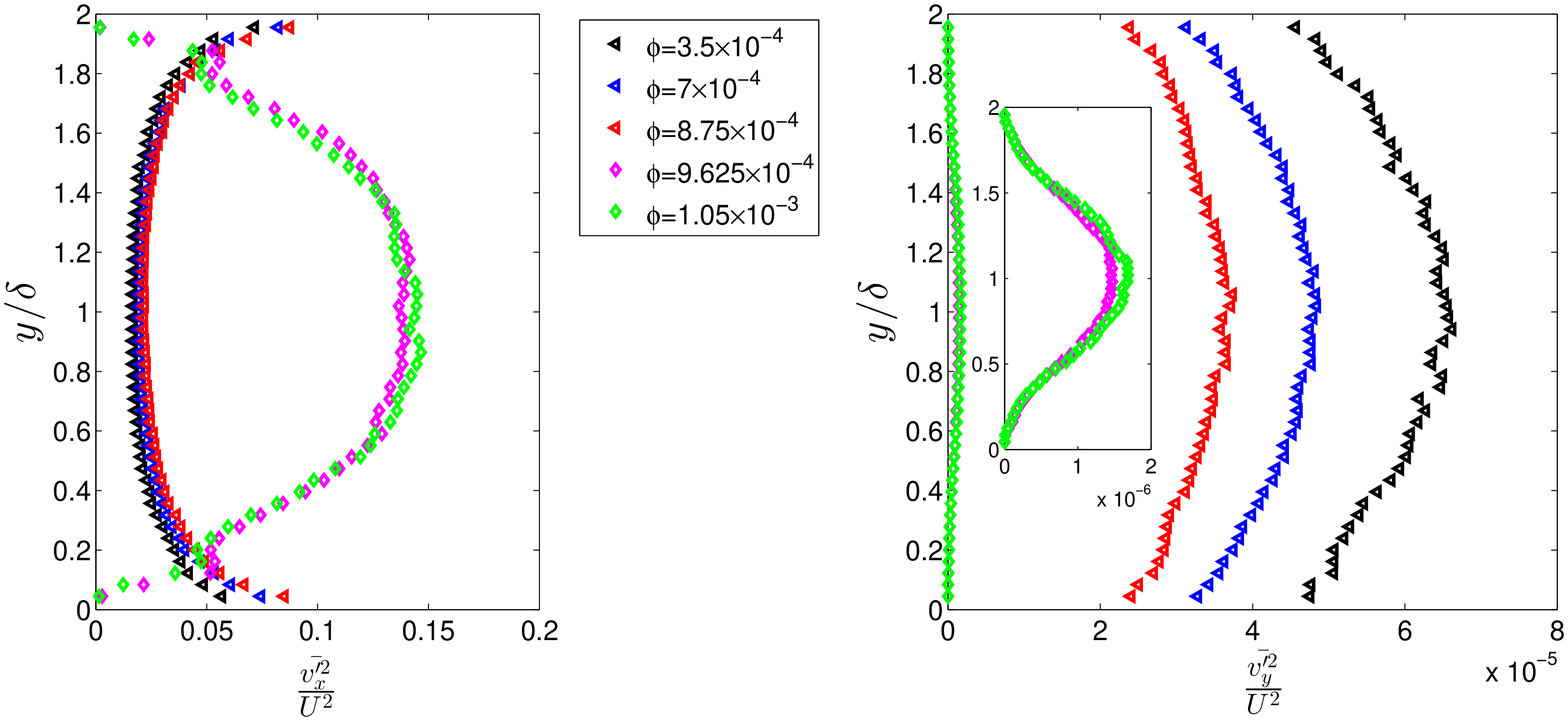}
			\caption{Effect of particle volume fraction on (a) stream-wise and (b) cross-stream mean square velocity of the particle phase in absence of inter-particle collisions}
			\label{fig:particle_phase_stat_2_zero_coll}
		\end{figure*}

The hypothetical study of switching off the inter-particle collisions, as discussed in section \ref{sec:zero_coll}, bring about a few interesting and different qualitative behaviour of particle phase statistics especially after the transition in turbulence. After the transition, unlike any cases discussed before, particle mean velocity do not show any slip at the walls along with a very high mean velocity gradient near the walls (Fig.\ref{fig:particle_phase_stat_1_zero_coll}(a)). The particles tend to accumulate more in the near-wall region for suspension above critical volume fraction, although very little variation of particle concentration is observed before transition (Fig. \ref{fig:particle_phase_stat_1_zero_coll}(b)). Due to the absence of inter-particle collisions the redistribution of particle x momentum to y and z direction gets drastically decreased. Hence we observe span-wise and cross-stream particle mean square velocities and covariance $\overline{v_x'v_y'}$ drastically decrease.  This decrease is about two-orders of magnitude for $\bar{v_y'^2}$ and $\bar{v_z'^2}$ and one-order of magnitude for $\overline{v_x'v_y'}$ with respect to the ideal collisional condition shown in section\label{sec:Particle Phase Statistics}. 
The most interesting trend is observed for $\bar{v_x'^2}$. Increase in $\bar{v_x'^2}$ values with a completely different trend, i.e. zero at the walls and maximum at the centre, occur after the transition in turbulence as a result of the  shear induced particle migration. Hence the different trends are found to be independent with small changes in volume fraction as well. \textcolor{black}{Following analysis suggests that the streamwise fluctuation of the particle phase in the absence of particle-particle collision originates from the wall normal migration of the particles under sheared velocity profile. In this mechanism, the magitude of streamwise velocity fluctuation ($v_x'$) induced, can be written as:
\begin{equation}
    v_x'\approx\sqrt{\overline{v_y'^2}}\tau_v \frac{d U}{d y}
\end{equation} 
For, $\phi=9.625\times10^{-4}$, at $y=\delta$, $\overline{v_x'^2}=1.468\times10^{-6}$; average mean velocity gradient $\frac{\partial U}{\partial y}\approx1.0$
and $\tau_v=367$ yields 
\begin{math}
v_x'\approx 0.448
\end{math}
or, 
\begin{math}
\overline{v_x'^2}\approx 0.19
\end{math}. 
This is of the same order of magnitude as observed from the Fig. \ref{fig:particle_phase_stat_2_zero_coll}(b) which is around 0.14.}

   Thus the the particle-phase statistics along with the fluid phase statistics take a very important role in understanding two-way coupled turbulent suspensions in absence of collisions. 

	\clearpage
		\begin{acknowledgments}
				We wish to acknowledge the financial support of SERB, DST, Government of INDIA. We would also like to thank Mr. N.S. Pradeep Muramalla for his invaluable technical inputs.  
			\end{acknowledgments}
		
		\section*{Data Availability Statement}
		The data that support the findings of this study are available
		from the corresponding author upon reasonable request.

\bibliography{Ghosh_goswami_arXiv3.bib}	

\begin{thebibliography}{27}%
\makeatletter
\providecommand \@ifxundefined [1]{%
 \@ifx{#1\undefined}
}%
\providecommand \@ifnum [1]{%
 \ifnum #1\expandafter \@firstoftwo
 \else \expandafter \@secondoftwo
 \fi
}%
\providecommand \@ifx [1]{%
 \ifx #1\expandafter \@firstoftwo
 \else \expandafter \@secondoftwo
 \fi
}%
\providecommand \natexlab [1]{#1}%
\providecommand \enquote  [1]{``#1''}%
\providecommand \bibnamefont  [1]{#1}%
\providecommand \bibfnamefont [1]{#1}%
\providecommand \citenamefont [1]{#1}%
\providecommand \href@noop [0]{\@secondoftwo}%
\providecommand \href [0]{\begingroup \@sanitize@url \@href}%
\providecommand \@href[1]{\@@startlink{#1}\@@href}%
\providecommand \@@href[1]{\endgroup#1\@@endlink}%
\providecommand \@sanitize@url [0]{\catcode `\\12\catcode `\$12\catcode
  `\&12\catcode `\#12\catcode `\^12\catcode `\_12\catcode `\%12\relax}%
\providecommand \@@startlink[1]{}%
\providecommand \@@endlink[0]{}%
\providecommand \url  [0]{\begingroup\@sanitize@url \@url }%
\providecommand \@url [1]{\endgroup\@href {#1}{\urlprefix }}%
\providecommand \urlprefix  [0]{URL }%
\providecommand \Eprint [0]{\href }%
\providecommand \doibase [0]{http://dx.doi.org/}%
\providecommand \selectlanguage [0]{\@gobble}%
\providecommand \bibinfo  [0]{\@secondoftwo}%
\providecommand \bibfield  [0]{\@secondoftwo}%
\providecommand \translation [1]{[#1]}%
\providecommand \BibitemOpen [0]{}%
\providecommand \bibitemStop [0]{}%
\providecommand \bibitemNoStop [0]{.\EOS\space}%
\providecommand \EOS [0]{\spacefactor3000\relax}%
\providecommand \BibitemShut  [1]{\csname bibitem#1\endcsname}%
\let\auto@bib@innerbib\@empty
\bibitem [{\citenamefont {Muramulla}\ \emph {et~al.}(2020)\citenamefont
  {Muramulla}, \citenamefont {Tyagi}, \citenamefont {Goswami},\ and\
  \citenamefont {Kumaran}}]{muramulla2020disruption}%
  \BibitemOpen
  \bibfield  {author} {\bibinfo {author} {\bibfnamefont {P.}~\bibnamefont
  {Muramulla}}, \bibinfo {author} {\bibfnamefont {A.}~\bibnamefont {Tyagi}},
  \bibinfo {author} {\bibfnamefont {P.}~\bibnamefont {Goswami}}, \ and\
  \bibinfo {author} {\bibfnamefont {V.}~\bibnamefont {Kumaran}},\ }\bibfield
  {title} {\enquote {\bibinfo {title} {Disruption of turbulence due to particle
  loading in a dilute gas--particle suspension},}\ }\href@noop {} {\bibfield
  {journal} {\bibinfo  {journal} {Journal of Fluid Mechanics}\ }\textbf
  {\bibinfo {volume} {889}} (\bibinfo {year} {2020})}\BibitemShut {NoStop}%
\bibitem [{\citenamefont {Marati}, \citenamefont {Casciola},\ and\
  \citenamefont {Piva}(2004)}]{marati2004energy}%
  \BibitemOpen
  \bibfield  {author} {\bibinfo {author} {\bibfnamefont {N.}~\bibnamefont
  {Marati}}, \bibinfo {author} {\bibfnamefont {C.}~\bibnamefont {Casciola}}, \
  and\ \bibinfo {author} {\bibfnamefont {R.}~\bibnamefont {Piva}},\ }\bibfield
  {title} {\enquote {\bibinfo {title} {Energy cascade and spatial fluxes in
  wall turbulence},}\ }\href@noop {} {\bibfield  {journal} {\bibinfo  {journal}
  {Journal of Fluid Mechanics}\ }\textbf {\bibinfo {volume} {521}},\ \bibinfo
  {pages} {191--215} (\bibinfo {year} {2004})}\BibitemShut {NoStop}%
\bibitem [{\citenamefont {Andrade}\ \emph {et~al.}(2018)\citenamefont
  {Andrade}, \citenamefont {Martins}, \citenamefont {Mompean}, \citenamefont
  {Thais},\ and\ \citenamefont {Gatski}}]{andrade2018analyzing}%
  \BibitemOpen
  \bibfield  {author} {\bibinfo {author} {\bibfnamefont {J.~R.}\ \bibnamefont
  {Andrade}}, \bibinfo {author} {\bibfnamefont {R.~S.}\ \bibnamefont
  {Martins}}, \bibinfo {author} {\bibfnamefont {G.}~\bibnamefont {Mompean}},
  \bibinfo {author} {\bibfnamefont {L.}~\bibnamefont {Thais}}, \ and\ \bibinfo
  {author} {\bibfnamefont {T.~B.}\ \bibnamefont {Gatski}},\ }\bibfield  {title}
  {\enquote {\bibinfo {title} {Analyzing the spectral energy cascade in
  turbulent channel flow},}\ }\href@noop {} {\bibfield  {journal} {\bibinfo
  {journal} {Physics of Fluids}\ }\textbf {\bibinfo {volume} {30}},\ \bibinfo
  {pages} {065110} (\bibinfo {year} {2018})}\BibitemShut {NoStop}%
\bibitem [{\citenamefont {Squires}\ and\ \citenamefont
  {Eaton}(1991)}]{squires1991preferential}%
  \BibitemOpen
  \bibfield  {author} {\bibinfo {author} {\bibfnamefont {K.~D.}\ \bibnamefont
  {Squires}}\ and\ \bibinfo {author} {\bibfnamefont {J.~K.}\ \bibnamefont
  {Eaton}},\ }\bibfield  {title} {\enquote {\bibinfo {title} {Preferential
  concentration of particles by turbulence},}\ }\href@noop {} {\bibfield
  {journal} {\bibinfo  {journal} {Physics of Fluids A: Fluid Dynamics}\
  }\textbf {\bibinfo {volume} {3}},\ \bibinfo {pages} {1169--1178} (\bibinfo
  {year} {1991})}\BibitemShut {NoStop}%
\bibitem [{\citenamefont {Fessler}, \citenamefont {Kulick},\ and\ \citenamefont
  {Eaton}(1994)}]{fessler1994preferential}%
  \BibitemOpen
  \bibfield  {author} {\bibinfo {author} {\bibfnamefont {J.~R.}\ \bibnamefont
  {Fessler}}, \bibinfo {author} {\bibfnamefont {J.~D.}\ \bibnamefont {Kulick}},
  \ and\ \bibinfo {author} {\bibfnamefont {J.~K.}\ \bibnamefont {Eaton}},\
  }\bibfield  {title} {\enquote {\bibinfo {title} {Preferential concentration
  of heavy particles in a turbulent channel flow},}\ }\href@noop {} {\bibfield
  {journal} {\bibinfo  {journal} {Physics of Fluids (1994-present)}\ }\textbf
  {\bibinfo {volume} {6}},\ \bibinfo {pages} {3742--3749} (\bibinfo {year}
  {1994})}\BibitemShut {NoStop}%
\bibitem [{\citenamefont {Pirozzoli}, \citenamefont {Bernardini},\ and\
  \citenamefont {Orlandi}(2011)}]{pirozzoli2011large}%
  \BibitemOpen
  \bibfield  {author} {\bibinfo {author} {\bibfnamefont {S.}~\bibnamefont
  {Pirozzoli}}, \bibinfo {author} {\bibfnamefont {M.}~\bibnamefont
  {Bernardini}}, \ and\ \bibinfo {author} {\bibfnamefont {P.}~\bibnamefont
  {Orlandi}},\ }\bibfield  {title} {\enquote {\bibinfo {title} {Large-scale
  motions and inner/outer layer interactions in turbulent couette--poiseuille
  flows},}\ }\href@noop {} {\bibfield  {journal} {\bibinfo  {journal} {Journal
  of fluid mechanics}\ }\textbf {\bibinfo {volume} {680}},\ \bibinfo {pages}
  {534--563} (\bibinfo {year} {2011})}\BibitemShut {NoStop}%
\bibitem [{\citenamefont {Wang}\ and\ \citenamefont
  {Richter}(2019)}]{wang2019modulation}%
  \BibitemOpen
  \bibfield  {author} {\bibinfo {author} {\bibfnamefont {G.}~\bibnamefont
  {Wang}}\ and\ \bibinfo {author} {\bibfnamefont {D.}~\bibnamefont {Richter}},\
  }\bibfield  {title} {\enquote {\bibinfo {title} {Modulation of the turbulence
  regeneration cycle by inertial particles in planar couette flow},}\
  }\href@noop {} {\bibfield  {journal} {\bibinfo  {journal} {Journal of Fluid
  Mechanics}\ }\textbf {\bibinfo {volume} {861}},\ \bibinfo {pages} {901--929}
  (\bibinfo {year} {2019})}\BibitemShut {NoStop}%
\bibitem [{\citenamefont {Yang}, \citenamefont {Zhao},\ and\ \citenamefont
  {Andersson}(2017)}]{yang2017preferential}%
  \BibitemOpen
  \bibfield  {author} {\bibinfo {author} {\bibfnamefont {K.}~\bibnamefont
  {Yang}}, \bibinfo {author} {\bibfnamefont {L.}~\bibnamefont {Zhao}}, \ and\
  \bibinfo {author} {\bibfnamefont {H.~I.}\ \bibnamefont {Andersson}},\
  }\bibfield  {title} {\enquote {\bibinfo {title} {Preferential particle
  concentration in wall-bounded turbulence with zero skin friction},}\
  }\href@noop {} {\bibfield  {journal} {\bibinfo  {journal} {Physics of
  Fluids}\ }\textbf {\bibinfo {volume} {29}},\ \bibinfo {pages} {113302}
  (\bibinfo {year} {2017})}\BibitemShut {NoStop}%
\bibitem [{\citenamefont {Goswami}\ and\ \citenamefont
  {Kumaran}(2010)}]{goswami2010particle1}%
  \BibitemOpen
  \bibfield  {author} {\bibinfo {author} {\bibfnamefont {P.~S.}\ \bibnamefont
  {Goswami}}\ and\ \bibinfo {author} {\bibfnamefont {V.}~\bibnamefont
  {Kumaran}},\ }\bibfield  {title} {\enquote {\bibinfo {title} {Particle
  dynamics in a turbulent particle--gas suspension at high stokes number. part
  1. velocity and acceleration distributions},}\ }\href@noop {} {\bibfield
  {journal} {\bibinfo  {journal} {Journal of Fluid Mechanics}\ }\textbf
  {\bibinfo {volume} {646}},\ \bibinfo {pages} {59--90} (\bibinfo {year}
  {2010})}\BibitemShut {NoStop}%
\bibitem [{\citenamefont {Muramulla}(2022)}]{Pradeep_2022}%
  \BibitemOpen
  \bibfield  {author} {\bibinfo {author} {\bibfnamefont {N.~S.~P.}\
  \bibnamefont {Muramulla}},\ }\emph {\bibinfo {title} {Turbulence modulation
  in particle-laden channel flow}},\ \href@noop {} {Ph.D. thesis},\ \bibinfo
  {school} {Dept. of Chem. Engg., IIT Bombay}, \bibinfo {address} {Maharashtra,
  India (based on personal communication)} (\bibinfo {year} {2022})\BibitemShut
  {NoStop}%
\bibitem [{\citenamefont {Ghosh}\ and\ \citenamefont
  {Goswami}(2022)}]{ghosh2022statistical}%
  \BibitemOpen
  \bibfield  {author} {\bibinfo {author} {\bibfnamefont {S.}~\bibnamefont
  {Ghosh}}\ and\ \bibinfo {author} {\bibfnamefont {P.}~\bibnamefont
  {Goswami}},\ }\bibfield  {title} {\enquote {\bibinfo {title} {A statistical
  analysis of velocity and acceleration fluctuations of inertial particles in
  particle-laden turbulent couette flow},}\ }\href@noop {} {\bibfield
  {journal} {\bibinfo  {journal} {Physics of Fluids}\ }\textbf {\bibinfo
  {volume} {34}},\ \bibinfo {pages} {015103} (\bibinfo {year}
  {2022})}\BibitemShut {NoStop}%
\bibitem [{\citenamefont {Kline}\ \emph {et~al.}(1967)\citenamefont {Kline},
  \citenamefont {Reynolds}, \citenamefont {Schraub},\ and\ \citenamefont
  {Runstadler}}]{kline1967structure}%
  \BibitemOpen
  \bibfield  {author} {\bibinfo {author} {\bibfnamefont {S.~J.}\ \bibnamefont
  {Kline}}, \bibinfo {author} {\bibfnamefont {W.~C.}\ \bibnamefont {Reynolds}},
  \bibinfo {author} {\bibfnamefont {F.}~\bibnamefont {Schraub}}, \ and\
  \bibinfo {author} {\bibfnamefont {P.}~\bibnamefont {Runstadler}},\ }\bibfield
   {title} {\enquote {\bibinfo {title} {The structure of turbulent boundary
  layers},}\ }\href@noop {} {\bibfield  {journal} {\bibinfo  {journal} {Journal
  of Fluid Mechanics}\ }\textbf {\bibinfo {volume} {30}},\ \bibinfo {pages}
  {741--773} (\bibinfo {year} {1967})}\BibitemShut {NoStop}%
\bibitem [{\citenamefont {Praturi}\ and\ \citenamefont
  {Brodkey}(1978)}]{praturi1978stereoscopic}%
  \BibitemOpen
  \bibfield  {author} {\bibinfo {author} {\bibfnamefont {A.~K.}\ \bibnamefont
  {Praturi}}\ and\ \bibinfo {author} {\bibfnamefont {R.~S.}\ \bibnamefont
  {Brodkey}},\ }\bibfield  {title} {\enquote {\bibinfo {title} {A stereoscopic
  visual study of coherent structures in turbulent shear flow},}\ }\href@noop
  {} {\bibfield  {journal} {\bibinfo  {journal} {Journal of Fluid Mechanics}\
  }\textbf {\bibinfo {volume} {89}},\ \bibinfo {pages} {251--272} (\bibinfo
  {year} {1978})}\BibitemShut {NoStop}%
\bibitem [{\citenamefont {Head}\ and\ \citenamefont
  {Bandyopadhyay}(1981)}]{head1981new}%
  \BibitemOpen
  \bibfield  {author} {\bibinfo {author} {\bibfnamefont {M.}~\bibnamefont
  {Head}}\ and\ \bibinfo {author} {\bibfnamefont {P.}~\bibnamefont
  {Bandyopadhyay}},\ }\bibfield  {title} {\enquote {\bibinfo {title} {New
  aspects of turbulent boundary-layer structure},}\ }\href@noop {} {\bibfield
  {journal} {\bibinfo  {journal} {Journal of fluid mechanics}\ }\textbf
  {\bibinfo {volume} {107}},\ \bibinfo {pages} {297--338} (\bibinfo {year}
  {1981})}\BibitemShut {NoStop}%
\bibitem [{\citenamefont {Sekimoto}, \citenamefont {Dong},\ and\ \citenamefont
  {Jim{\'e}nez}(2016)}]{sekimoto2016direct}%
  \BibitemOpen
  \bibfield  {author} {\bibinfo {author} {\bibfnamefont {A.}~\bibnamefont
  {Sekimoto}}, \bibinfo {author} {\bibfnamefont {S.}~\bibnamefont {Dong}}, \
  and\ \bibinfo {author} {\bibfnamefont {J.}~\bibnamefont {Jim{\'e}nez}},\
  }\bibfield  {title} {\enquote {\bibinfo {title} {Direct numerical simulation
  of statistically stationary and homogeneous shear turbulence and its relation
  to other shear flows},}\ }\href@noop {} {\bibfield  {journal} {\bibinfo
  {journal} {Physics of Fluids}\ }\textbf {\bibinfo {volume} {28}},\ \bibinfo
  {pages} {035101} (\bibinfo {year} {2016})}\BibitemShut {NoStop}%
\bibitem [{\citenamefont {Dong}\ \emph {et~al.}(2017)\citenamefont {Dong},
  \citenamefont {Lozano-Dur{\'a}n}, \citenamefont {Sekimoto},\ and\
  \citenamefont {Jim{\'e}nez}}]{dong2017coherent}%
  \BibitemOpen
  \bibfield  {author} {\bibinfo {author} {\bibfnamefont {S.}~\bibnamefont
  {Dong}}, \bibinfo {author} {\bibfnamefont {A.}~\bibnamefont
  {Lozano-Dur{\'a}n}}, \bibinfo {author} {\bibfnamefont {A.}~\bibnamefont
  {Sekimoto}}, \ and\ \bibinfo {author} {\bibfnamefont {J.}~\bibnamefont
  {Jim{\'e}nez}},\ }\bibfield  {title} {\enquote {\bibinfo {title} {Coherent
  structures in statistically stationary homogeneous shear turbulence},}\
  }\href@noop {} {\bibfield  {journal} {\bibinfo  {journal} {Journal of Fluid
  Mechanics}\ }\textbf {\bibinfo {volume} {816}},\ \bibinfo {pages} {167--208}
  (\bibinfo {year} {2017})}\BibitemShut {NoStop}%
\bibitem [{\citenamefont {Papavassiliou}\ and\ \citenamefont
  {Hanratty}(1997)}]{papavassiliou1997interpretation}%
  \BibitemOpen
  \bibfield  {author} {\bibinfo {author} {\bibfnamefont {D.~V.}\ \bibnamefont
  {Papavassiliou}}\ and\ \bibinfo {author} {\bibfnamefont {T.~J.}\ \bibnamefont
  {Hanratty}},\ }\bibfield  {title} {\enquote {\bibinfo {title} {Interpretation
  of large-scale structures observed in a turbulent plane couette flow},}\
  }\href@noop {} {\bibfield  {journal} {\bibinfo  {journal} {International
  journal of heat and fluid flow}\ }\textbf {\bibinfo {volume} {18}},\ \bibinfo
  {pages} {55--69} (\bibinfo {year} {1997})}\BibitemShut {NoStop}%
\bibitem [{\citenamefont {Tsukahara}, \citenamefont {Kawamura},\ and\
  \citenamefont {Shingai}(2006)}]{tsukahara2006dns}%
  \BibitemOpen
  \bibfield  {author} {\bibinfo {author} {\bibfnamefont {T.}~\bibnamefont
  {Tsukahara}}, \bibinfo {author} {\bibfnamefont {H.}~\bibnamefont {Kawamura}},
  \ and\ \bibinfo {author} {\bibfnamefont {K.}~\bibnamefont {Shingai}},\
  }\bibfield  {title} {\enquote {\bibinfo {title} {Dns of turbulent couette
  flow with emphasis on the large-scale structure in the core region},}\
  }\href@noop {} {\bibfield  {journal} {\bibinfo  {journal} {Journal of
  Turbulence}\ ,\ \bibinfo {pages} {N19}} (\bibinfo {year} {2006})}\BibitemShut
  {NoStop}%
\bibitem [{\citenamefont {Hwang}\ and\ \citenamefont
  {Cossu}(2010)}]{hwang2010amplification}%
  \BibitemOpen
  \bibfield  {author} {\bibinfo {author} {\bibfnamefont {Y.}~\bibnamefont
  {Hwang}}\ and\ \bibinfo {author} {\bibfnamefont {C.}~\bibnamefont {Cossu}},\
  }\bibfield  {title} {\enquote {\bibinfo {title} {Amplification of coherent
  streaks in the turbulent couette flow: an input--output analysis at low
  reynolds number},}\ }\href@noop {} {\bibfield  {journal} {\bibinfo  {journal}
  {Journal of Fluid Mechanics}\ }\textbf {\bibinfo {volume} {643}},\ \bibinfo
  {pages} {333--348} (\bibinfo {year} {2010})}\BibitemShut {NoStop}%
\bibitem [{\citenamefont {Hamilton}, \citenamefont {Kim},\ and\ \citenamefont
  {Waleffe}(1995)}]{hamilton1995regeneration}%
  \BibitemOpen
  \bibfield  {author} {\bibinfo {author} {\bibfnamefont {J.~M.}\ \bibnamefont
  {Hamilton}}, \bibinfo {author} {\bibfnamefont {J.}~\bibnamefont {Kim}}, \
  and\ \bibinfo {author} {\bibfnamefont {F.}~\bibnamefont {Waleffe}},\
  }\bibfield  {title} {\enquote {\bibinfo {title} {Regeneration mechanisms of
  near-wall turbulence structures},}\ }\href@noop {} {\bibfield  {journal}
  {\bibinfo  {journal} {Journal of Fluid Mechanics}\ }\textbf {\bibinfo
  {volume} {287}},\ \bibinfo {pages} {317--348} (\bibinfo {year}
  {1995})}\BibitemShut {NoStop}%
\bibitem [{\citenamefont {Richter}\ and\ \citenamefont
  {Sullivan}(2013)}]{richter2013momentum}%
  \BibitemOpen
  \bibfield  {author} {\bibinfo {author} {\bibfnamefont {D.~H.}\ \bibnamefont
  {Richter}}\ and\ \bibinfo {author} {\bibfnamefont {P.~P.}\ \bibnamefont
  {Sullivan}},\ }\bibfield  {title} {\enquote {\bibinfo {title} {Momentum
  transfer in a turbulent, particle-laden couette flow},}\ }\href@noop {}
  {\bibfield  {journal} {\bibinfo  {journal} {Physics of Fluids}\ }\textbf
  {\bibinfo {volume} {25}},\ \bibinfo {pages} {053304} (\bibinfo {year}
  {2013})}\BibitemShut {NoStop}%
\bibitem [{\citenamefont {Richter}\ and\ \citenamefont
  {Sullivan}(2014)}]{richter2014modification}%
  \BibitemOpen
  \bibfield  {author} {\bibinfo {author} {\bibfnamefont {D.~H.}\ \bibnamefont
  {Richter}}\ and\ \bibinfo {author} {\bibfnamefont {P.~P.}\ \bibnamefont
  {Sullivan}},\ }\bibfield  {title} {\enquote {\bibinfo {title} {Modification
  of near-wall coherent structures by inertial particles},}\ }\href@noop {}
  {\bibfield  {journal} {\bibinfo  {journal} {Physics of Fluids}\ }\textbf
  {\bibinfo {volume} {26}},\ \bibinfo {pages} {103304} (\bibinfo {year}
  {2014})}\BibitemShut {NoStop}%
\bibitem [{\citenamefont {Richter}(2015)}]{richter2015turbulence}%
  \BibitemOpen
  \bibfield  {author} {\bibinfo {author} {\bibfnamefont {D.~H.}\ \bibnamefont
  {Richter}},\ }\bibfield  {title} {\enquote {\bibinfo {title} {Turbulence
  modification by inertial particles and its influence on the spectral energy
  budget in planar couette flow},}\ }\href@noop {} {\bibfield  {journal}
  {\bibinfo  {journal} {Physics of Fluids}\ }\textbf {\bibinfo {volume} {27}},\
  \bibinfo {pages} {063304} (\bibinfo {year} {2015})}\BibitemShut {NoStop}%
\bibitem [{\citenamefont {Hinze}(1975)}]{Hinze}%
  \BibitemOpen
  \bibfield  {author} {\bibinfo {author} {\bibfnamefont {J.}~\bibnamefont
  {Hinze}},\ }\href@noop {} {\emph {\bibinfo {title} {Turbulence}}},\
  McGraw-Hill series in mechanical engineering\ (\bibinfo  {publisher} {New
  York : McGraw-Hill},\ \bibinfo {year} {1975})\BibitemShut {NoStop}%
\bibitem [{\citenamefont {Dennis}(2015)}]{dennis2015coherent}%
  \BibitemOpen
  \bibfield  {author} {\bibinfo {author} {\bibfnamefont {D.~J.}\ \bibnamefont
  {Dennis}},\ }\bibfield  {title} {\enquote {\bibinfo {title} {Coherent
  structures in wall-bounded turbulence},}\ }\href@noop {} {\bibfield
  {journal} {\bibinfo  {journal} {Anais da Academia Brasileira de
  Ci{\^e}ncias}\ }\textbf {\bibinfo {volume} {87}},\ \bibinfo {pages}
  {1161--1193} (\bibinfo {year} {2015})}\BibitemShut {NoStop}%
\bibitem [{\citenamefont {Zhao}, \citenamefont {Andersson},\ and\ \citenamefont
  {Gillissen}(2010)}]{zhao2010turbulence}%
  \BibitemOpen
  \bibfield  {author} {\bibinfo {author} {\bibfnamefont {L.}~\bibnamefont
  {Zhao}}, \bibinfo {author} {\bibfnamefont {H.~I.}\ \bibnamefont {Andersson}},
  \ and\ \bibinfo {author} {\bibfnamefont {J.}~\bibnamefont {Gillissen}},\
  }\bibfield  {title} {\enquote {\bibinfo {title} {Turbulence modulation and
  drag reduction by spherical particles},}\ }\href@noop {} {\bibfield
  {journal} {\bibinfo  {journal} {Phys. Fluids}\ }\textbf {\bibinfo {volume}
  {22}},\ \bibinfo {pages} {081702} (\bibinfo {year} {2010})}\BibitemShut
  {NoStop}%
\bibitem [{\citenamefont {Mito}\ and\ \citenamefont
  {Hanratty}(2006)}]{mito2006effect}%
  \BibitemOpen
  \bibfield  {author} {\bibinfo {author} {\bibfnamefont {Y.}~\bibnamefont
  {Mito}}\ and\ \bibinfo {author} {\bibfnamefont {T.~J.}\ \bibnamefont
  {Hanratty}},\ }\bibfield  {title} {\enquote {\bibinfo {title} {Effect of
  feedback and inter-particle collisions in an idealized gas--liquid annular
  flow},}\ }\href@noop {} {\bibfield  {journal} {\bibinfo  {journal}
  {International journal of multiphase flow}\ }\textbf {\bibinfo {volume}
  {32}},\ \bibinfo {pages} {692--716} (\bibinfo {year} {2006})}\BibitemShut
  {NoStop}%
\end{thebibliography}%
\end{document}